\documentclass[10pt]{elsarticle}
\journal{}
\usepackage{graphicx}
\usepackage{amsmath,amsfonts,amsthm,bm,mathtools} % Math packages
\usepackage{lipsum}
\usepackage{float}
\usepackage{pgfplots}
\usepackage{graphicx,amsmath,upgreek,bm,amssymb}
\usepackage{subcaption}
\usepackage{siunitx}
\usepackage{makecell}
\usepackage{booktabs}
\usepackage{multicol}
\usepackage{xcolor,colortbl}
\usepackage{geometry}
\usepackage{afterpage}
\usepackage{longtable}
\usepackage[tableposition=top]{caption}
\usepackage{multirow}

\usepackage[font=small,labelfont=bf,labelfont=sf,
justification=justified,
format=plain]{caption} 
\newlength\fwidth
\usepackage{natbib}
\definecolor{best_acc}{rgb}{0,0.5,0}
\definecolor{refcol}{rgb}{.25,0,1}
\definecolor{mygray}{gray}{0.85}
\definecolor{mycite}{gray}{0.0}
\definecolor{rev1}{rgb}{0,0,0}

\usepackage[colorlinks=blue]{hyperref}
\hypersetup{
	colorlinks=false,
	pdfborder={0 0 0},
}
\hypersetup{
	colorlinks=true,
	linkcolor=blue,
	citecolor= blue,
	filecolor=magenta,      
	urlcolor=blue,
}
\usepackage[misc]{ifsym}
\usepackage{fontawesome}
\definecolor{mygray}{gray}{0.85}
\usepackage{ragged2e}

\usepackage{todonotes}
\usepackage{setspace}
\usepackage{lineno}
\begin{document}
	%---------------------------------------------
	% 		      Authors 	
	%---------------------------------------------
	\begin{frontmatter}
		\title{
			A Modal Decomposition Approach to Topological Wave Propagation 
		}		
		\author{Joshua R.~Tempelman}
		%		\ead{jrt7@illinois.edu}
		\author{Alexander F.~Vakakis}
		%	\ead{avakakis@illinois.edu}
		\author{Kathryn H.~Matlack}
		%		\ead{kmatlack@illinois.edu}
		\address{Department of Mechanical Science and Engineering, University of Illinois at Urbana-Champaign, Urbana, IL, USA}
		
		%---------------------------------------------
		% 		    ABSTRACT 	
		%--------------------------------------------
		\begin{abstract}
			Mechanical topological insulators have attracted extensive attention over the past several years. 
			%%%%
			\textcolor{rev1}{
				The characteristics of topologically protected wave propagation is typically predicted via the band structure of the primitive unit cell, using Berry curvature to predict localized interface or boundary states (as well as their degree of localization), and the dispersion relation to predict propagating group velocity.}
			%%%%
			However, practical systems are finite in size and are driven by excitations with a finite bandwidth as guaranteed by the Fourier uncertainty principle. 
			Hence, the dynamics predicted by the ideal infinite systems driven at a given frequency deviate from practical topological systems driven across a finite bandwidth.
			In this work, we demonstrate that the propagating topological waves in a valley Hall system can be interpreted using the underlying linear degenerate modal basis.
			%%%%
			\textcolor{rev1}{
				We show that only a small subset of closely spaced modes with the appropriate phase differences comprise the topological waves, which enables the construction of analytical reduced-order models that accurately capture the topological wave. This result is related to the sparse density of the modal spectrum inside the topological band gap, which allows for the energy to be carried nearly completely by a narrow band of spectrally isolated eigenmodes that are spatially localized along the domain boundary.}
			%Furthermore, the relationships between the lattice contrast parameter, driving frequency, and driving bandwidth is presented.
			%These results are then employed to accurately predict the true group velocity of the topological wave with respect to input signal bandwidth and frequency. 
			This finding is subsequently employed to predict the true group velocity of propagating topological waves with respect to input signal bandwidth and frequency by matching each active mode to corresponding band locations of the supercell dispersion diagram describing the semi-infinite system.
			Moreover, we demonstrate that damped topological waves may be studied within the framework of modal superposition by utilizing the damped modal spectrum to characterize the variation topological wave group velocity and predict edge-to-bulk transitions. Hence,  this work establishes a framework for accurately characterizing undamped and damped topological wave propagation in quantum valley Hall systems using the machinery of classical dynamics.
		\end{abstract}
		
	\end{frontmatter}
	%---------------------------------------------

	% ---------------------------------------------------
	\section{Introduction} 
	\label{sec:introduction}
	% ---------------------------------------------------
	
	The adaptation of topological band theories to classical systems has opened the door to new paradigms of exotic wave behavior in mechanical and acoustical systems~\cite{Ma2019}. Originally envisioned for quantum mechanical systems which obey the Schr\"{o}dinger equation~\cite{Hasan2010}, the underlying band topology that enables topological wave phenomena can be extrapolated to classical systems with the same effective Hamiltonian symmetries~\cite{Ma2019,Suesstrunk2016}. 
	This realization has enabled rich topological phenomena to be realized in classical dynamical systems, such as systems governed by magnetic~\cite{Vila2019}, elastic~\cite{Pal2017,Chen2018,Vila2017,Huo2021}, acoustic~\cite{Yang2015,Kliewer2021,Fleury2016},  and gyroscopic~\cite{Suesstrunk2015,Wang2015,Nash2015} laws. 
	%%%
	While classical domains are commonly used as a proving ground for quantum-based band topological band theories, there are emergent applications of these phenomena in classical mechanics such as  topologically protected wave guides~\cite{Miniaci2018,Babaa2020}, programmable mechanical logic gates~\cite{Xia2018}, wave steering~\cite{Sirota2021,Tian2020}, signal splitting~\cite{Zhuo2021}, and signal processing~\cite{ZangenehNejad2019}.

	Topologically insulated waves in the classical domain manifest as either standing or traveling waves, depending on the nature of the lattice symmetries. Stationary insulators include interface and edge modes of 1-D systems based on the \textcolor{rev1}{Su-Schrieffer-Heeger} chain~\cite{Su1979,Yin2018,Tempelman2021,Hu2022} or  higher-order corner states of 2- and 3-D systems~\cite{Chen2021a,SerraGarcia2018}, whereas traveling-wave insulators are most commonly enabled in 2D hexagonal lattices with the insulated wave traveling across edges or interfaces~\cite{Chen2018,Miniaci2018,Wang2015,Pal2017,Chen2021}.
	In standing wave insulators, the topologically insulated wave corresponds to a single stationary vibrational mode of the finite system. 
	This finite mode is considered as the product of bulk-boundary correspondence and therefore reaps the benefits of topological band theory such as robustness to perturbations. 
	Indeed, this has been an important aspect that researchers have utilized to study the finite system, especially when mechanical nonlinearity is at play and nonlinear normal mode analysis is applicable~\cite{Tempelman2021,Rosa2022,Chaunsali2021,Pal2018}.
	In contrast, traveling wave insulators are commonly enabled in classical systems by leveraging the quantum valley Hall (QVH)~\cite{Pal2017}, quantum spin Hall (QSH)~\cite{Suesstrunk2015}, quantum anomalous Hall (QAH)~\cite{Wang2015} effect, \textcolor{rev1}{or even a combination of the QVH and QSH effects~\cite{Miniaci2019,Wang2019}}, all of which correspond to different band topologies which are measured by the Berry curvature across closed contours of the reciprocal space~\cite{Ma2019}.  
	Unlike stationary topological modes by their very nature, traveling topological waves cannot be regarded as single modes.
	
	%%%%%%%%%%%%%%%%%%%%%%%%%%%%%%%
	% ADDING REFS TO OTHER WORKS
	%%%%%%%%%%%%%%%%%%%%%%%%%%%%%%%
	\textcolor{rev1}{Herein we will focus exclusively on the QVH effect. We refer the reader to~\cite{ZangenehNejad2020} for descriptions of other topological phenomena, and to~\cite{Miniaci2021} for a particularly relevant tutorial regarding their implementation in mechanical systems.}
	The QVH effect was originally discovered in electronic graphene~\cite{Neto2009} and manifests hexagonal lattices with broken inversion symmetry and preserved $\mathcal{C}_3$ symmetry~\cite{Pal2017}. 
	%%%%%%%%%%%%%%%%%%%%%
	% To enable the QVH phenomena in an elastic system, Pal and Ruzzene configured a mechanical analog to graphene with the necessary dispersion characteristics in the effective Hamiltonian~\cite{Pal2017}.
	%%%%%%%%%%%%%%%%%%%%%%%%%%
	% SHORTEN UP TOPOLOGY INTRO
	%%%%%%%%%%%%%%%%%%%%%%%%%%
	\textcolor{rev1}{
		The key band phenomenon is a Dirac cone which forms at the $K/K'$ points of the reciprocal space. 
		Opening this degeneracy results in time-reversed modes at the $K$ and $K'$ valleys which carry opposite Berry curvature. Accordingly, locally integrating the Berry Curvature around the $K/K'$ regions of the band structure~\cite{Berry1984,Thouless1982}; this recovers a quantity known as the valley Chern number~\cite{ZangenehNejad2020}.}
	A non-zero valley Chern number ensures that two lattices with inverted symmetry form a topologically protected wave at the interface which is immune to backscatter and lattice perturbations~\cite{Rudner2013}.
	
	The unique topological characteristics of QVH systems has recently instigated an interest in topological-based wave control. For instance, predicting and controlling the wave propagation along QVH interfaces can enable applications such as wave field focusing~\cite{Zhuo2021} and route reconfiguration~\cite{Tian2020}.
	Moreover, there has been an emergence of interest in the optimization of these systems from a perspective of energy localization~\cite{Du2020,Chen2021,Christiansen2019} and effective topological bandwidth~\cite{Du2020,Zhang2020,Ma2021}.
	%%%%%%%%%%%%%%%%%%%%%%%%%%%%%%%%%%%%%%%%%%%%%%%%%%
	However, these studies rely on the dispersion characteristics of ideal infinite systems to determine  the performance criteria.
	%%%%%%%%%%%%%%%%%%%%%%%%%%%%%%%%%%%%%%%%%%%%%%%%%%%%%%%%%%
	In practice, topological systems are finite in nature, and there is certainly a degree of error introduced due to the truncation of the infinite system.
	%%%%%%%%%%%%%%%%%%%%%%%%%%%%%%%%%%%%%%%%%%%%%%%%%%%%%%%%%%
	Moreover, these approaches consider a single point in frequency on the dispersion curves. 
	For traveling wave systems, however, the spectrum of the excitation cannot be captured by a single frequency since a wave packet typically excites the topological wave. 
	%%%%%%%%%%%%%%%%%%%%%%%%%%%%%%%%%%%%%%%%%%%%%%%%%%%%%%%%%%
	Wave packet excitations are proportional to a Gaussian curve in Fourier space with a finite bandwidth due to the Fourier uncertainty principle, and this will undoubtedly cause theoretical predictions based on a single point of the infinite band behavior to deviate when the finite system is analyzed.
	%%%%%%%%%%%%%%%%%%%%%%%%%%%%%%%%%%%%%%%%%%%%%%%%%%%%%%%%%%
	\textcolor{rev1}{
		Evaluating the practical performance of a finite topological insulator subject to finite bandwidth excitations typically requires direct numerical integration of a large system which imposes a great computational constraint on the design and tuning of insulators.}
	%%%%%%%%%%%%%%%%%%%%%%%%%%%%%%%%%%%%%%%%%%%%%%%%%%%%%%%%%%
	% CHALLENGES OF USING DIRECT SIMULATIONS
	%%%%%%%%%%%%%%%%%%%%%%%%%%%%%%%%%%%%%%%%%%%%%%%%%%%%%%%%%%
	\textcolor{rev1}{However, the finite modal spectrum of any phononic system can be mapped directly to the infinite dispersion relation~\cite{Miniaci2021}. Considering that the predicted \textit{spatially localized} topological modes of QVH systems exist in the band-gap and are therefore also \textit{spectrally localized}, a reasonable hypothesis is that the topological wave propagation can be described accurately by relatively few eigenmodes of the finite system located in the band gap and with mode shapes describing the spatial extent of the insulators domain boundaries, even for relatively broad-band excitations.
		%%%%%%%%%%%%%%%%%%%%%%%%%%%%%%%%%%%%%%%%%%%%%%%%%%%%%%%%%%
		Hence, it is logical to look at the modal spectrum of the finite system to uncover the dynamics from a perspective of modal superposition and gain insight to topological wave propagation of the finite system with a computationally inexpensive reduced order analytical solution.}
	%%%%%%%%%%%%%%%%%%%%%%%%%%%%%%%%%%%%%%%%%%%%%%%%%%%%%%%%%%
	To this point, the authors are unaware of any such studies in the current literature.
	%%%%%%%%%%%%%%%%%%%%%%%%%%%%%%%%%%%%%%%%%%%%%%%%%%%%%%%%%%
	
	The underlying modal spectrum may offer valuable insights for the case of damped topological wave propagation as well.
	\textcolor{rev1}{Whereas the presence of damping violates the required Hamiltonian symmetries for insulation in quantum systems~\cite{Zeuner2015}, its effects are inherently present in any practical mechanical topological insulator~\cite{Suesstrunk2015,Vila2017,Zhu2018}.}
	Moreover, while it has been shown that the mechanical analog of a damped vibrational lattice preserves gauge invariance and allows for robust edge states to exist over a damped spectrum in QAH systems~\cite{Xiong2016}, little quantitative analysis has been presented regarding the practical effects of damping in QVH systems, as it is typically interpreted in terms of energy transmission loss~\cite{Pal2017}.
	Because  solutions for proportionally damped systems may be written as a superimposition of exponentially decaying eigenmodes, the employment of the modal basis may naturally be extended to the damped system. This may in turn be used to explain the transmission loss and delocalization in damped topological wave travel through the lens damped modal spectrum.
	%%
	%However, solutions for proportionally damped systems may be written as a superimposition of exponentially decaying eigenmodes. Hence, the relative energies of the underlying modal basis may offer valuable insight to the evolution of topological wave propagation in a damped system 
	%the transmission loss and even mode delocaliation of damped topological waves can be explained in terms of the instantaneous relative energies of edge versus bulk eigenmodes. 
	%The effect of damping on topological wave propagation must be considered also, as this is inherently present in any practical mechanical topological insulator~\cite{Suesstrunk2015,Vila2017,Zhu2018}.
	%Whereas the presence of damping is violates the required Hamiltonian symmetries for insulation in quantum systems~\cite{Zeuner2015}, it has been shown that the mechanical analog of a damped vibrational lattice preserves gauge invariance and allows for robust edge states to exist over a damped spectrum in QAH systems~\cite{Xiong2016}. However, little quantitative analysis has been presented regarding the practical effects of damping in QVH systems, as it is typically interpreted in terms of energy transmission loss~\cite{Pal2017}.
	
	In this work, we propose an approach for understanding QVH wave propagation in terms of the modal basis of the finite system. 
	It is well-understood in classical dynamics that the response of any linear, time-invariant mechanical  system can be exactly represented as a superposition of linearly independent modal oscillators by invoking the modal expansion theorem.
	%%%%%%%%%%%%%%%%%%%%%%%%%%%%%%%
	% CLARIFYING PHASE DIFFERENCES
	%%%%%%%%%%%%%%%%%%%%%%%%%%%%%%%
	\textcolor{rev1}{
		We leverage this result to study a QVH system from the context of classical dynamics. Namely, we demonstrate that topological wave propagation is comprised of the superimposition of a limited subset of closely spaced modes inside of the band gap, possessing the appropriate phase differences to produce localized topological propagating wave packets across domain boundaries.}
	%%%%%%%%%%%%%%%%%%%%%%%%%%%%%%%
	% COMPARISON TO TRIVIAL WAVES
	%%%%%%%%%%%%%%%%%%%%%%%%%%%%%%%
	\textcolor{rev1}{This is shown to hold true even for systems with complex (zigzagged) wave paths, which is a result of the sparsity of the modal spectrum within the topological band gap, as well as the immunity to backscatter around corners as the wave packet propagates.} 
	%\textcolor{rev1}{We demonstrate that this is in contradiction to bulk wave propagation and to localized wave propagation about zigzagged trivial waveguides, which require many more modes to accurately reconstruct the traveling wave.}
	%%%%%%%%%%%%%%%%%%%%%%%%%%%%%%%
	Next, we formulate a solution based on modal-expansion to compute group velocity in finite systems driven by finite bandwidth inputs in a method analogous to the Normal Mode Expansion Technique~\cite{Auld1971,Ditri1994,Auld}.
	Lastly, we show that the inequivalence of damping rates between the modal coordinates of the finite system causes a transition of the instantaneous relative energy from edge states to bulk states that is predictable by the modal damping spectrum of the finite system.
	
	Accordingly, the organization of this paper is as follows. Section~\ref{sec:Desc} provides a description of the band topology as well as the discrete, finite dynamical systems considered in this work. The modal decomposition of topological waves and analytical solution is given in Section~\ref{Subsec: Modal Decomp}. Section~\ref{Sec:VG} discusses the issue of estimating the group velocity of propagating topological waves in finite QVH lattices and presents a ``mode-band matching" technique to better predict this quantity based on modal expansion. The effects of damping are discussed in section~\ref{Sec:Damping}. Lastly, section~\ref{Sec:Conclusion} provides concluding remarks.

	% ---------------------------------------------------	
	\section{Modal decomposition of propagating topological wave} 			
	\label{sec:Desc} 										
	% ---------------------------------------------------	
	\textcolor{black}{We begin with a discrete mechanical graphene model which is known to have a nontrivial valley Chern number. Analytical solutions for its response based on the well known modal expansion theorem are then presented and confirmed  with direct numerical solutions. This result is then used to show that the propagating topological  wave can be expressed in terms of a modal superimposition of closely spaced modes with nontrivial phase difference.
		\textcolor{rev1}{
			This framework is then extended to explore the relationship of interface energy concentration and modal composition of the propagating wave with respect to the frequency and bandwidth of the excitation source.}}
	% ---------------------------------------------------
	\begin{figure}[t!]
		\centering
		\includegraphics[width=\textwidth]{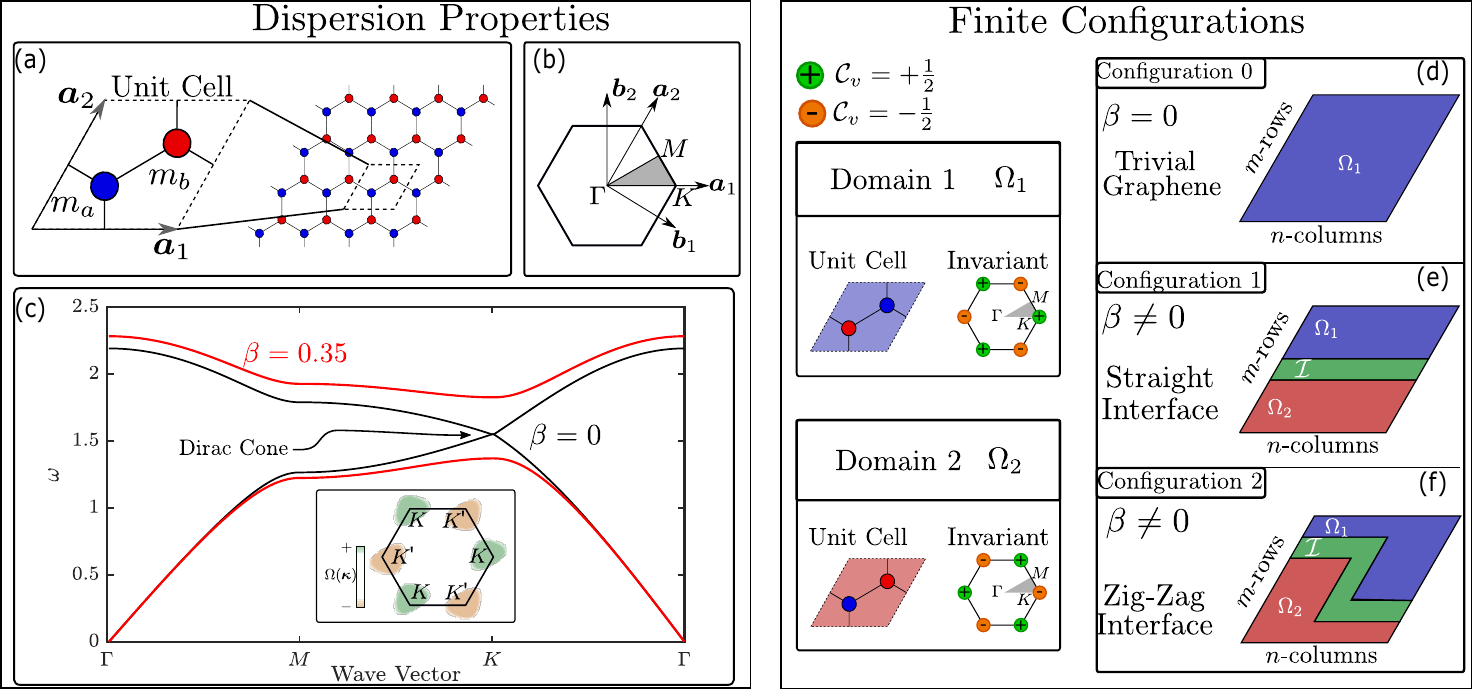}
		\caption{The system (linear phononic structure) and corresponding band topology.. (a) Unit cell depiction showing the basis vectors as well as a tessellation demonstrating a finite system. (b) The irreducible Brillouin zone shaded in gray with high symmetry points $\Gamma$-$M$-$K$-$\Gamma$ labeled and reciprocal basis vectors $\bm{b}_{1,2}$. (c) The band structure depicting the dispersion curves for $\beta = 0$ in black and $\beta = 0.35$ in red with the Berry curvature at the $K/K'$ points depicted in the inset. \textcolor{rev1}{Schematics of (d) the trivial finite graphene model and (e,f)  the two topological configurations each incorporating two distinct sub-domains with opposite valley Chern numbers.}}
		\label{FIG:SysDescription}
	\end{figure}
	
	% ---------------------------------------------------	
	\subsection{System Description}							
	% ---------------------------------------------------	
	We consider the di-atomic hexagonal unit cell that is tessellated into the periodic phononic system discussed in~\cite{Pal2017} (Fig~\ref{FIG:SysDescription}(a)). 
	The stiffness is assumed to be unity for all springs, $k = 1$, and the mass components $m_a$ and $m_b$ are related by the nominal mass $m = 1.25$ and contrast parameter $\beta$ as $m_a = m+\beta$ and $m_b = m-\beta$. %The position of each unit cell in the tesselated sytstem can be related by the basis vectors $\bm{a}_1$ and $\bm{a}_2$ as $\bm{r}^{p,q} = p\bm{a}_1 + q\bm{a}_2$
	The equations of motion for an interior unit cell located in the $p-th$ row $q-th$ column for the cell components $u_a^{p,q}$ and $u_b^{p,q}$ read,
	\begin{equation}
		\begin{aligned}
			&\ddot{u}_a^{p,q}=\frac{k}{m_a}\left(u_b^{p,q} + u_b^{p,q-1} + u_b^{p-1,q} -3u_a^{p,q}\right)\\
			&\ddot{u}_b^{p,q}=\frac{k}{m_b}\left(u_a^{p,q} + u_a^{p,q+1} + u_a^{p+1,q} -3u_b^{p,q}\right),
		\end{aligned}
		\label{EQ:eom}
	\end{equation}
	where a $\dot{(\ \ )}$ denotes a time derivative.
	The dispersion relation for the infinite system is obtained by assuming a Bloch wave ansatz
	$\bm{u}(\bm{x},\bm{\kappa};t)= \tilde{\bm{u}}(\bm{x},\bm{\kappa})e^{-i(\omega t-\bm{\kappa}^{\intercal}\bm{r} )}$ where $\bm{r} = p\bm{a}_1 + q\bm{a}_2$ denotes the position of cell ($p$,$q$) with respect to the basis vectors $\bm{a}_1 = [1 \ 0]^{\intercal}$ and $\bm{a}_2 = [\cos\frac{\pi}{3} \ \sin\frac{\pi}{3}]^{\intercal}$ and $\bm{\kappa} = \kappa_1\bm{b}_1 + \kappa_2\bm{b}_2$ is the wave vector in terms of the reciprocal basis vectors $\bm{b}_1$ and $\bm{b}_2$ (Fig~\ref{FIG:SysDescription}(b)). 
	Inserting the Bloch ansatz into Eq~\eqref{EQ:eom} yields two dispersion surfaces which can be traced along the $\Gamma$-$M$-$K$-$\Gamma$ boundary of the irreducible Brillouin Zone (IBZ) (Fig~\ref{FIG:SysDescription}(c)). 
	
	The dispersion relation encodes the underlying band topology which predicts whether or not protected interface waves will manifest in the finite system. 
	In the Graphene model, the topology is locally observed near the Dirac cone (e.g., the $K$/$K$' point, Fig~\ref{FIG:SysDescription}(c)). 
	%%%%%%%%%%%%
	Low energy excitations occur at the $K/K'$ points due to the contrast parameter $\beta$ which is referred to as the pseudo-spin degree of freedom; manipulation of this degree of freedom opens a topologically nontrivial band gap~\cite{Xin2020}.
	%%%%%%%%%%%%
	When $\beta\neq0$, the inversion symmetry is broken while the $\mathcal{C}_3$ symmetry is preserved. 
	The breaking of inversion symmetry causes the Dirac cone to open and for Berry curvature of opposite sign to locally accumulate at the $K$ and $K'$ valleys, leading to a locally defined topological invariant.
	The invariant is mathematically described in terms of the Berry connection,
	%%%%%%%%%%%%%%%%
	$\mathcal{A}(\bm{\kappa)} = -i\langle \tilde{\bm{u}}_n|\nabla_{\bm{\kappa}}|\tilde{\bm{u}}_n^\dagger\rangle$.
	%%%%%%%%%%%%%%%%
	Locally integrating the Berry connection over the $K$ and $K'$ valleys returns the valley Chern number, 
	\begin{eqnarray}
		2\pi \mathcal{C}_v
		= \int_{\upsilon} \nabla\times\mathcal{A}(\bm{\kappa})\rm{d}\bm{\kappa}
		=  \int_{\upsilon}\Omega(\bm{\kappa})\rm{d}\bm{\kappa}
	\end{eqnarray}
	%%%%%%%%%%%%%%
	where $\Omega(\bm{\kappa}) = \partial_{\kappa_x}\mathcal{A}_y - \partial_{\kappa_y}\mathcal{A}_x$ is the Berry curvature (Fig~\ref{FIG:SysDescription}(c)).
	If the domain of integration $\upsilon$ is taken over the entire IBZ, then $\mathcal{C}_{v} = 0$. 
	However, $ \mathcal{C}_v=+(-)\frac{1}{2}$ when locally integrated around the $K(K')$ points of reciprocal space. 
	%%%%%%%%%%%%%%%%%%%%%%%%
	If two systems are interfaced with opposite Berry curvature (e.g., inverted symmetry), then an integer invariant exists between the two systems and the bulk-boundary correspondence principle ensures that a topological wave exists at the interface~\cite{Rudner2013}.
	%%%%%%%%%%%%%%%

	%%%%%%%%%%%%%%%
	% DISCUSSING THE FINITE MODELS
	%%%%%%%%%%%%%%%
	\textcolor{rev1}{We consider two configurations of systems that satisfy the bulk-boundary-correspondence principle.
		Namely, the straight-edge interface depicted in Fig~\ref{FIG:SysDescription}(e) which will be referred to as configuration 1, and the zigzagged interfaced depicted in Fig~\ref{FIG:SysDescription}(f) which will be referred to as configuration 2. }
	\textcolor{rev1}{
		The reason for considering both configurations is to demonstrate that the modal decomposition approach presented herein works even for complex domain boundaries, as a hallmark trait of QVH wave propagation is the ability to traverse corners without backscatter. 
		%%%
		In both configurations, two domains are denoted by $\Omega_1$ and $\Omega_2$ which correspond to lattices with inverted symmetry and opposite topological character, and the interface region (denoted $\mathcal{I}$) where the two domains meet; this is the region where the topological wave is realized which is also referred to as the domain boundary. The entire domain of the study is taken as $\mathcal{D} = \Omega_1\cup\Omega_2\cup\mathcal{I}$.
		All exterior boundaries are grounded with a stiffness $k$.
		An additional configuration possessing a double zigzag domain boundary is discussed in~\ref{APX: doubleZigZag} to further confirm indifference to wave path, but is not discussed herein for brevity.} 
	\textcolor{rev1}{Moreover, we also consider a trivial graphene model depicted in Fig~\ref{FIG:SysDescription}(d)  with no domain boundary  and with $\beta = 0$ in order to juxtapose bulk wave propagation to topological wave propagation in the context of the methods discussed in section~\ref{Subsec: Modal Decomp}.}
	%%
	% ---------------------------------------------------
	
	% ---------------------------------------------------
	\subsection{Finite Lattice Modal Decomposition}
	\label{Subsec: Modal Decomp}
	% ---------------------------------------------------
	The finite lattice system is comprised of $m$ rows by $n$ columns with a total of $N =2mn$ degrees of freedom. 
	We leave the nominal contrast parameter as $\beta = 0.3$ unless otherwise stated. 
	The equations of motion for all oscillators can be written in matrix form as follows,
	%%%%%%%%%%%%%%%%%%%%%%%%%%%%%%%%%%%%%%%%%%%%%
	\begin{equation}
		\textbf{M}\ddot{\bm{u}} + \textbf{K}\bm{u}
		= \bm{f}(t),
		\label{EQ:EOM_standard}
	\end{equation}
	%%%%%%%%%%%%%%%%%%%%%%%%%%%%%%%%%%%%%%%%%%%%%
	where $\textbf{K}$ is the stiffness matrix, $\textbf{M}$ the mass matrix, $\bm{u}$ the column vector of displacements, and $\bm{f}$ a non-homogeneous forcing term.
	We consider a windowed tone burst excitation with center frequency $\Omega$ at the oscillator site $\xi$ which is selected as an exterior element of $\mathcal{I}$,
	%%%%%%%%%%%%%%%%%%%%%%%%%%%%%%%%%%%%%%%%%%%%%
	\begin{equation}
		{f}_i(t) = 	\delta_{i\xi}A
		\left[H(t)-H\left(t-\frac{2\pi}{\Omega}N_{cyc}\right)\right]
		\left[1-\cos\left(\frac{\Omega t}{N_{cyc}}\right)\right]
		\sin(\Omega t),
		\label{EQ:forcing}
	\end{equation}
	%%%%%%%%%%%%%%%%%%%%%%%%%%%%%%%%%%%%%%%%%%%%%
	where $H(\cdot)$ denotes the heaviside function, $A$ is the amplitude, $\Omega$ the driving frequency, and $N_{cyc}$ the (finite) number of excitation periods in the window.
	%%%%%%%%%%%%%
	%Direct numerical simulation of system~\eqref{EQ:EOM_standard} allows for straightforward confirmation of the topological wave at the band-gap frequency of $\Omega = 1.62$ (Fig~\ref{FIG:modal_decomp}(a)) which excites the topological wave~\cite{Pal2017}. 
	%The frequency of $\Omega = 1.5$ corresponds to a band gap frequency and a non-zero slope of the topological branch of the infinite-strip dispersion~\cite{Pal2017}.
	
	The propagating topological wave can be decomposed into the system's eigenmodes via the modal transformation matrix which is derived from the governing eigenvalue problem,
	%%%%%%%%%%%%%%%%%%%%%%%%%%%%%%%%%%%%%%%%%%%%%
	\begin{equation}
		\left[\textbf{K}-\omega^2\textbf{M}\right]\tilde{\bm{\varphi}} = \bm{0},\ \ \
		\bm{\Phi} =\begin{bmatrix}
			\vline&&\vline\\
			{\bm{\varphi}}_1&\dots &	{\bm{\varphi}}_n\\
			\vline&&\vline\\
		\end{bmatrix} \in \mathbb{R}^{N \times N}
		\label{EQ:eval_prob}
	\end{equation}
	%%%%%%%%%%%%%%%%%%%%%%%%%%%%%%%%%%%%%%%%%%%%%
	\textcolor{rev1}{
		where $\bm{\varphi} = \left(\tilde{\bm{\varphi}}_i^{\intercal}\textbf{M}\tilde{\bm{\varphi}}_i\right)^{-1/2}\tilde{\bm{\varphi}}_i$ ensures that the orthornamility conditions $\bm{\varphi}_i^{\intercal}\textbf{M}\bm{\varphi}_i = 1$ and  $\bm{\varphi}_i^{\intercal}\textbf{K}\bm{\varphi}_i = \omega_i^2$ are satisfied.}
	The modal responses are obtained by the linear transformation $\bm{u}=\bm{\Phi}\bm{\eta}$ which converts the solutions of system~\eqref{EQ:EOM_standard} into a modal basis. 
	%%%%%%%%%%%%%%%%%%%%%%
	%%%%%%%%%%%%%%%%%%%%%%
	% In deriving (4) the orthonormality relations satisfied by the modes of the system were taken into account.
	%%%%%%%%%%%%%%%%%%%%%%
	%%%%%%%%%%%%%%%%%%%%%%
	Moreover, the modal transformations may be directly inserted into eq~\eqref{EQ:EOM_standard} and premultiplied by $\bm{\Phi}^{\intercal}$ to produce the decoupled system of modal equations,
	%%%%%%%%%%%%%%%%%%%%%%%%%%%%%%%%%%%%%%%%%%%%%
	\begin{equation}
		\ddot{\bm{\eta}} + \bm{\omega}^2\bm{\eta} = \bm{q}(t),
		\label{EQ:eom_modal}
	\end{equation}
	%%%%%%%%%%%%%%%%%%%%%%%%%%%%%%%%%%%%%%%%%%%%%
	where $\bm{\omega}^2 = {\rm diag}[\omega_1^2,\ \dots,\ \omega_N^2]$ is a diagonal matrix of eigenfrequencies and $\bm{q}(t)= \bm{\Phi}^{\intercal}\bm{f}(t)$ the modal forcing function.
	Solutions to Eqs~\eqref{EQ:EOM_standard} can then be written as a superimposition of normal modes $\bm{\varphi}$ modulated in amplitude by the modal coordinate solutions of Eqs~\eqref{EQ:eom_modal} through the well known expansion theorem,
	%%%%%%%%%%%%%%%%%%%%%%%%%%%%%%%%%%%%%%%%%%%%%
	\begin{equation}
		\bm{u}(t) = \sum_{i=1}^N\bm{\varphi}_i\eta_i(t) = \bm{\Phi}\bm{\eta}(t).
		\label{EQ:expansion}
	\end{equation}
	%%%%%%%%%%%%%%%%%%%%%%%%%%%%%%%%%%%%%%%%%%%%%
	% ---------------------------------------------------
	\begin{figure}[t!]
		\centering
		%	\fbox{
		\begin{subfigure}{.475\linewidth}	
			\includegraphics[width=\textwidth]{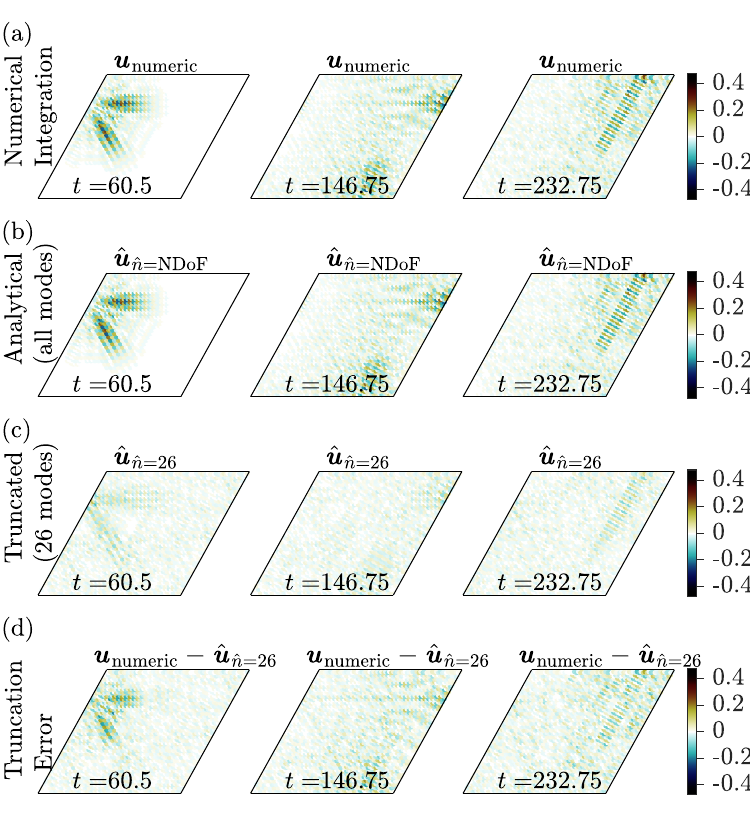}
		\end{subfigure} \hfill
		\begin{subfigure}{.475\linewidth}
			\includegraphics[width=\textwidth]{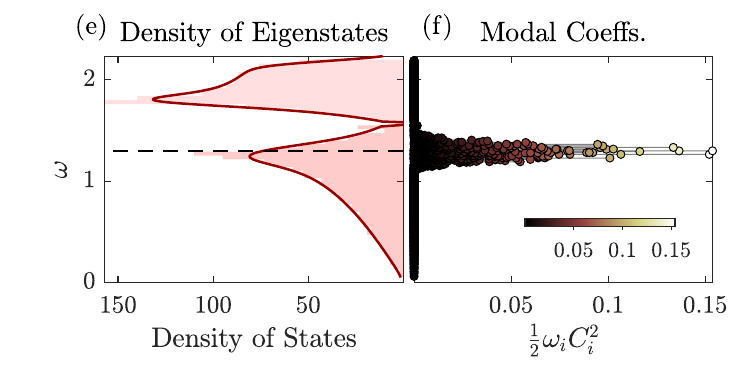}
			\includegraphics[width=\textwidth]{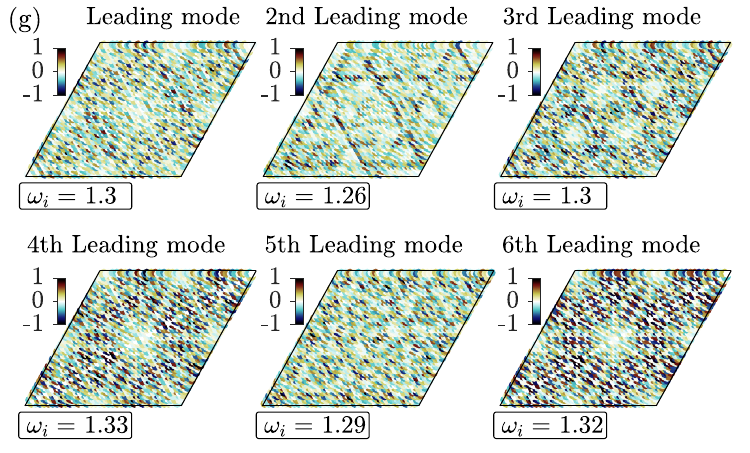}
		\end{subfigure}
		%}
		\caption{\textcolor{rev1}{The modal decomposition of a trivial wave in the hexagonal lattice of dimension $36\times36$. (a) The numerical solution for the finite system, as well as (b) the analytical solution using all modes, and (c) a truncated analytical solution using only 26 modes; (d) depicts the difference between the numerical and truncated solutions.
				(e) The histogram of eigenfrequencies denoting the ``density'' of the eigenstates across the spectrum with a black dashed line denoting excitation frequency, and (f) the energy associated with each eigenmode computed with solution~\eqref{EQ:expansion} with respect to their placement in the spectrum.
				%(e) The modal velocities of the numerical solution with envelop color depicting contribution to the dynamics. Also shown is (f) the modal frequency versus mode number of the analytical solution~\eqref{EQ:expansion} with color depicting contribution to the dynamics. 
				(g) The leading 6 eigenmodes used to construct the propagating wave as determined by their relative contributions in modal energy.}}
		\label{FIG:modal_decomp_trivial}
	\end{figure}
	% ---------------------------------------------------
	% ---------------------------------------------------
	\begin{figure}[t!]
		\centering
		\begin{subfigure}{.475\linewidth}
			\includegraphics[width=\textwidth]{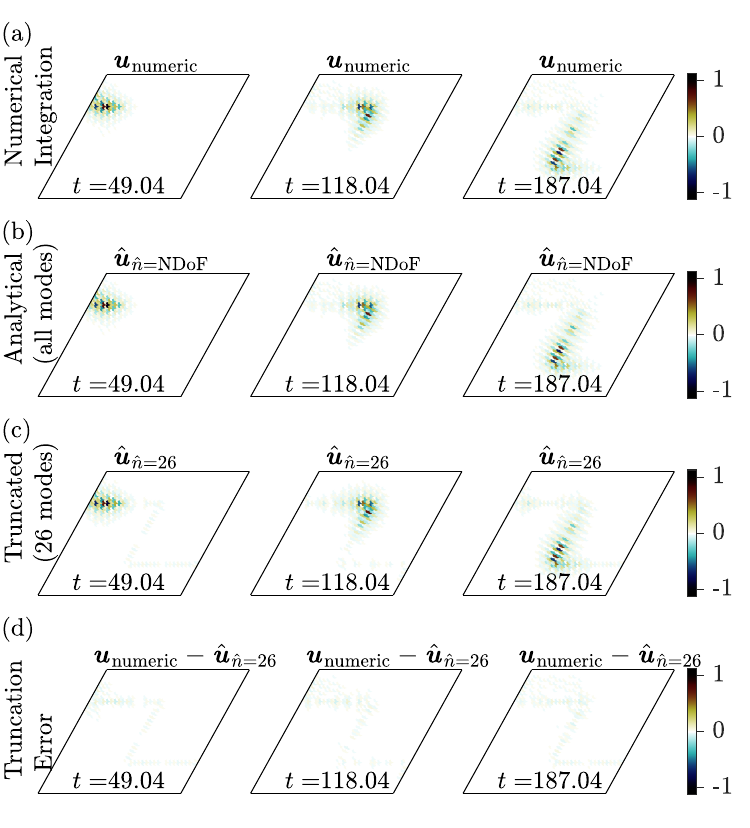}
		\end{subfigure} \hfill
		\begin{subfigure}{.475\linewidth}
			\includegraphics[width=\textwidth]{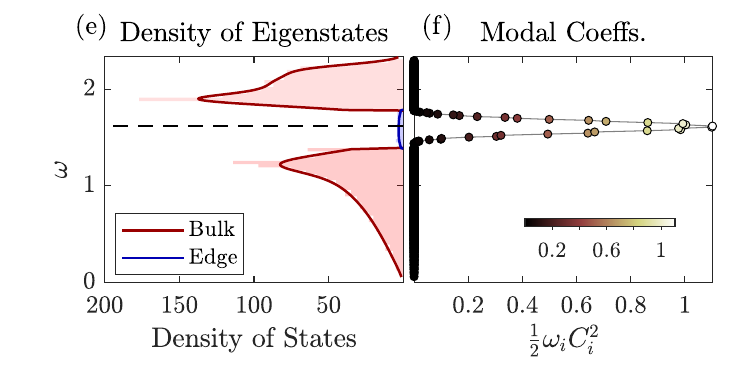}
			\includegraphics[width=\textwidth]{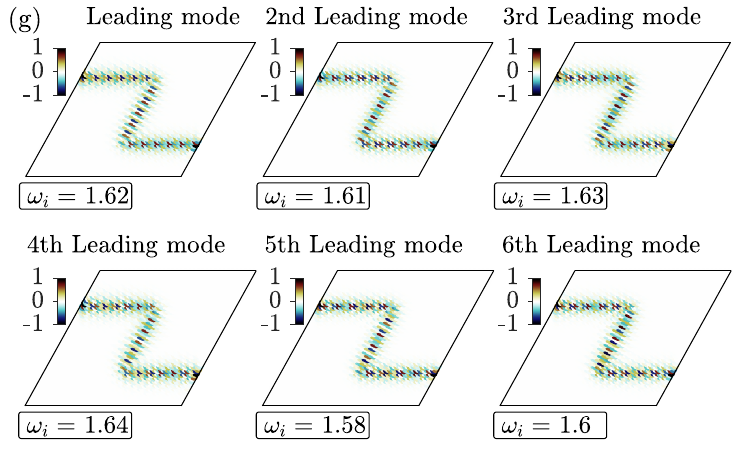}
		\end{subfigure}
		\caption{The modal decomposition of a topological wave in the hexagonal lattice of dimension $36\times36$. (a) The numerical solution for the finite system, as well as (b) the analytical solution using all modes, and (c) a truncated analytical solution using only 26 modes; (d) depicts the difference between the numerical and truncated solutions.
			\textcolor{rev1}{(e) The histogram of eigenfrequencies denoting the ``density'' of the eigenstates across the spectrum with a black dashed line denoting excitation frequency, and (f) the energy associated with each eigenmode computed with solution~\eqref{EQ:expansion} with respect to their placement in the spectrum.}
			%(e) The modal velocities of the numerical solution with envelop color depicting contribution to the dynamics. Also shown is (f) the modal frequency versus mode number of the analytical solution~\eqref{EQ:expansion} with color depicting contribution to the dynamics. 
			(g) The leading 6 eigenmodes used to construct the propagating topological wave as determined by their relative contributions in modal energy.}
		\label{FIG:modal_decomp}
	\end{figure}
	% ---------------------------------------------------

	The solutions to Eqs~\eqref{EQ:eom_modal} are written over two time intervals.
	The first interval encompasses the excitation period $0<t<T$ whereby the modal solutions are based on the initial conditions in modal space $(\bm{\eta}(0), \ \dot{\bm{\eta}}(0))  %=(\bm{\Phi}^{-1} \bm{u}(0),\ \bm{\Phi}^{-1} \dot{\bm{u}}(0)) 
	= (\bm{0},\bm{0})$ and subject to the non-homogeneous forcing term $\bm{q}(t)$.
	After forcing ends at time $t = {2\pi N_{cyc}}/{\Omega} \coloneqq T$, the modal solutions are stationary with amplitudes determined by the modal initial conditions  at $t = T$. For $t>T$, the solution can be thought of as an initial value problem whereby $\bm{\eta}(T)$ and $\dot{\bm{\eta}}(T)$ serve as the initial conditions for a system shifted in time by $T$. Hence, the analytical modal solutions are written explicitly as
	%%%%%%%%%%%%%%%%%%%%%%%%%%%%%%%%%%%%%%%%%%%%%
	\begin{equation}
		\begin{aligned}
			&\eta_i(t) = \frac{1}{\omega_i} \int_{0}^{t} q_i(\tau)\sin(\omega_i(t-\tau)){\rm d}\tau, \ \ \ &t \in [0, \ T]\\
			&\eta_i(t) = C_i \cos(\omega_i(t-T)-\psi_i(T)), \ \ \ &t \in (T,\infty]
		\end{aligned}
		\label{EQ:undamped_modal_sol}
	\end{equation}
	%%%%%%%%%%%%%%%%%%%%%%%%%%%%%%%%%%%%%%%%%%%%%
	where $C_i = \left[\eta_i^2(T) + \dot{\eta}_i^2(T)/\omega_i^2\right]^{1/2}$ and $\psi_i = \tan^{-1}\left(\frac{\dot{\eta}_i(T)}{\omega_i\eta_i(T)}\right) - \frac{2\pi N_{cyc}\omega_i}{\Omega} $  are the modal coefficient and phase of mode $i$, respectively. 
	Note that the addition to the phase  of ${2\pi N_{cyc}\omega_i}/{\Omega}$ accounts for the time-shift of the solution for $t>T$. 
	The analytical expressions for $\eta(T)$ and $\dot{\eta}(T)$ are readily recovered using computer algebra packages such as the \texttt{symbolic toolbox} in \texttt{\textsc{matlab}}.

	The modal solution may be employed to construct a reduced order solution,  $\hat{\bm{u}}$, which is taken as the sum of the leading $\hat{n}$ modal coefficients computed at $t = T$,
	\begin{equation}
		\hat{\bm{u}}(t) = 
		\sum_{i=1}^{\hat{n}}\bm{\varphi}_i\eta_i(t)  = \hat{\bm{\Phi}}\hat{\bm{\eta}}(t), \ \ \
		\hat{\bm{\Phi}} \in  \mathbb{R}^{N\times\hat{n}}, \ \ \hat{\bm{\eta}}\in\mathbb{R}^{\hat{n}\times 1}.
		%=\sum_{i=1}^{\hat{n}}\bm{\varphi}_iC_i(T)\cos(\omega_i(t-T)-\psi_i(T)) .\
		\label{EQ:reconstruction}
	\end{equation}
	%%%%%%%%%%%%%%%%%%%%%%%%%%%%%%%%%%%%%%%
	% IMPORTANCE OF REDUCED SYSTEM
	%%%%%%%%%%%%%%%%%%%%%%%%%%%%%%%%%%%%%%%
	\textcolor{rev1}{The truncated solutions of Eq~\eqref{EQ:reconstruction} may provide critical insights into the modal dynamics at play during topological wave propagation. 
		While the entirety of the modal basis is required to \textit{perfectly} capture the wave propagation in the entire spatio-temporal domain (assuming that $q_i\neq 0 \ \forall i$), accurate estimations of the localized topological wave transmission can be described by fewer eigenmodes.}

	%%%%%%%%%%%%%%%%%%%%%%%%%%%%%%%%%%%%%%%
	% TRIVIAL WAVE DECOMPOSITION
	%%%%%%%%%%%%%%%%%%%%%%%%%%%%%%%%%%%%%%%
	\textcolor{rev1}{We begin by analyzing a trivial bulk wave for a baseline comparison.
		%To explore the number of modes required to adequately reconstruct the system propagating topological wave, 	
		A 36$\times$36 trivial finite system described by configuration 0 of Fig~\ref{FIG:SysDescription}(e) is considered for an acoustic branch frequency ($\Omega = 1.3$). Fig.~\ref{FIG:modal_decomp_trivial}(a) depicts the direct numerical solution, Fig.~\ref{FIG:modal_decomp_trivial}(b) depicts the complete analytical modal solution ($\hat{\bm{u}}_{\hat{n} = N}$), and Fig~\ref{FIG:modal_decomp_trivial}(c) depicts the the truncated solution~\eqref{EQ:reconstruction} using 1\% of the modes ($\hat{n}= 26$) for a 36$\times$36 dimensional lattice. It is clear that the truncated solution of the bulk wave does not adequately capture the propagation, as the error of the truncated solution depicted in Fig~\ref{FIG:modal_decomp_trivial}(d) is of the same order as the numerical and full analytical solutions. 
		For a numerical evaluation of the truncated reconstruction, we consider the normalized error of the truncated solution to be
		$||\hat{\bm{u}}_{\hat{n}=26}-\hat{\bm{u}}_{\hat{n}={\rm NDoF}}||/||\hat{\bm{u}}_{\hat{n}={\rm NDoF}}||$, which is on the order of 100\% for Fig~\ref{FIG:modal_decomp_trivial}(d). 
		This is of no surprise, however, since it is well known that bulk wave propagation requires many modes for accurate construction. 
		%%%
		Fig~\ref{FIG:modal_decomp_trivial}(e) depicts the density of modes versus frequency, and it is apparent that there are many eigenmodes comprising the bulk acoustic state as the modal density is high. Moreover, Fig~\ref{FIG:modal_decomp_trivial}(f) depicts the analytical modal coefficients computed with Eq~\eqref{EQ:undamped_modal_sol} with respect to the modal frequencies. While a relatively narrow bandwidth of modes receiving substantial energy in the frequency, the high density of modes in this bandwidth results in many more than 1\% being required for accurate reconstruction, which explains the significant error of Fig~\ref{FIG:modal_decomp_trivial}(d). 
		Finally, Fig~\ref{FIG:modal_decomp_trivial}(g) depicts the mode shapes of the leading six bulk modes which are confirmed to be energetic across the spatial extent of the lattice.}
	
	%To explore the number of modes required to adequately reconstruct the system propagating topological wave, the modes are rank ordered with respect to the energy associated with each mode ($\frac{1}{2}\omega_i^2C_i^2$). 
	
	%%%%%%%%%%%%%%%%%%%%%%%%%%%%%%%%%%%%%%%
	% TOPOLOGICAL WAVE DECOMP
	%%%%%%%%%%%%%%%%%%%%%%%%%%%%%%%%%%%%%%%
	%\hrule
	\textcolor{rev1}{
		The same analysis was then applied to a propagating topological wave.
		%Fig~\ref{FIG:modal_decomp}(b) depicts the complete analytical modal solution ($\hat{\bm{u}}_{\hat{n} = N}$), while Fig~\ref{FIG:modal_decomp}(c) depicts the the truncated solution~\eqref{EQ:reconstruction} using 1\% of the modes ($\hat{n}= 26$) for a 36$\times$36 dimensional lattice of configuration 2 (zigzag boundary).
		%%%%
		Fig~\ref{FIG:modal_decomp} depicts the numerical and modal solutions for a 36$\times$36 dimensional lattice of configuration 2 (zigzag boundary) excited at the band gap frequency $\Omega = 1.62$.
		The difference between the complete solution (2,592 modes) and the reduced order solution of the 26 most energetic modes results in only a 10\% reconstruction error.}
	Remarkably, this is achieved using only 1\% of the modal basis demonstrating that the topological wave is comprised of a narrow subset of the modal space.
	%%%%%%%%
	%%%%%%%%%
	%This is confirmed with Fig~\ref{FIG:modal_decomp}(e) where it is shown that the amplitudes of the leading modes is much higher the for topological wave (which is plotted to the same scale as Fig~\ref{FIG:modal_decomp_trivial}(e)) for comparison), which results in part from the relatively loclized 
	%%%%%%%%%
	\textcolor{rev1}{Looking to Fig~\ref{FIG:modal_decomp}(e), the density of eigenmodes is clearly sparse inside of the topological band gap, especially when compared to the bulk spectrum; it is thus expected that excitation in the topological band gap will be described by far fewer modes. This reliance on a limited subset of modes is confirmed by the analytically computed modal coefficients depicted in Fig~\ref{FIG:modal_decomp}(f) which shows that energy is concentrated on the narrow band of spatially localized interface modes corresponding to band gap frequencies in the finite spectrum. 
		Moreover, the distribution of modal energy follows very closely to a smooth Gaussian curve which emulates precisely the form of the windowed tone burst excitation in the Fourier domain.}
	%%%
	%Looking to Fig~\ref{FIG:modal_decomp}(e), the modal velocity envelopes of the transformed numerical solution show that indeed only a small set of modes are excited by the input force (which is plotted to the same scale as Fig~\ref{FIG:modal_decomp_trivial}(e) to emphasize the increase in energy of the leading modes as compared to bulk waves).
	%This reliance on a limited subset of modes is confirmed by the analytically computed modal coefficients depicted in Fig~\ref{FIG:modal_decomp}(f) which shows that energy is concentrated in a narrow band of modes corresponding to band-gap frequencies of the finite spectrum, \textcolor{rev1}{which are relatively localized in the spectrum as compared to bulk modes, as is also confirmed by the spectral density calculations given in~\ref{APX: DOS}.}
	%
	\textcolor{rev1}{Lastly, Fig~\ref{FIG:modal_decomp}(g) depicts the
		mode shapes of the six leading (most energetic) eigensolutions of the topological system where it is apparent that each corresponds to a mode concentrated at the domain boundary.}

	\textcolor{rev1}{
		%is contradictory to typical bulk wave travel in mechanical domains whereby many modes are required to construct traveling wave solutions, and it 
		This ability to so accurately describe the traveling wave packet using such a narrow subspace of the finite spectrum is exclusively due to the symmetries (degeneracies) of the 2D periodic lattice system.}
	\textcolor{rev1}{
		Moreover, this finding is not solely the result of the spatial localization of the topological eigenmodes (as it is shown in~\ref{APX: Trivial} and~\ref{APX: DOS} that localized propagation in a trivial waveguide cannot be described to the same degree of accuracy), as the effects of backscatter in trivial systems greatly complicates the modal reconstruction. 
		Therefore, the spectral sparsity of the the topological band gap, combined with the immunity to backscatter around corners (a feature which is unique to topological wave propagation), is a key component for the reconstruction the transient topological wave packet across complex domain boundaries using very few vibrational modes.} 
	%\textcolor{rev1}{Moreover, the band topology of QVH waveguides provides \textit{predictive} localization} whereby one can know a-priori the number of edge modes is equal to the number of columns in the interface system, and the remaining eigenmodes correspond to bulk mode solutions (i.e., on the dispersion diagram of the infinite system). 
	Hence, an interesting observation can be deduced:
	\textit{the topological wave traveling across a hexagonal lattice can be described by the superimposition of  few modes that are closely spaced in frequency within the band gap, concentrated at the interface, and possessing the appropriate phase differences to produce a traveling wave solution when superimposed, \textcolor{rev1}{
			and that the same reconstruction across complex boundaries would not be achievable without the unique properties of topological insulation.}}

	%\includegraphics[width=\linewidth]{GraphicalAbstractLightR1}
	%\hrule
	
	% ---------------------------------------------------
	\begin{figure}[t!]
		\centering
		\includegraphics[width=\textwidth]{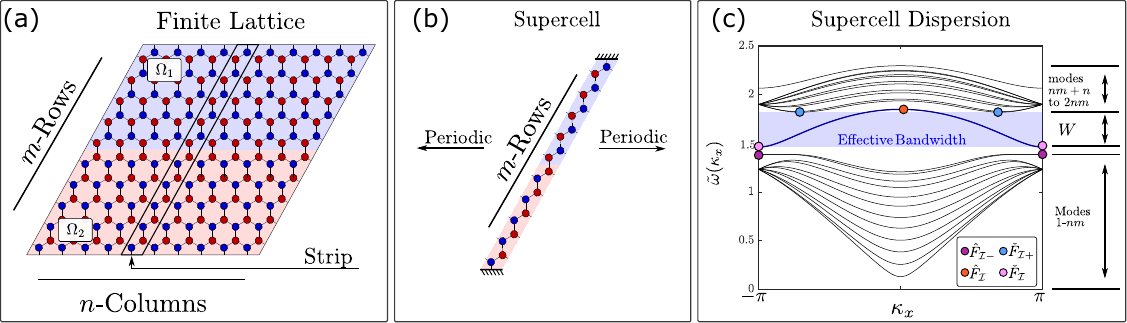}
		\caption{A depiction of a finite lattice consisting of $n$ columns and $m$ rows with a highlighted region depicting a \textit{supercell}. (b) The supercell is depicted with periodic boundary conditions in the $x$ direction. (c) The band structure of the supercell with the topological branch in blue with the shaded region depicting the band width $W$ and labels depicting the band width boundaries.}
		\label{FIG:infinite_strip_dgm}
	\end{figure}
	% ---------------------------------------------------

	% ---------------------------------------------------
	\subsection{Modal reconstructions for narrow- and broad-band inputs}
	% ---------------------------------------------------
	The proposition that the propagation topological wave is the result of a few key eigenmodes (Fig~\ref{FIG:modal_decomp}) may be further tested by studying the relationship of the modal decomposition to changes in topological wave behavior.
	Namely, interplay between modal dynamics and topological wave propagation may be furthered by establishing the relationships between interface energy concentration and the underlying modal spectrum.
	%%%%%%%%%%%%%%%%%%%%%%%%%%
	% BANDWIDTH AN BERRY CURVATURE
	%%%%%%%%%%%%%%%%%%%%%%%%%%
	\textcolor{rev1}{
		It is well established that the amount of energy retained on the interface of a topological insulator is in part determined by the contrast parameter.
		When the contrast is raised, greater localization occurs at domain boundaries corresponding to increased spectral localization of the edge states, but at the cost of diminishing the concentration of Berry curvature at the valleys~\cite{Qian2018,Eisenberg2019} and increasing the prominence of backscatter due to intervalley mixing~\cite{Zhu2018}.
		The other determining factor is the bandwidth of the input excitation. As the number of forcing periods in the windowed tone burst increases, the corresponding Gaussian curve in the frequency domain narrows and a more concentrated spectral subspace receives the majority of the excitation energy~\cite{Ditri1994}.}

	\textcolor{rev1}{
		Since the topological mode exists in the band-gap of the 2D system, one may utilize the semi-infinite \textit{supercell} dispersion to gain inference regarding the admissible spectrum of the topological wave.}
	The supercell dispersion curve is derived from a column-strip of the configuration 1 interface lattice (Fig~\ref{FIG:infinite_strip_dgm}(a)).
	%%%%%%%%
	Applying periodic boundary conditions on the left and right boundaries, fixing the top and bottom boundaries, and inserting an assumed Bloch solution in the $x$-direction (Fig~\ref{FIG:infinite_strip_dgm}(b)), the dispersion relation for the supercell is recovered.
	%%%%%%%%%%%%%%%%%%%%%%%%%%%%%%%%%
	% RETURN TO ME
	%%%%%%%%%%%%%%%%%%%%%%%%%%%%%%%%%
	The supercell dispersion relation will be denoted as $\tilde{\omega} = \tilde{\omega}(\kappa_x)$ (Fig~\ref{FIG:infinite_strip_dgm}(c)). 
	The key result is that a band now exists in the band gap which corresponds to the topological band of the finite system which can be used to define the \textit{effective topological bandwidth} as~\cite{Du2020},
	\begin{equation}
		W = \min\left\{\check{F}_{\mathcal{I}}, \hat{F}_{\mathcal{I+}} \right\}
		- \max\left\{\check{F}_{\mathcal{I}}, \hat{F}_{\mathcal{I-}} \right\},
		\label{EQ:BW}
	\end{equation}
	where $\hat{F}_{\mathcal{I}}$ and $\check{F}_{\mathcal{I}}$ are the upper and lower limits of the topological dispersion band of the interface supercell, whereas $\hat{F}_{\mathcal{I}+}$ and $\check{F}_{\mathcal{I}-}$ are the maximum and minimum frequencies of the adjacent bulk-bands (Fig~\ref{FIG:infinite_strip_dgm}(c)).

	% ---------------------------------------------------
	\begin{figure}[t!]
		\centering
		\includegraphics[width=.5\textwidth]{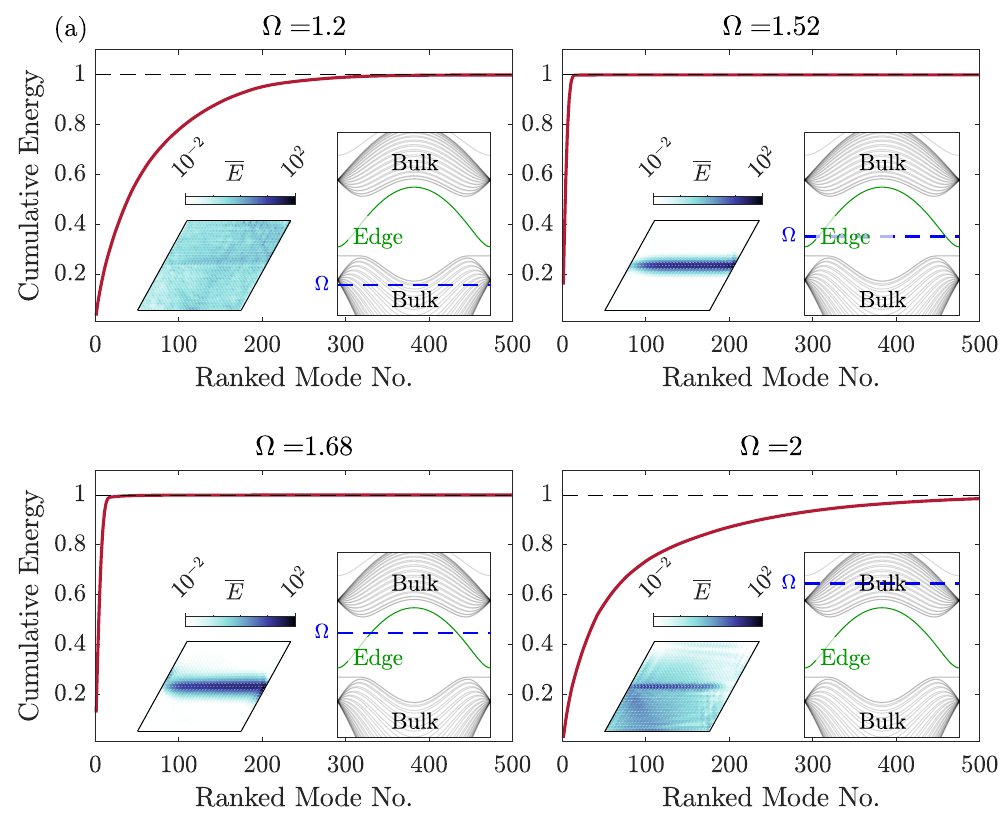}%
		\includegraphics[width=.5\textwidth]{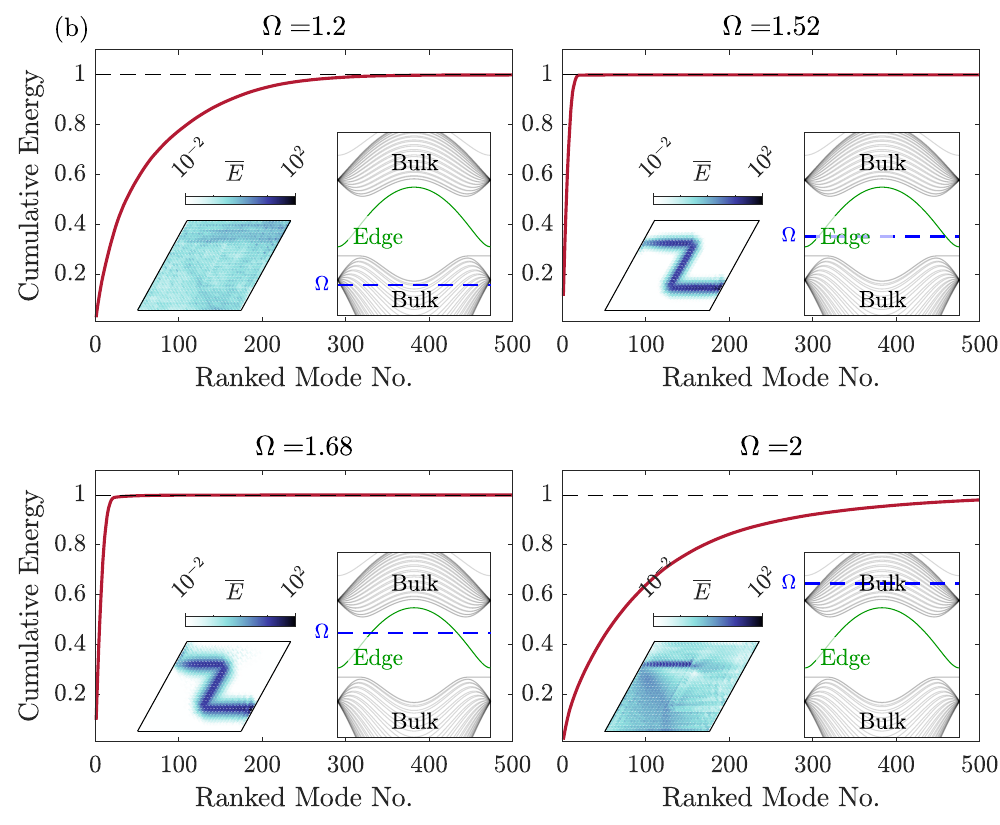}
		\includegraphics[width=.5\textwidth]{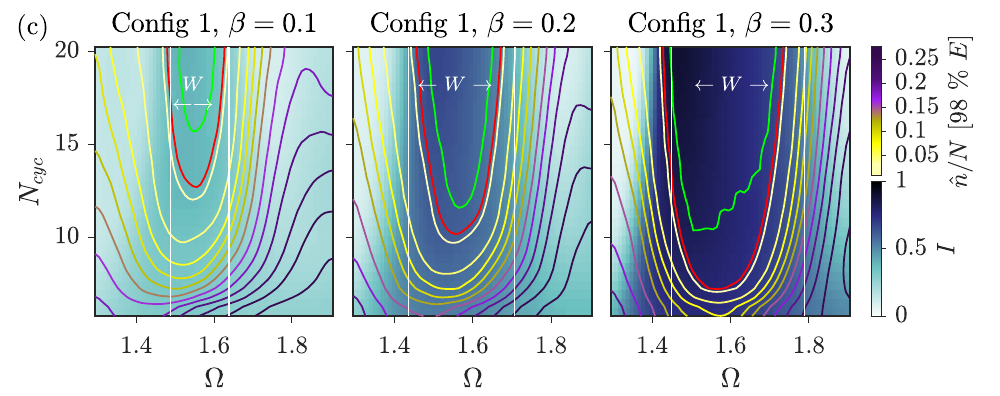}%
		\includegraphics[width=.5\textwidth]{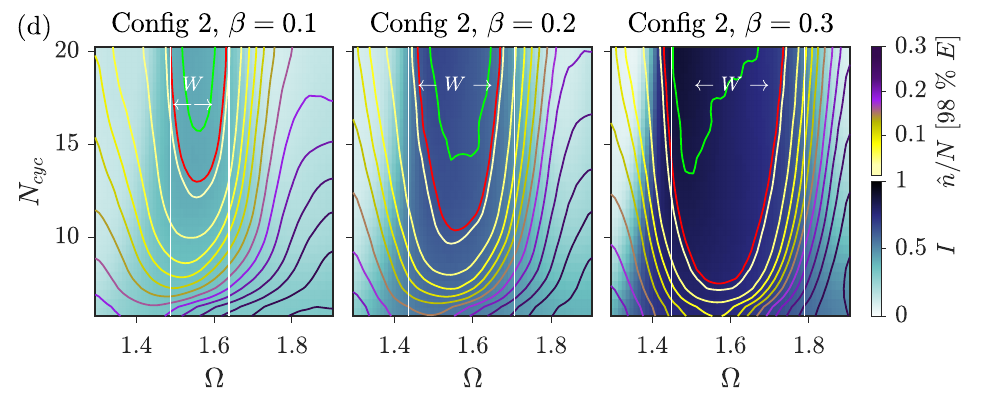}
		%%%%%%%%%%%%%%%%%%%%
		\caption{\textcolor{rev1}{The cumulative variance of the ordered modal solution with respect to total modal energy captured by the leading 500 eigenmodes  for excitation frequencies below ($\Omega<1.45$), inside ($1.45<\Omega<1.79$), and above ($\Omega>1.79$) the effective insulator bandwidth for a lattice with $\beta = 0.35$ with results depicted for (a) configuration 1 and (b) configuration 2, respectively.
				%%%%
				The energy quantities (a) and (b) are normalized with respect to the total energy of the system.  The left insets depict the time-averaged energy for each oscillator. 
				The right insets depict a zoomed in view of the supercell band structure with a green line denoting the topological band, gray lines denoting bulk bands, and a dashed blue line denoting the location of the excitation frequency with respect to the band structure.
				(c) and (d): The energy intensities at the interface for various $\beta$ between 0.1 and 0.35 as a function of input signal bandwidth and frequency for configuration 1 and configuration 2, respectively. The superimposed contours depict the number of bulk modes of the modal decomposition  required  to reconstruct 98\% of the wave propagation. Green lines of (c,d) depict the parameter domain for which 1\% of modes account for the reconstruction, while red lines corresponding to  2\% of modes accounting for the reconstruction.}}
		\label{FIG:E_vs_BW}
	\end{figure}
	% ---------------------------------------------------
	
	We study the changes in topological wave propagation with respect to the input bandwidth and center frequency inside and outside of the effective topological bandwidth over a range of contrast parameters with the aim of relating the performance of the propagating wave to the modal decomposition of the system. The performance of topological wave propagation for each parameter configuration is computed by considering the distribution of energy in the finite system.
	This energy distribution is summarized by using the \textit{interface concentration ratio} which is denoted as $I$. 
	Here, we take $I$ to represent the time-averaged energy $\overline{E}$ of oscillators at the interface ($\mathcal{I}$) as compared to the energy across the entire domain ($\mathcal{D}$),
	\begin{equation}
		I = \frac{\sum_{(p,q,\alpha)\in \mathcal{I}}\overline{E}_\alpha^{p,q}(t)}
		{\sum_{(p,q,\alpha)\in \mathcal{D}}\overline{E}_\alpha^{p,q}(t)}.
	\end{equation}
	\textcolor{rev1}{While the idealization of the domain boundary is a 1D line, we consider the interface domain $\mathcal{I}$ as the unit cells of domains 1 and 2 which border one-another in a similar fashion to other works concerned with spatial energy localization~\cite{Du2020}.}
	Details pertaining to the calculation of instantaneous energy for each oscillator are given in~\ref{APX:energy_table}.
	
	%The closed form solutions~\eqref{EQ:undamped_modal_sol} may now be used to directly relate the concentration ratio to the modal solution.
	\textcolor{rev1}{
		The interface concentration may be related to the closed form modal solutions~\eqref{EQ:undamped_modal_sol} of the propagating topological wave by considering the relation of $I$ to the required number of modes to accurately reconstruct the wave propagation.}
	%%%%%
	\textcolor{rev1}{
		To do this, the number of rank-ordered modes ($\hat{n}$) required to reconstruct a given portion of the total modal energy are compared directly to the concentration ratio for various excitation profiles.}
	%%%%%
	Figs~\ref{FIG:E_vs_BW}(a) \textcolor{rev1}{and (b)} depict the cumulative modal energy of the acoustics accounted for by the leading 500 eigenmodes \textcolor{rev1}{for finite systems of configuration 1 and configuration 2, respectively}.
	The insets depict the time-averaged energy for each oscillator across a range of $\Omega\in(1.2,2)$ for a system with a contrast of $\beta = 0.3$ corresponding to an effective topological bandwidth of $W = \{\Omega: 1.45<\Omega<1.79\}$. 
	It is apparent that the energy retained at the interface and the required number of eigenmodes necessary to accurately capture the topological wave propagation are closely related. When the excitation frequency is inside of the effective topological bandwidth, the propagating wave is effectively excited and convergence to the total modal energy is reached rapidly when compared to excitations outside of the effective topological bandwidth; this is expected since it has been shown that the topological wave is described by the superimposition of a narrow band of closely spaced eigenfunctions.

	Fig~\ref{FIG:E_vs_BW}(c) depicts the interface concentration as a function of forcing frequency and number of excitation cycles for a selection of contrast parameters between $\beta = 0.1$ and $\beta = 0.3$ for a configuration 1 lattice (straight-edge interface), and the same is shown in Fig~\ref{FIG:E_vs_BW}(d) for a configuration 2 lattice (zigzagged-interface).
	%%%%%%%%%%%%%
	The superimposed white lines of Fig~\ref{FIG:E_vs_BW}(c,d) show the boundaries of the effective bandwidth as computed by Eq~\eqref{EQ:BW}.  
	%%%%%%%%%%%%
	The higher the number of excitation cycles is, the lower the input bandwidth is, which in turn leads to a narrow-band edge wave within the topological bandwidth, as expected.
	Moreover, higher contrast parameters lead to much greater energy concentration which is explainable by \textcolor{rev1}{increased spacing between the topological modes and the bulk spectrum.}
	\textcolor{rev1}{
		The relation between the excitation profile and modal solution is further explored by the superimposed contours of Figs~\ref{FIG:E_vs_BW}(c,d) which shows the relations between the number of bulk-modes required to reconstruct 98\% of the lattice acoustics based on the analytical modal solution for the interface concentration, excitation frequency, and excitation bandwidth.}
	%%%%%
	\textcolor{rev1}{The green contour lines indicate the domain of the excitation parameters for which only 1\% of the modes are required to account for 98\% of the energy.}
	%%%%%
	
	For frequencies outside of the topological bandwidth, a large number of bulk modes is required to reconstruct the topological wave as the energy leaks from the interface to the bulk spectrum.
	Frequencies within the topological bandwidth require far fewer bulk modes for the reconstruction, as expected.
	%%%
	As the excitation bandwidth increases at frequencies well within the effective topological bandwidth, the number of bulk modes falls drastically down from the order of hundreds to only the modes corresponding to the interface dynamics, \textcolor{rev1}{and near total reconstruction is achievable with just 1\% of the modal spectrum.}
	%When only the interface modes are utilized, the dimenisonality of the propagating wave is reduced to the number of columns ($n$) since $n$ interface modes exist for a $m\times n$ lattice.
	%%
	\textcolor{rev1}{
		Moreover, the number of modes required to construct 98\% of the modal energy converges to zero much faster as the contrast parameter grows for a straight-edge domain boundary, whereas this convergence is rapid in the zigzagged model since the decrease in Berry curvature inevitably leads to greater backscatter, and hence a more complicated wave packet reconstruction around corners. Nevertheless, even for $\beta = 0.3$, sufficiently narrow excitation bandwidths still lead to wave propagation reproducible by 1\% of the modal spectrum with negligible error.}
	%%%
	This further confirms that the topological wave propagation is enabled largely by the closely spaced edge modes with nontrivial phase differences in the limit of perfect excitation of the topological wave.
	\textcolor{rev1}{
		Moreover, by considering finite systems with straight (Fig~\ref{FIG:E_vs_BW}(c)) and zigzagged (Fig~\ref{FIG:E_vs_BW}(d)) domain boundaries, we confirm that this framework is indifferent to the path that the topological wave traverses.}
	\textcolor{rev1}{This is further supported by the results of~\ref{APX: doubleZigZag} which consider an additional double zigzagged configuration (not presented here). Moreover, these findings are applicable to lattices of arbitrary size (so long as the system is large enough to be valid for the Bloch theorem), as shown in~\ref{APX: sizeComp}.}

	\textcolor{rev1}{A key implication worth noting is that this reduced order solution can be computed in a fraction of the time required for numerical simulations. 
		Modal superposition is typically considered inefficient for resolving propagating waves since many modes are typically required to resolve transient wave packets, and performing a full modal superposition can be very costly for large-scale acoustic waveguides. 
		Accordingly, a majority of computational studies involving topologically protected waves have relied on numerical integration to study the finite system. 
		However, if only a small subset of the modal spectrum is required, then it is in fact much more efficient to employ modal superposition. 
		By leveraging the result that very few modes can describe the topological wave propagation, numerical linear algebra techniques may be employed to locally solve a small subset of the modal spectrum centered about the excitation frequency in order to compute a truncated solution which captures the main portion of the energy of the wave. Doing so reduces the computation time by an order of 100, as discussed in~\ref{APX: CompTime} in greater detail. }

	%For a given forcing profile, the set of modal coefficients Eq~\eqref{EQ:undamped_modal_sol} delivers a set of modal coefficients 

	% ---------------------------------------------------
	\section{Group Velocity of the Propagating Topological Wave in the Finite System} 
	\label{Sec:VG}
	% ---------------------------------------------------
	%%
	By leveraging the results of section~\ref{sec:Desc}, we now investigate how the group velocity of the topological wave may be better predicted by leveraging the modal solution for topological systems of finite size subject to excitations with finite bandwidths. 
	%% 
	%	To do this, we first compare estimates of the group velocity derived from the modal solution to the standard theoretical calculation of group velocity used ubiquitously in mechanical insulator studies. These measures are then compared to empirically measured velocity of the propagating wave's energy centroid. Next, we formulate an estimate based on the underlying modal solution termed the 'mode-band matching' technique.
	%%
	%	The modal solution can be used to accurately predict the  group velocity of the propagating topological mode in the finite system.
	%%
	The standard measure for the group velocity of a propagating wave in an infinite  2D lattice system at a driving frequency $\Omega$ is given as $v_g = \nabla_{\bm{\kappa}}\omega(\bm{\kappa})|_{\omega = \Omega}$, where $\omega(\bm{\kappa})$ is the dispersion relation.
	%%%%%%%
	For topological systems where the propagating wave are realized in the band gap of the phononic lattice, the supercell dispersion of Fig~\ref{FIG:infinite_strip_dgm} may be used so that the group velocity in the $x$ direction at the excitation frequency $\Omega$ could be analytically approximated, at least in principle, as:
	\begin{equation}
		v_g(\Omega) = \partial_{\kappa_x}\tilde{\omega}(\kappa_x)\big|_{\tilde{\omega} =\Omega} .
		\label{EQ:vg_theory}
	\end{equation}

	While Eq~\eqref{EQ:vg_theory} yields a convenient theoretical formula, it is not likely that this quantity will be seen in practice for finite excitation windows. 
	\textcolor{rev1}{Indeed, the Fourier uncertainty principle ensures that the excitation profile will excite a finite spectrum of modes, which, when superimposed, construct the band-limited wave at the center frequency $\Omega$; this has been well discussed in prior work in the context of propagating modes in elastic solids~\cite{Ditri1994}.} Hence, the selection of any single frequency value along the curve $\tilde{\omega}(\kappa_x)$ is not likely to deliver accurate velocity estimates as the input bandwidth grows or as the center frequency moves from the center of the effective topological bandwidth, since these input variations will surely excite regions of the supercell band structure not accounted for in Eq~\eqref{EQ:vg_theory}. 
	%%%%%%
	To investigate this, we first compare theoretical group velocity estimations of a topological system to measured estimates of group velocity values to demonstrate the inaccuracies that may arise in topological systems. We subsequently employ the modal solution to develop a modal-based group velocity estimate termed the \textit{mode-band matching} method and compare it directly with Eq~\eqref{EQ:vg_theory}, and lastly we confirm the accuracy of the mode-band matching method over an array of contrast parameters and excitation profiles.

	% ---------------------------------------------------
	\subsection{Estimations of true group velocity} 
	%	\label{sec:Desc}
	% ---------------------------------------------------
	We consider the progression of the energy centroid of the propagating topological wave versus time to recover estimates of the \textit{true} group velocity of the finite system subject to finite excitation. 
	The instantaneous energy of each oscillator and it's corresponding $x$-coordinate can be used to compute the instantaneous energy centroid in the $x$ direction,
	\begin{equation}
		\bar{E}_x(t)  = 
		\sum_{(p,q,\alpha)\in\mathcal{D}} x_\alpha^{p,q}E_\alpha^{p,q}(t) \Bigg/ \sum_{(k,j,\alpha)\in\mathcal{D}}E_\alpha^{k,j}(t)  
		%\frac{1}{\sum_{k,j,\alpha}E_\alpha^{k,j}(t)  } 
		%\sum_{p,q,\alpha} x_\alpha^{p,q}E_\alpha^{p,q},  
		\label{EQ:Ecentroid}
	\end{equation}
	where $E_\alpha^{p,q}$ is the instantaneous energy of point mass $\alpha$ of unit cell $(p,q)$, and $x_\alpha^{p,q}$ the corresponding $x$ coordinate.
	By computing $\bar{E}_x(t)$ across times $T<t<T_{rb}$ where $T_{rb}$ denotes the time taken for the wave to reach the right boundary, we can estimate the true group velocity by simply taking the difference of position with respect to time, $v_g^{\rm rec} = \Delta \bar{E}_x(t)/\Delta t$.

	% ---------------------------------------------------
	\begin{figure}[t!]
		\centering
		\includegraphics[width=\textwidth]{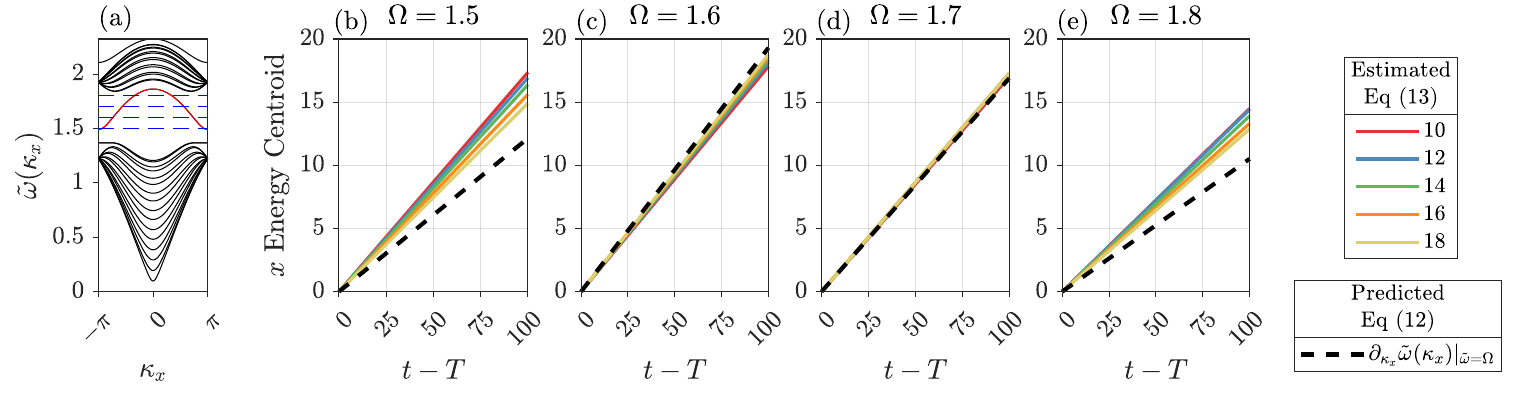}
		\caption{The comparison of the energy centroid versus shifted time for a 12$\times$24 lattice with $\beta = 0.35$. (a) The range of forcing frequencies with the effective topological bandwidth as shown by the blue-dashed lines of the corresponding super cell dispersion. 
			(b-e) The solid lines correspond to the numerically measured quantities recovered via Eq~\eqref{EQ:Ecentroid} whereas the black-dashed line corresponds to the theoretical velocity of Eq~\eqref{EQ:vg_theory}. Line colors depict different forcing bandwidths as denoted by the number of cycles $N_{cyc}$ of Eq~\eqref{EQ:forcing}.}
		\label{FIG:VG_error_ex}
	\end{figure}
	% ---------------------------------------------------
	
	The estimated quantities of group velocity (Eq~\eqref{EQ:Ecentroid}) can be compared directly the the theoretical prediction (Eq~\eqref{EQ:vg_theory}) for an assortment of excitation profiles.
	Typically, topological systems are studied near the center of the effective bandwidth where the dispersion curve is steepest and furthest from bulk-band frequencies. 
	However, in practice, topological protection may persist across various spectral regions of the effective bandwidth. To study the effectiveness of Eq~\eqref{EQ:vg_theory} for finite systems across the topological bandwidth we compute solutions to system~\eqref{EQ:EOM_standard} using solution~\eqref{EQ:undamped_modal_sol} for a  configuration 1 lattice system with $m=12$ rows and $n=24$ columns. We select $\beta = 0.35$, which corresponds to the supercell dispersion diagram of Fig~\ref{FIG:infinite_strip_dgm}(c). The effective bandwidth for this system spans $W = \{\Omega: 1.48 \leq\Omega\leq1.82\}$, and we select four frequencies between $\Omega = 1.5$ and $\Omega = 1.8$ to sample. Moreover, multiple selections of $N_{cyc}$ are considered in order to investigate how the input signal bandwidth influences the group velocity. Fig~\ref{FIG:VG_error_ex} shows the supercell dispersion relation along with the plots of temporal evolution of the energy as observed using Eq~\eqref{EQ:Ecentroid} and as predicted using Eq~\eqref{EQ:vg_theory}. While there is good agreement for frequencies that correspond to steep-sloped regions of the topological band that are well separated from bulk states, it is clear that near the boundaries of the effective topological bandwidth the effectiveness of Eq~\eqref{EQ:vg_theory} breaks down. Moreover, for frequencies not centered on the effective bandwidth, the group velocity clearly depends on $N_{cyc}$ for a given excitation frequency.
	As we will show in the following section, these changes in group velocity can be explained from the perspective of modal superposition.

	% ---------------------------------------------------
	\subsection{Mode-band matching for group velocity predictions}
	% ---------------------------------------------------
	Because a finite-band input corresponds to a finite spectrum of excitation, it also holds that a finite spectrum of the modal solutions must be excited. 
	This correlates with the fact that a traveling wave cannot exist from a single standing mode alone but rather is a superposition of a set of modes with appropriate (non-trivial) phase differences. 
	This fact has been leveraged in previous works to account for the contribution of propagating modes in continuous elastic systems with respect to the profile of input sources~\cite{Auld1971,Ditri1994,Rose2000} via the normal mode expansion technique.
	Here, we formulate an analogous mode-band matching methodology to predict the propagating group velocity of discrete topological systems based on the spectrum of excited stationary modes.
	
	Given that a mechanically consistent system delivers a denumerable set of eigenfrequenices, any wave propagating in the spatio-temporal domain cannot be represented by a single frequency, such as in Eq~\eqref{EQ:vg_theory} which assumes a near zero-band input.
	\textcolor{rev1}{However, we have now demonstrated that any finite input applied to the linear lattice results in modal coefficients in the finite system's spectrum.}
	\textcolor{rev1}{Moreover, it is well-known that the vibrational modal spectrum of a discrete periodic lattice can be mapped directly to a sequence of wavenumbers across the Brillouin Zone of the band structure~\cite{Miniaci2021}.} 
	Accordingly, the modal expansion method needs to be adapted in order to refine the single frequency prediction~\eqref{EQ:vg_theory}. 
	%\textcolor{rev1}{Since we are interested in topological w
	This is achievable by mapping each eigenmode to a unique location on the dispersion curves \textcolor{rev1}{to estimate the dispersion characteristics of the propagating wave as a combination of those associated with the most energetic modes.}
	%
	%To investigate the use of the modal expansion solution in recovering a more accurate prediction of topological wave group velocity in the finite system, we return to the infinite strip dispersion relation.
	\textcolor{rev1}{Because no band describing the topological wave  exits in the hexagonal lattice's dispersion, a method for associating each mode to the supercell dispersion curves must be established.}
	This can be done by recognizing that dimension of the supercell for a finite lattice of dimension $2m\times n$ is $2m \times 1$ (since each unit cell possesses 2-DoF); the number of dispersion curves generated by the supercell is thus equal to twice the number of rows of the finite system, that is, $2m$. Hence, each band of the supercell dispersion can be related back to the $2m\times n$-dimensional modal spectrum of the corresponding finite system by recognizing that each branch corresponds to an equivalent number of modes as there are columns in the finite system.
	That is, modes 1 to $m\times n$ correspond to the acoustic bulk states, modes $m\times n+1$ to $m\times n+n$ to the topological states inside the band gap,  lastly modes $m\times n+n+1$ to $2m\times n$ to the optical bulk states. 
	
	% ---------------------------------------------------
	\begin{figure}[t!]
		\centering
		\includegraphics[width=\textwidth]{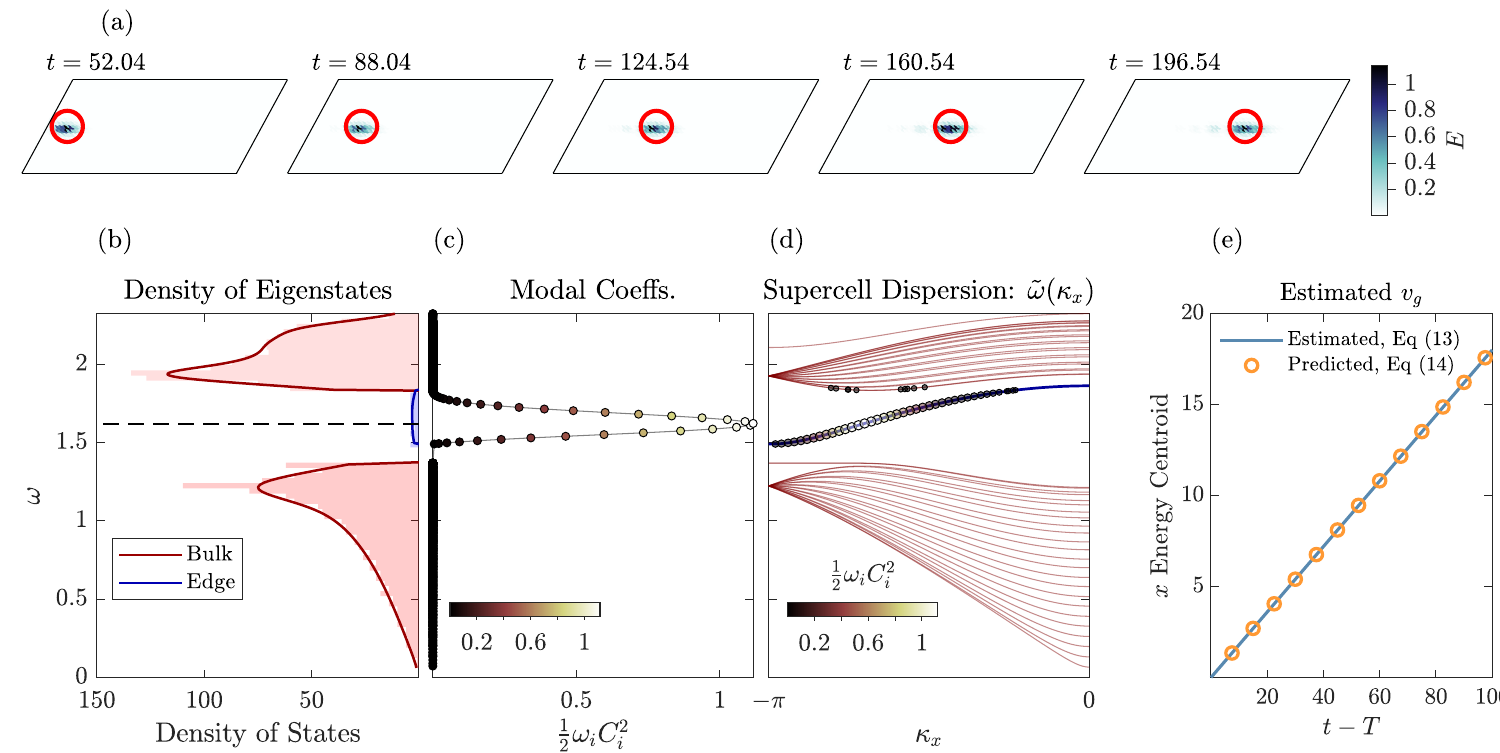}
		\caption{The comparison between estimated and predicted group velocity. (a) The evolution of the energy centroid in time with the centroid depicted with a red circle, 
			\textcolor{rev1}{(b) The density of modes for the 40$\times$20 finite lattice with a black dashed line denoting excitation frequency; (c) energy associated with each eigenmode in the finite spectrum, and (d) the corresponding locations mapped to the supercell dispersion diagram with color depicting contribution to group velocity. (f) The propagation of the energy centroid as estimated by Eq~\eqref{EQ:Ecentroid} and predicted by Eq~\eqref{EQ:PredVG}.}}
		\label{FIG:VG_pred_ex}
	\end{figure}
	% ---------------------------------------------------
	% ---------------------------------------------------
	\begin{figure}[t!]
		\centering
		\includegraphics[width=\textwidth]{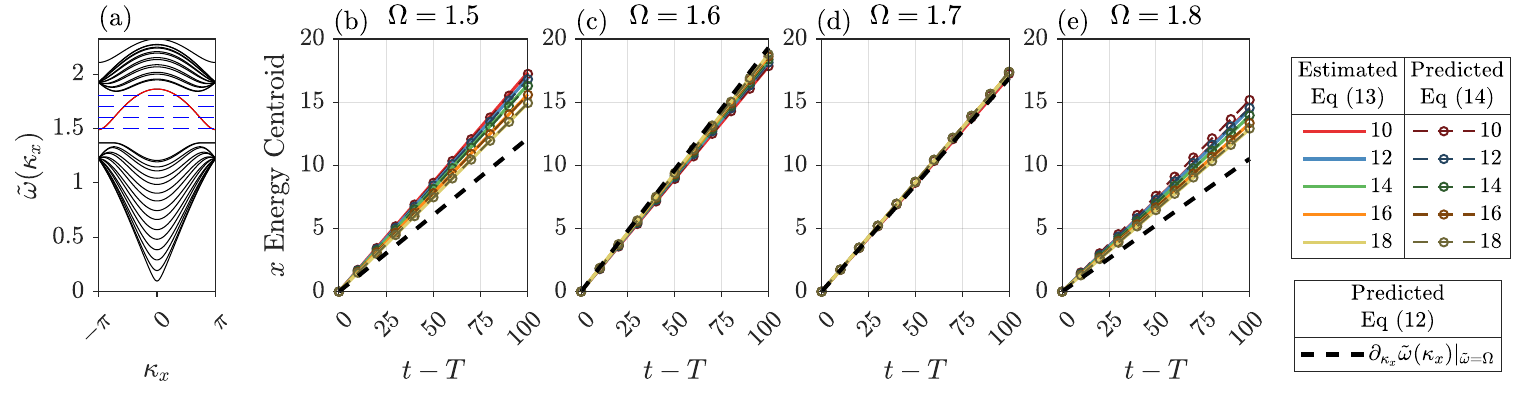}
		\caption{The same as Fig~\ref{FIG:VG_error_ex}(a-e) but now with dotted lines superimposed corresponding to predictions rendered using mode-band matching (Eq~\eqref{EQ:PredVG}).}
		\label{FIG:VG_pred_plt}
	\end{figure}
	% ---------------------------------------------------
	
	By considering the topological wave as a superposition of modal components that can be mapped to the supercell dispersion curves, the group velocity of the propagating topological wave can be predicted based on the leading $\hat{n}$ modal coefficients and their corresponding eigenfunctions in the finite modal spectrum.
	Using the finite eigenfrequncy of each mode as a point of reference, the modal components are matched to a corresponding group-velocity based on the supercell dispersion diagram, $\tilde{\omega}({\kappa}_x)$, computed at the corresponding value of $\kappa_x$. The value of $\partial_{\kappa_x}\tilde{\omega}(\kappa_x)$ taken at the matched point in $\kappa_x$-$\tilde{\omega}$ space for a given branch of the supercell dispersion gives the group velocity associated with the $i$-$th$ mode. 
	%%%%%%%%
	The square of the modal coefficients ($C_i$) computed via Eq~\eqref{EQ:undamped_modal_sol} are then used to weight the velocity estimates.
	Specifically, the modal-based group velocity estimates per the mode-band matching methodology are computed as %root-summing weighted group velocity contributions of the leading $\hat{n}$ eigenmodes, 
	% and the square root these weighted estimates are summed  averaged to render a predicted value for $\hat{v}_g$ based on the modal solution and infinite strip dispersion,
	\begin{equation}
		\hat{v}_g = \frac{\left[\sum_i^{\hat{n}} C_i^2\frac{\partial \tilde{\omega}_\gamma}{\partial \kappa_x}\Big|_{\tilde{\omega_\gamma} = \omega_i}\right]^{1/2}}{
			\left[\sum_j^{\hat{n}} C_j^2\right]^{1/2}},  
		\label{EQ:PredVG}
	\end{equation}
	where $\tilde{\omega}_\gamma$ refers to the $\gamma$-$th$ branch of the supercell dispersion where $\gamma = \lfloor 2n/k  \rfloor$, $n$ is the number of columns in the finite system, and $k$ is the rank of the $i$-$th$ mode sorted by frequency value.
	This ensures that the weighted contribution of each group velocity component corresponds to the correct dispersion branch, since the $2m$-branches map back of the $2m\times n$ finite modal solutions which contribute to the wave travel of the finite system. 
	The summation of Eq~\eqref{EQ:PredVG} is taken over the leading $\hat{n}$ eigenmodes required to reconstruct a desired portion of the total lattice energy. While $\hat{n}$ can be taken over all available modes if a complete reconstruction is desired, we note that convergence of $\hat{v}_g$ is typically achieved using far fewer modes since the contribution of modal coefficients outside of the band-gap tends to zero when the topological wave is excited (cf.~Figs~\ref{FIG:E_vs_BW}(e,f)).

	% ---------------------------------------------------
	\begin{figure}[t!]
		\centering
		\includegraphics[width=\textwidth]{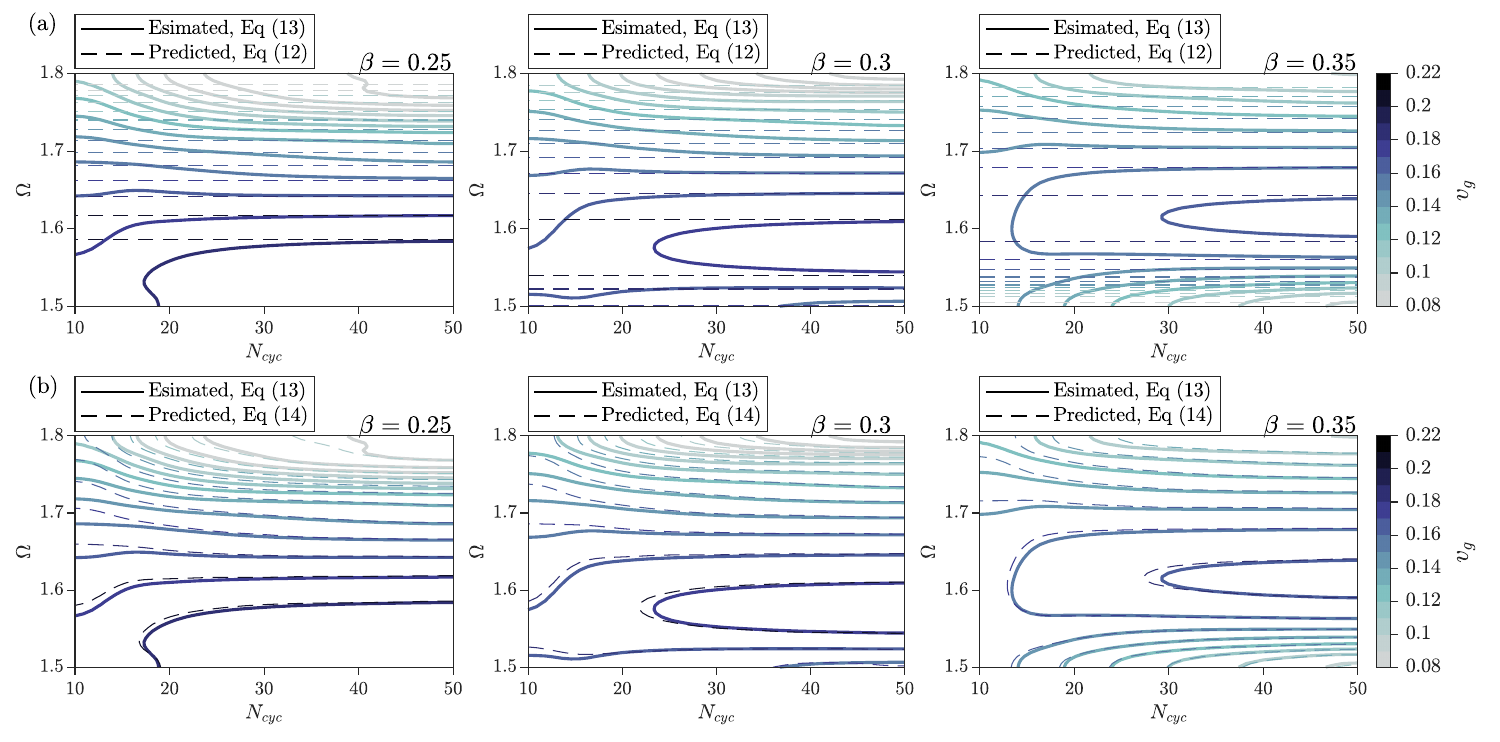}
		\caption{Contours of the estimated group velocity of the propagating topological wave  based on Eq~\eqref{EQ:Ecentroid} for various excitation frequencies $\Omega$ and various excitation bandwidths for contrast paramters $\beta = 0.25,\ 0.30,$ and 0.35. This is compared with (a) the predicted values based on the theoretical quantity of Eq~\eqref{EQ:PredVG}, and (b) the predictive values based on mode-band matching (Eq~\eqref{EQ:vg_theory}).}
		\label{FIG:VG_surf}
	\end{figure}
	% ---------------------------------------------------
	
	Fig~\ref{FIG:VG_pred_ex}(a-e) provides an example of the mode-band matching routine to predict group velocity for a 20$\times$40 lattice with $\beta = 0.35$ and $N_{cyc} = 12$ driven at $\Omega = 1.62$. The energy centroid moves along the interface domain as computed by Eq~\eqref{EQ:Ecentroid} and is depicted as a red circle (Fig~\ref{FIG:VG_pred_ex}(a)). 
	%%%%%%%%%
	\textcolor{rev1}{
		The sparsity of the topological band gap is given in Fig~\ref{FIG:VG_pred_ex}(b), and the contribution of each mode of the finite system is given in Fig~\ref{FIG:VG_pred_ex}(c).
		These modal contributions are mapped to the supercell dispersion diagram in Fig~\ref{FIG:VG_pred_ex}(d) where marker colors indicate the relative modal contributions across the dispersion diagram.
		Fig~\ref{FIG:VG_pred_ex}(e) depicts the exceptional match between the estimated and predicted group velocities by applying Eq~\eqref{EQ:PredVG} with $\hat{n}$ taken as the number of modes required for the truncated solution to account for 98\% of the total modal energy. 
		To compare directly with Fig~\ref{FIG:VG_error_ex}, the same range of excitation bandwidths and frequencies are predicted per Eq~\eqref{EQ:PredVG}, and Fig~\ref{FIG:VG_pred_plt} shows that the modal-based velocity estimates are greatly improved in comparison to Eq~\eqref{EQ:vg_theory}.}

	To further demonstrate the generality of the mode-band matching method, we consider a range of contrast parameters for $\beta\in(0.25,0.35)$, excitation frequencies $\Omega \in (1.5,1.8)$, and excitation bandwidths corresponding to $N_{cyc}\in(10,50)$. %Fig~\ref{FIG:stripdispall} shows the dispersion diagram for each contrast parameter along with the corresponding group velocities, and it is clear that the range of $\Omega$ considered includes frequencies above the effective bandwidth for lower contrast parameters. 
	The predicted and observed group velocities are shown in Fig~\ref{FIG:VG_surf}. The estimated group velocities computed via Eq~\eqref{EQ:Ecentroid} match remarkably well with the predicted group velocities computed via Eq~\eqref{EQ:PredVG} as compared to the more error prone theoretical predictions rendered by~\eqref{EQ:vg_theory}. 
	Furthermore, the effects of the spectral spreading resulting from the uncertainty principle are captured by the modal superposition method as well, whereas the theoretical estimates based on of Eq~\eqref{EQ:vg_theory} obviously cannot account for such effects.
	\textcolor{rev1}{
		Moreover, the mode-band matching method of Eq~\eqref{EQ:PredVG} performs well also for frequencies outside of the effective topological bandwidth, which implies that this method can provide accurate group velocity estimations over a wide spectral range of the phononic systems considered and may not be specific to topological waveguides.}

	% ---------------------------------------------------
	\section{Proportionally Damped Lattice} 
	\label{Sec:Damping}
	% ---------------------------------------------------
	With the framework established to analyze topological wave propagation through the lens of the finite modal basis for conservative systems,
	we may approach problem of \textcolor{rev1}{viscously} damped topological wave propagation by considering the damped wave as a superposition of decaying modes. We do so by incorporating a proportional damping matrix $\textbf{C}$ into Eq~\eqref{EQ:EOM_standard},
	\begin{equation}
		\textbf{M}\ddot{\bm{u}} +\textbf{C}\dot{\bm{u}}  + \textbf{K}\bm{u}= 
		\textbf{M}\ddot{\bm{u}} + \left[\lambda\textbf{K}+\varepsilon\textbf{M}\right]\dot{\bm{u}}  + \textbf{K}\bm{u}
		= \bm{f}(t),
	\end{equation}
	where $\lambda$ and $\varepsilon$ are taken as scalar constants so that the relation~\eqref{EQ:eval_prob} provides valid eigenmodes for the damped system. Applying the modal coordinate transformation, the system may be decoupled as before,
	\begin{equation}
		\ddot{\bm{\eta}} + \hat{\textbf{C}}\dot{\bm{\eta}}  + \bm{\omega}^2\bm{\eta}
		= \bm{\Phi}^{\intercal}\bm{f}(t),
	\end{equation}
	where $\hat{\textbf{C}} = \bm{\Phi}^{\intercal}\textbf{C}\bm{\Phi} = {\rm diag}\begin{bmatrix}2\zeta_1\omega_1 & \dots & 2\zeta_n\omega_n\end{bmatrix}$ is a matrix of modal damping coefficients where $\zeta_i$ is the modal damping ratio of the $i$-th eigenmode. The damped modal solutions are
	\begin{equation}
		\begin{aligned}
			&\eta_i(t) = \int_{0}^{t} \bm{\varphi}_i^{\intercal}\bm{f}(\tau)
			\exp\left[-\zeta_i\omega_i(t-\tau)\right] \sin\omega_{i,d}(t-\tau){\rm d}\tau  \ \ \ t\in [0,T]  \\
			&\eta_i(t) = C_{i,d} \exp\left[-\zeta_i\omega_i(t-T)\right]\cos(\omega_{i,d}t-\psi_{i,d}-\omega_{i,d}T) \ \ \  t  \in (T,\infty]
		\end{aligned}
		\label{EQ:modal_sol_dam}
	\end{equation}
	where $\omega_{i,d} = \omega_i\sqrt{1-\zeta_i^2}$ is the damped natural frequency of the $i$-th mode, $C_{i,d} = (\eta_i^2(T)\omega_i^2 + \dot{\eta}_i^2(T)+2\eta_i(T)\dot{\eta}_i(T)\zeta_i\omega_i)^{1/2}/\omega_{i,d}$, and $\psi_{i,d} = \arctan\left( \frac{\dot{\eta}_i(T)+\zeta_i\omega_i\eta_i(T)}{\eta_i(T)\omega_{d,i}(T)}\right)$.
	The same modal expansion solution~\eqref{EQ:expansion} and truncated solution~\eqref{EQ:reconstruction} apply, however we note a few minor differences from the fully analytical solutions of Eq~\eqref{EQ:undamped_modal_sol}.
	The convolution integrals of Eq~\eqref{EQ:modal_sol_dam} are too complicated for standard computer algebra, and numerical integration must be employed to compute solutions. The velocity terms $\dot{\eta}_i$ may be obtained by recognizing that the following relation holds, $\frac{\rm d}{{\rm d}t}\int_0^{t}f(\tau)g(t-\tau){\rm d}\tau = \int_0^{t} \dot{f}(t)g(t-\tau){\rm d}\tau$.
	Moreover, in contrast to the undamped system where $C_i = \eta_i(T)$ remain valid after excitation, the solutions of the damped system are nonstationary for $t>T$ and the modal coefficients $C_i$ no longer represent the modal contribution in time. The time-dependent quantities $\eta_i(t)$ must be used instead.
	
	% ---------------------------------------------------
	\subsection{Edge-to-bulk transitions}
	% ---------------------------------------------------
	% ---------------------------------------------------
	\begin{figure}[t!]
		\centering
		\includegraphics[width=.5\textwidth]{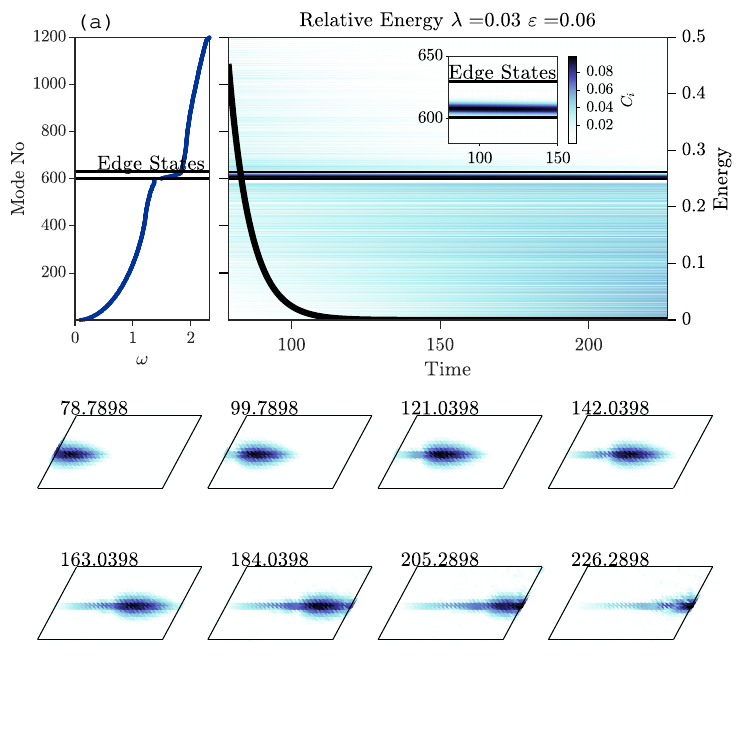}%
		\includegraphics[width=.5\textwidth]{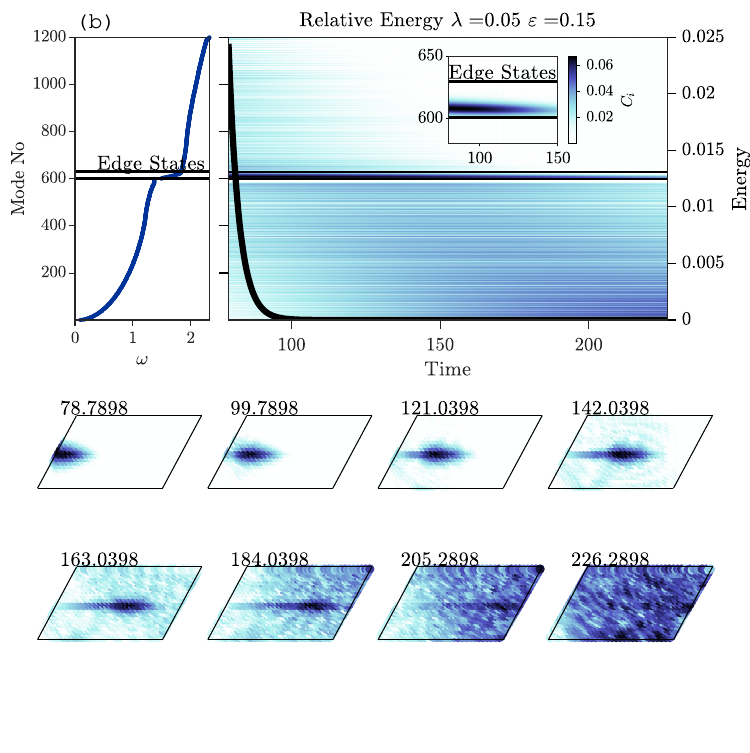}
		\caption{The finite band structure, evolution of the modal coefficients in time due to damping, and the evolution of energy across a 30$\times$20 topological lattice in time for (a) a damping configuration that preserves the edge wave and (b) a damping configuration that disperses the edge wave into a bulk wave. Insets of the color plots are zoomed in to show the evolution of the interface modal coefficients during the first seconds of decay.}
		\label{FIG:damp_ex}
	\end{figure}
	% ---------------------------------------------------
	
	The effect of various damping levels on topological wave propagation may now be studied from the perspective of modal reconstruction. 
	At the beginning of the topological wave propagation, the total energy of the lattice system is primarily trapped in a finite subset of closely spaced interface modes assuming that  the topological band is excited (cf.~section~\ref{Subsec: Modal Decomp}). 
	However, as time progresses, the instantaneous energy distribution changes throughout the modal spectrum. 
	While it is impossible for linear systems with proportional damping to exchange energy between eigenmodes, the rate at which energy dissipates in each mode differs based on the modal damping $\zeta_i$ and natural frequency $\omega_i$. 
	\textcolor{rev1}{
		For mechanical systems, mass and stiffness proportional damping is generally an accurate assumption, and therefore higher modes typically dissipate energy much quicker than lower modes.} 
	Hence, the relative energy captured by the lower modes is expected to dominate the dynamics for topological systems with sufficient size and damping.
	
	Fig~\ref{FIG:damp_ex} shows the damped responses of a configuration 1 system of size $20\times30$ lattice driven by an input wave packet of $\Omega = 1.6$ and $N_{cyc} = 20$.
	We consider two variations of proportional damping: a moderately damped system $(\lambda,\varepsilon) = (0.03,0.06)$ and a heavily damped system $(\lambda,\varepsilon) = (0.05,0.15)$.
	%%%%
	The color plots depict the instantaneous energy in each mode, whereas the black line corresponds to the right $y$-axis which depicts the total instantaneous energy in the damped lattice.
	Although the dissipated energy on a linear scale is seemingly comparable between the two systems, as is the transmitted energy, the dynamics vary greatly. 
	%%%%%
	While dissipation throughout the moderately damped system is quicker in the higher modes (and hence the lower modes gain relative energy in time), the energy primarily remains in the interface modes throughout the propagation of the wave from the left to right boundary, and hence the wave packet propagates along the interface until it reaches the right boundary. In contrast, for the strongly damped system, the modal damping factors of the edge states are substantially higher than those of the acoustic states. As a result, the relative instantaneous energy transitions from the edge states to the acoustics states and causes an \textit{edge-to-bulk} transition of the wave before the right boundary is reached.
	%%%
	Hence, the disbandment of a topological valley-Hall wave due to dissipation can be interpreted in a conventional modal decomposition and modal decay framework.
	%%%%%%%

	To anticipate whether or not an edge wave will remain on the interface of a given finite system, one may consider the evolution of the modal amplitudes by computing the modal damping factors.
	The transitions from edge to bulk wave are a product of the damping ratio and lattice size. 
	Surely, for a larger system, the time taken for the wave to reach the right boundary ($T_{rb}$) is larger, and thus the effect of the damping over the time interval $\tau$ is greater.
	If $T_{rb}$ is greater than the time required for the energy of the bulk spectrum to overtake the total energy of edge states, then the topological wave will remain at the interface. 
	%While it is typically not practical to engineering modal damping ratios, one may consider the influence of topol
	
\begin{figure}[t!]
		\centering
		\includegraphics[width=\textwidth]{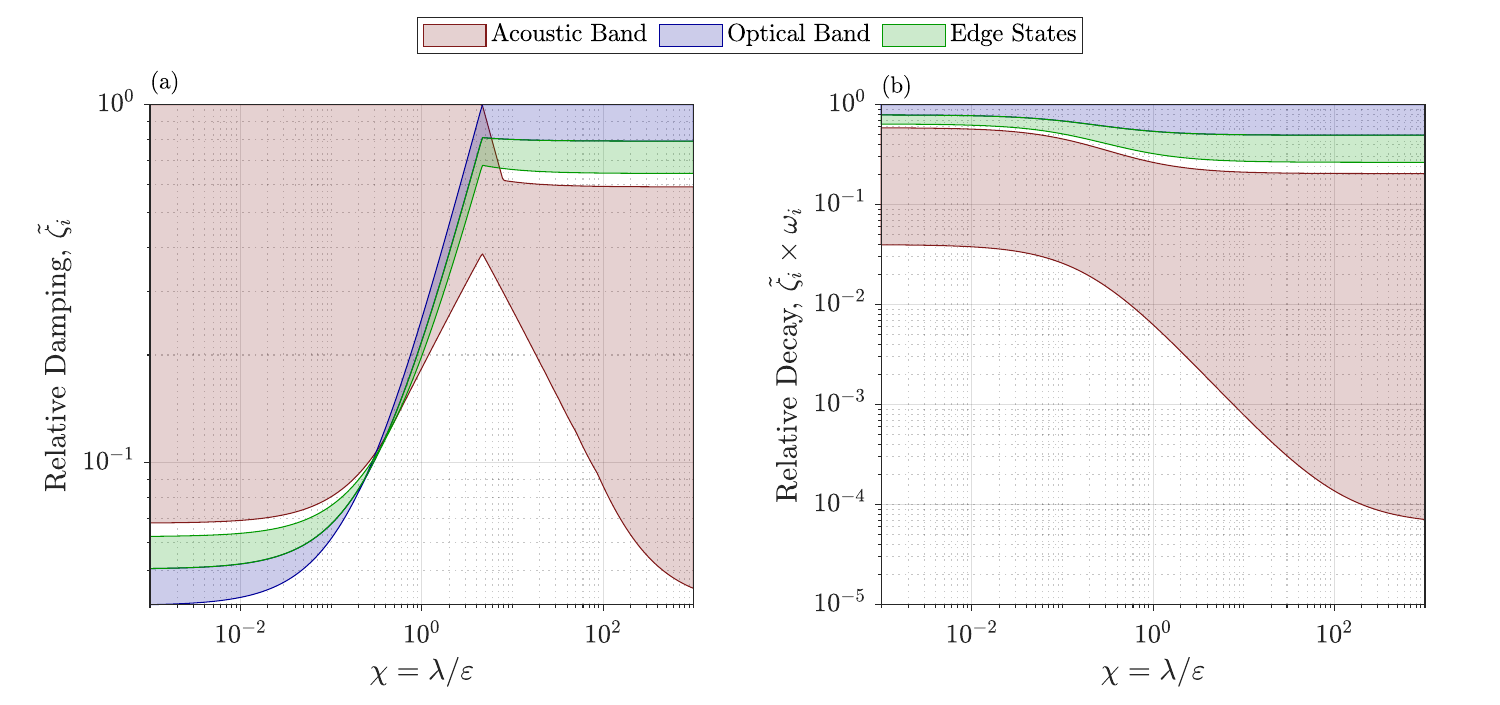}
		\caption{(a) The relative damping ratios ($\tilde{\zeta}_i$)  and (b) the relative 	dissipation rates s ($\tilde{\zeta}_i\times\omega_i$)  for each eigenmode with respect the ratio of stiffness-vs-mass proportional damping $\chi = \lambda/\varepsilon$ varying between $\mathcal{O}(10^{-3})$ to  $\mathcal{O}(10^{3})$.}
		\label{FIG:zeta_v_chi}
	\end{figure}

	The edge-to-bulk transitions will always occur from the edge states to the acoustic states for the proportionally damped lattice.
	Unless a damping matrix is \textit{engineered} such that $\zeta_{i\in{\rm edge}}\ll\zeta_{ i\in {\rm bulk}}$, the energy decay rates of edge modes will always be higher than those of acoustic modes.
	Fig~\ref{FIG:zeta_v_chi} shows the variation of the normalized damping ratios $\tilde{\zeta}_i$  versus the ratio of mass-vs-stiffness to construct the proportional damping, $\chi = \lambda/\varepsilon$. The normalized damping ratios of Fig~\ref{FIG:zeta_v_chi} are normalized so that for each $\tilde{\zeta}_i$ the maximum value is one.  While it is indeed seen in Fig~\ref{FIG:zeta_v_chi}(a) that values $\chi<\approx0.4$  permit lower damping ratios in the optical band than the acoustic one, and that there even exists a narrow range where the edge modes have the lowest damping ratio, the effective decay rate for each mode ($\tilde{\zeta}_i\times\omega_i$) is always sorted such that the acoustic modes have the lowest decay rate in the system
	(Fig~\ref{FIG:zeta_v_chi}(b)). Hence, for any  practical damping consideration, the energy will dissipate from the edge and optical modes more rapidly (cf.~Eq~\eqref{EQ:modal_sol_dam}).

	% ---------------------------------------------------
	% ------------------------------------------------------------------------------------------------------

	% ---------------------------------------------------
	\subsection{Instantaneous Group Velocity}
	% ---------------------------------------------------
	The variation of relative energy over the modal spectrum in time will certainly influence the group velocity of a damped propagating topological traveling wave (cf.~section~\ref{Sec:VG}). 
	Hence, the group velocity becomes a nonstationary function of time and damping ratio of the modes. 
	%%%%%%%
	Accounting for this non-stationarity may predict the \textit{instantaneous} group velocity of a damped traveling wave by applying Eq~\eqref{EQ:PredVG} in time with the damped coefficients of Eq~\eqref{EQ:modal_sol_dam}. Moreover, we implement Eq~\eqref{EQ:PredVG} using the damped dispersion  relation as presented in~\cite{Hussein2010} as an alternative to the standard supercell dispersion $\tilde{\omega}(\kappa_x)$ of section~\ref{Sec:VG}.
	
\begin{figure}[t!]
		\centering
		\includegraphics[width=.65\textwidth]{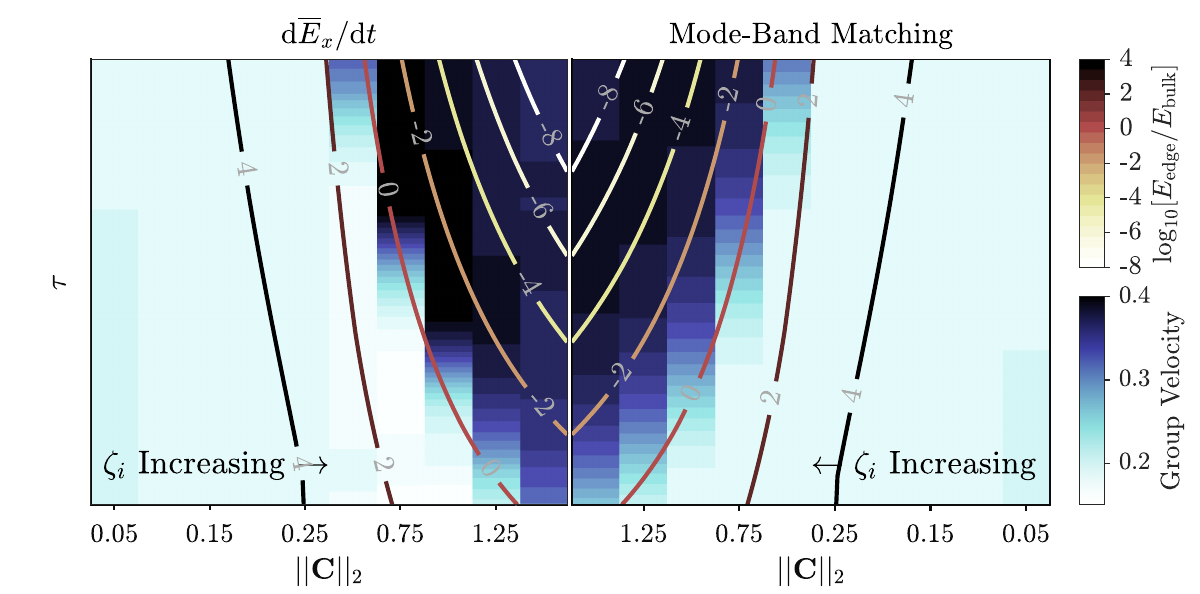}
		\caption{The instantaneous group velocity of a configuration 1 lattice subject to a damping matrix with various norms. The Left subplot depicts the observed group velocity whereas the right subplot depicts the predicted group velocity based on mode-band matching for each instant in the normalized time $\tau = \{t:T<t<T_{rb} \}$. The overlaying contours refers to the portion of edge-versus-bulk modes required to reconstruct 90\% of the wave energy. 
		}
		\label{FIG:Damped_VG_ex}
	\end{figure}

	Fig~\ref{FIG:Damped_VG_ex} depicts an example of the instantaneous group velocity as computed by the damped mode-band matching routine (Eq~\eqref{EQ:PredVG}) and as estimated per the energy centroid (Eq~\eqref{EQ:Ecentroid}) for a configuration 1 lattice of dimension $12\times30$ subject to a wave packet with $\Omega = 1.5$ and $N_{cyc} = 25$. 
	%%%%%%%%%%
	Proportional damping is considered with $\lambda = 2\varepsilon$ and scaled by the norm of the damping matrix, $||\textbf{C}||_2$, and we consider values of $||\textbf{C}||_2$ between 0.05 to 1.25. 
	%%%%%%%%%%
	Fig~\ref{FIG:Damped_VG_ex} shows the evolution of group velocity with respect to the normalized time $\tau = \{t:T<t<T_{rb} \}$ where the normalization is necessary since the variation of group velocity causes $T_{rb}$ to be different for systems with equivalent force but nonequivalent damping.
	Contours depict the proportion of energy retained in bulk modes with respect to the energy retained in edge modes at each snap shot in time.
	When this number crosses the value of 1/2 (which corresponds to the contour of $\log_{10}[E_{\rm edge}/E_{\rm bulk} = 0]$), there is a stark increase in both observed and predicted group velocity as the bulk waves of the acoustic band become the dominant modes contributing to the propagation (which account for much higher group velocities, see, e.g.,~Fig~\ref{FIG:infinite_strip_dgm}). This interpretation is supported by the color plots of Fig~\ref{FIG:damp_ex} which clearly show the transition of energy from interface modes to acoustic band modes for the strongly damped systems. Moreover, the recorded and predicted group velocities peak at approximately $v_g = 0.4$ for strongly damped systems. The topological band is not steep enough at any point to support wave propagation faster than $v_g = 0.2$, and so this stark rise in group velocity is in direct correspondence with the edge-to-bulk transition time. The mode-band matching of Eq~\eqref{EQ:PredVG} is hence proven an effective tool not just for topological modes, but also for providing remarkably accurate wave velocity predictions during edge-to-bulk transitions.
	
	%	% ---------------------------------------------------
	% ------------------------------------------------------------------------------------------------------
	%	% ---------------------------------------------------
	\begin{figure}[t!]
		\centering
		\begin{subfigure}{.5\linewidth}
			\includegraphics[width=\textwidth]{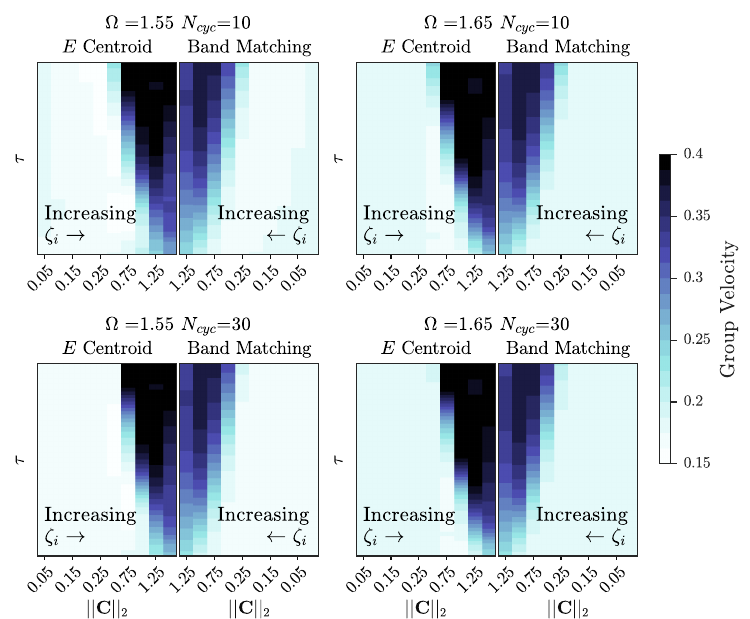}
			\caption{$\lambda = 2\varepsilon$}
			\label{FIG:damped_vg_all_rat2}
		\end{subfigure}%
		\begin{subfigure}{.5\linewidth}
			\includegraphics[width=\textwidth]{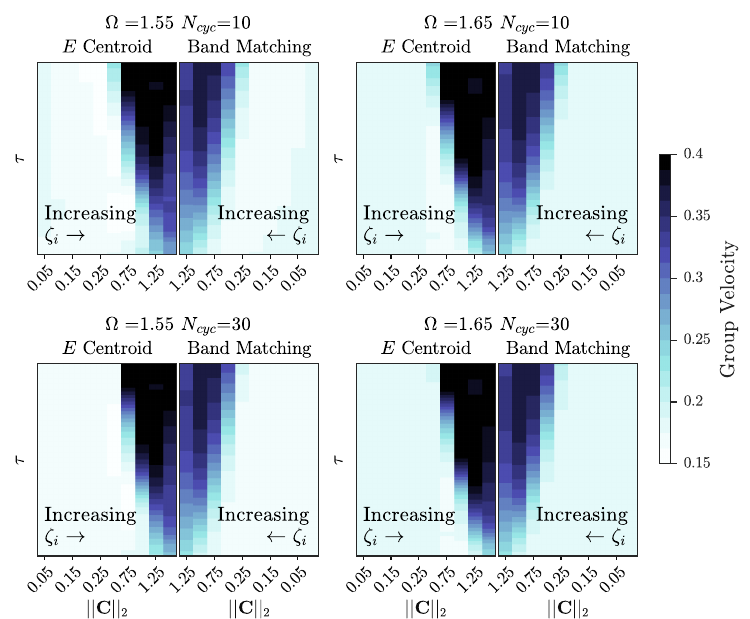}
			\caption{$2\lambda = \varepsilon$}
			\label{FIG:damped_vg_all_rat.5}
		\end{subfigure}
		\caption{The instantaneous group velocity of a configuration 1 lattice subject to a damping matrix with various norms  and damping proportions $\lambda = 2\varepsilon$. For each subplot, the  left plot depicts the observed group velocities whereas the right plot depicts the predicted group velocity based on mode-band matching for each instant in the normalized time $\tau = \{t:T<t<T_{rb} \}$.}
		\label{FIG:damped_vg_all_rat}
	\end{figure}
	% ----------------------------------------------------------------------------------------------
	
	To demonstrate the generality of this approach, we consider several variations of excitation frequency and input bandwidth for a system with $\beta = 0.35$; in addition, we consider the influence of mass versus stiffness proportionality of the proportional damping  coefficients $(\lambda,\varepsilon)$. The magnitude of damping on the system is varied by scaling $||\textbf{C}||_2$ between 0.05 and 1.25.  Figs~\ref{FIG:damped_vg_all_rat2} and \ref{FIG:damped_vg_all_rat.5} depict the instantaneous group velocities for systems with proportional damping constants $\lambda = 2\varepsilon$ and $2\lambda = \varepsilon$, respectively.
	When damping is sufficiently small (e.g., $||\textbf{C}||_2\leq.15$), the propagating topological waves maintain a near constant group velocity and most of the energy remains at the interface for nearly all excitation profiles; hence the topological wave is preserved. 
	%%%
	When the damping norms grow higher, however, steep increases in the group velocity are shown for finite times $t<T_{rb}$ which correspond directly to the dominant acoustics transitioning from the interface to acoustic bulk states.
	This is observed for both $\lambda = 2\varepsilon$ and $2\lambda = \varepsilon$ damping proportions with similar time-scales observed for bulk-to-edge transitions to occur.
	%%%
	Moreover, there is good agreement between predicted and observed group velocities for each considered excitation profile and damping configuration. 
	%%%
	While the numerical values of the group velocity differ to some degree when $v_g>0.3$, the transition from edge-to-bulk state (denoted by the sudden spike in group velocity) is accurately predicted by the instantaneous application of Eq~\eqref{EQ:PredVG} for each excitation profile and damping configuration shown. Hence, the modal solution is proven to be a valuable and predictive tool for predicting the group velocity and edge-to-bulk transitions of a damped valley Hall propagating topological wave.

	\section{Concluding Remarks} 
	\label{Sec:Conclusion}
	% ---------------------------------------------------
	In this work, we studied a valley-Hall topological interface lattice through the lens of modal superposition. 
	Employing the well known modal expansion theorem, analytically computed modal coefficients were recovered for each finite mode required to reconstruct the wave dynamics.
	\textcolor{rev1}{
		Moreover, the truncated modal solutions  demonstrated that only a narrow subset of spatially and spectrally localized eigenmodes is required to accurately  reconstruct the topological wave even when it propagates across complex zigzagged domains.}
	This contradicts the modal expansion of trivial wave propagation which requires many modes to reconstruct a traveling wave, \textcolor{rev1}{as is the case for bulk waves or trivial waveguides which inevitably backscatter at corners.}
	\textcolor{rev1}{ 
		This demonstrates that topological wave propagation is based on the excitation of a limited set of closely spaced modes within the band gap, oscillating with the appropriate phase differences to construct a localized propagating wave packet along the prescribed domain boundary of the finite lattice.
		In fact, it is well known that such degenerate closely spaced modes with appropriate phase differences may yield traveling waves in systems with symmetry, such as circular or square plates. 
		Furthermore, this is a linear effect (as are the topological lattices considered herein).}
	
	By employing the results of modal decomposition, a formula was provided to accurately predict the true group velocity of propagating topological waves in finite 2D lattice systems, referred to as the mode-band matching method. While the standard (classical) definition group velocity cannot account for the finite modal spectrum excited by finite-band excitations, the mode-band matching method developed herein can, yielding greatly improved group velocity predictions across the topological bandwidth. Moreover, the same method was proven to be effective for damped topological wave propagation, and it was shown that edge-to-bulk transitions corresponding to sudden spikes in group velocity occur due to the relative energy shifting to the acoustic band for strongly damped systems. 
	
	This alternative modal based perspective to topological propagation has many implications on the topological insulation community which has traditionally relied predominantly on dispersion characteristics to predict or evaluate topological insulator performance. 
	%%%%
	\textcolor{rev1}{Every finite linear system is governed by the same principle of modal superposition.} Hence, we have developed a tool that can greatly refine the computational predictions of topological protection and be employed for optimization routines similar to those of~\cite{Chen2021,Du2020,Christiansen2019}. 
	%%%
	\textcolor{rev1}{Moreover, given the drastic speed up in computation time that the truncated modal solution offers, such iterative optimization efforts could be rapidly accelerated if explored in the finite domain.}
	%%%
	\textcolor{rev1}{In addition, it should be studied how the topology of the dispersion relations are related to the convergence of the modal decompositions, so that optimization of the lattice structure could be proposed.}
	%%%%
	For example, how can one influence the density of the underlying closely space modes and their phases, since these affect greatly the propagating topological wave.
	%%%%
	%%%%
	\textcolor{rev1}{Future works could explore the same modal based methodologies applied to more realistic 3D elastic domains governed by the balance of linear momentum. While one would expect the findings of section~\ref{Subsec: Modal Decomp} to apply to other band topologies that are capable of producing spectrally and spatially localized edge states in the band gap, future studies could explore this directly by studying helical waves produced by spin-Hall and combined models from the perspective of modal superposition.}
	%%%
	Additional extensions of this methodology could include topics such as optimization of topological wave performance across the entirety of the effective topological bandwidth, optimization based on stochastic or unknown excitation profiles, damping matrix optimization for group velocity control, and nonlinear continuation for traveling waves using a greatly reduced basis of modes.
	
	% --------------------------------------------------- 
	\vspace{14pt}
	\noindent\textbf{Acknowledgments}\\
	\noindent This work is supported in part by the National
	Science Foundation Graduate Research Fellowship Program, Grant No.~DGE – 1746047. Any opinions, findings, and
	conclusions or recommendations expressed in this material
	are those of the authors and do not necessarily reflect the
	views of the National Science Foundation.
	\textcolor{rev1}{
		This work made use of the Illinois Campus Cluster, a computing resource that is operated by the Illinois Campus Cluster Program (ICCP) in conjunction with the National Center for Supercomputing Applications (NCSA) and which is supported by funds from the University of Illinois at Urbana-Champaign. The authors would like to acknowledge the anonymous reviewers for their valuable comments which has greatly improved this work.}
	
	\appendix
	
	\newpage
	\section{Energy distribution in the oscillators}
	\label{APX:energy_table}
	The energy of each oscillator is computed by summing the kinetic and potential terms.
	For springs that couple oscillators, the potential energy of the coupling springs is assumed to be distributed equally between two oscillators that share the spring. 
	The  resulting energy expressions are listed in Table~\ref{TAB:Energy}
	where $\partial\mathcal{D}_B$, $\partial\mathcal{D}_T$, $\partial\mathcal{D}_L$, $\partial\mathcal{D}_R$, refer to the bottom, top, left, and right boundaries of the finite lattice, respectively.  
	
	\begin{table}[h!]
		\centering
		\caption{Expressions for energy at each location}
		\label{TAB:Energy}
		\begin{tabular}{l|l}\hline\hline
			Domain&	$E_{\alpha}^{p,q}(t)$ \\ \hline
			$\alpha = a, \ (p,q)\in\partial\mathcal{D}_B$ &$\frac{m_a}{2}(\dot{u}_{a}^{p,q})^2 + \frac{k}{2}({u}_{a}^{p,q})^2 +  \frac{k}{4}\left[({u}_{a}^{p,q}-{u}_{b}^{p,q})^2 + ({u}_{a}^{p,q}-{u}_{b}^{p-1,q})^2 \right]$\\
			$\alpha = a, \ (p,q)\in\partial\mathcal{D}_T$ &$\frac{m_a}{2}(\dot{u}_{a}^{p,q})^2 + \frac{k}{4}\left[({u}_{a}^{p,q}-{u}_{b}^{p,q})^2 + ({u}_{a}^{p,q}-{u}_{b}^{p,q-1})^2+({u}_{a}^{p,q}-{u}_{b}^{p-1,q})^2 \right]$\\
			$\alpha = a, \ (p,q)\in\partial\mathcal{D}_L$ &$\frac{m_a}{2}(\dot{u}_{a}^{p,q})^2 + \frac{k}{2}({u}_{a}^{p,q})^2 +  \frac{k}{4}\left[({u}_{a}^{p,q}-{u}_{b}^{p,q})^2 + ({u}_{a}^{p,q}-{u}_{b}^{p-1,q})^2 \right]$ \\
			$\alpha = a, \ (p,q)\in\partial\mathcal{D}_R$ &$\frac{m_a}{2}(\dot{u}_{a}^{p,q})^2 +  \frac{k}{4}\left[({u}_{a}^{p,q}-{u}_{b}^{p,q})^2 + ({u}_{a}^{p,q}-{u}_{b}^{p-1,q})^2 +({u}_{a}^{p,q-1}-{u}_{b}^{p-1,q})^2 \right]$\\
			$\alpha = a, \ (p,q)\in \mathcal{D}\setminus \partial\mathcal{D}$ &$\frac{m_a}{2}(\dot{u}_{a}^{p,q})^2 + \frac{k}{4}\left[({u}_{a}^{p,q}-{u}_{b}^{p,q})^2 + ({u}_{a}^{p,q}-{u}_{b}^{p-1,q})^2 +({u}_{a}^{p,q-1}-{u}_{b}^{p-1,q})^2 \right]$\\
			$\alpha = b, \ (p,q)\in\partial\mathcal{D}_B$ &$	\frac{m_b}{2}(\dot{u}_{b}^{p,q})^2 +  \frac{k}{4}\left[({u}_{b}^{p,q}-{u}_{a}^{p,q})^2 + ({u}_{b}^{p,q}-{u}_{a}^{p+1,q})^2+({u}_{b}^{p,q}-{u}_{a}^{p,q+1})^2 \right]$\\
			$\alpha = b, \ (p,q)\in\partial\mathcal{D}_T$ &$\frac{m_b}{2}(\dot{u}_{b}^{p,q})^2 + \frac{k}{2}({u}_{b}^{p,q})^2 +  \frac{k}{4}\left[({u}_{b}^{p,q}-{u}_{a}^{p,q})^2 +({u}_{b}^{p,q}-{u}_{a}^{p+1,q})^2 \right]$ \\
			$\alpha = b, \ (p,q)\in\partial\mathcal{D}_L$ &$\frac{m_b}{2}(\dot{u}_{b}^{p,q})^2 +   \frac{k}{4}\left[({u}_{b}^{p,q}-{u}_{a}^{p,q})^2 + ({u}_{b}^{p,q}-{u}_{a}^{p+1,q})^2+({u}_{b}^{p,q}-{u}_{a}^{p,q+1})^2 \right]$ \\
			$\alpha = b, \ (p,q)\in\partial\mathcal{D}_R$ &$\frac{m_b}{2}(\dot{u}_{b}^{p,q})^2 + \frac{k}{2}({u}_{b}^{p,q})^2 +  \frac{k}{4}\left[({u}_{b}^{p,q}-{u}_{a}^{p,q})^2 + ({u}_{b}^{p,q}-{u}_{a}^{p,q+1})^2 \right]$\\
			$\alpha = b, \ (p,q)\in \mathcal{D}\setminus \partial\mathcal{D}$ &$	\frac{m_b}{2}(\dot{u}_{b}^{p,q})^2 + \frac{k}{4}\left[({u}_{b}^{p,q}-{u}_{a}^{p,q})^2 + ({u}_{b}^{p,q}-{u}_{a}^{p-1,q})^2 +({u}_{b}^{p,q+1}-{u}_{a}^{p+1,q})^2 \right]$\\ \hline\hline
		\end{tabular}
	\end{table}
	
	\newpage

	% ---------------------------------------------------
	\clearpage
	\newpage
	\section{Effects of Size on Modal Decomposition}
\begin{figure}[h!]
		\includegraphics[width=\linewidth]{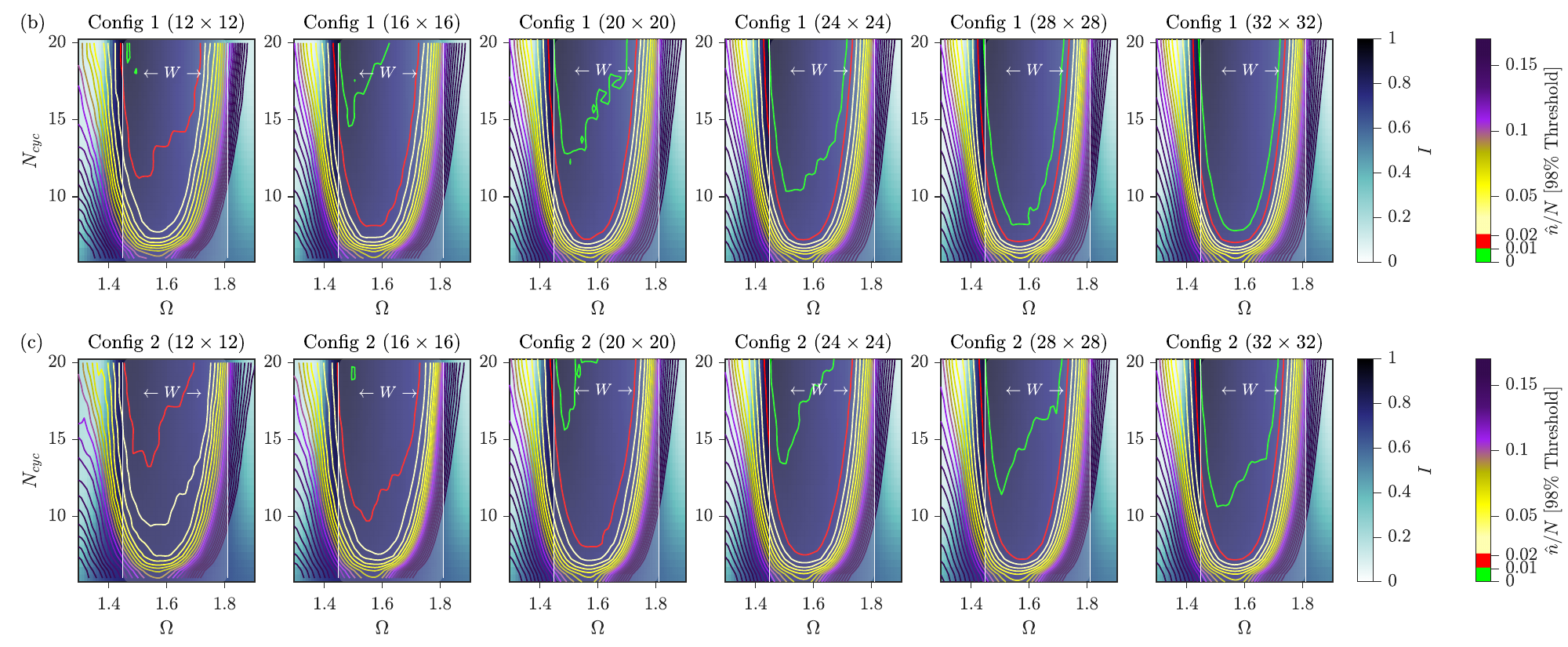}
		%%%%
		\caption{\textcolor{rev1}{The effect of lattice size on interface concentration and number of modes required for 98\% cumulative modal energy for (a) configuration 1 and (b) configuration 2 with $\beta = 0.3$. The colormap indicates interface energy concentration whereas the superimposed contours depict the number of modes of the modal decomposition  required  to reconstruct 98\% of the topological wave propagation. Green lines depict the parameter domain for which 1\% of modes account for the reconstruction, while red lines depict corresponding domain where 2\% of modes account for the reconstruction.}}
		\label{FIG:EvBW_sizeComp}
	\end{figure}

	\textcolor{rev1}{A main theme of this work is the interplay between the infinite dispersion of a topological system and the modal spectrum of the corresponding finite-size truncation.
		Therefore, the results presented in section~\ref{Subsec: Modal Decomp} must be verified across an array of finite system sizes in order to ensure the applicability of the discussed methods to finite lattices of arbitrary dimension.
		Of course, a limiting constraint is that the lattice should be sufficiently large so that the Bloch theorem is applicable, as the band properties which predict the localized wave travel obviously hinges on the assumption of an \textit{effectively} infinite periodic medium. 
		Therefore, we consider an extension of the study of decomposition contour plots of Fig~\ref{FIG:E_vs_BW} but with a fixed contrast parameter $\beta = 0.3$ and lattice dimensions ranging from $12\times12$ to $32\times32$.
		The results were generated for lattices of both configurations 1 and 2 of Fig~\ref{FIG:SysDescription}.  
		Fig~\ref{FIG:EvBW_sizeComp} depicts the results of this study where it is seen that even for the smallest considered configuration, only 2\% of the modal spectrum is required to accurately capture the topological wave propagation up to an energy reconstruction threshold of 98\%.
		The number of modes required to accurately construct the zigzagged domain of configuration 2 is seemingly higher than for the straight-edge domain of configuration 1 for all sizes, however this difference is not significant.
		As the lattice dimension grows, the relative number of required modes decreases, indicating a direct relationship between the size of the finite lattice and the relative number of modes describing the topological wave propagation; this is expected as the interface modes are spatially localized, so increasing the size of the bulk should not increase the number of modes required to reconstruct a traveling wave packet if the proposition holds that topological wave transport is accurately described by relatively few spatially and spectrally localized modes of the finite lattice system. 
	}
	\label{APX: sizeComp}
	% ---------------------------------------------------
	
	% ---------------------------------------------------
	\clearpage
	\newpage
	\section{Additional Example: Double zigzag}
	\label{APX: doubleZigZag}
\begin{figure}[h!]
		\begin{subfigure}{.475\linewidth}
			\includegraphics[width=\textwidth]{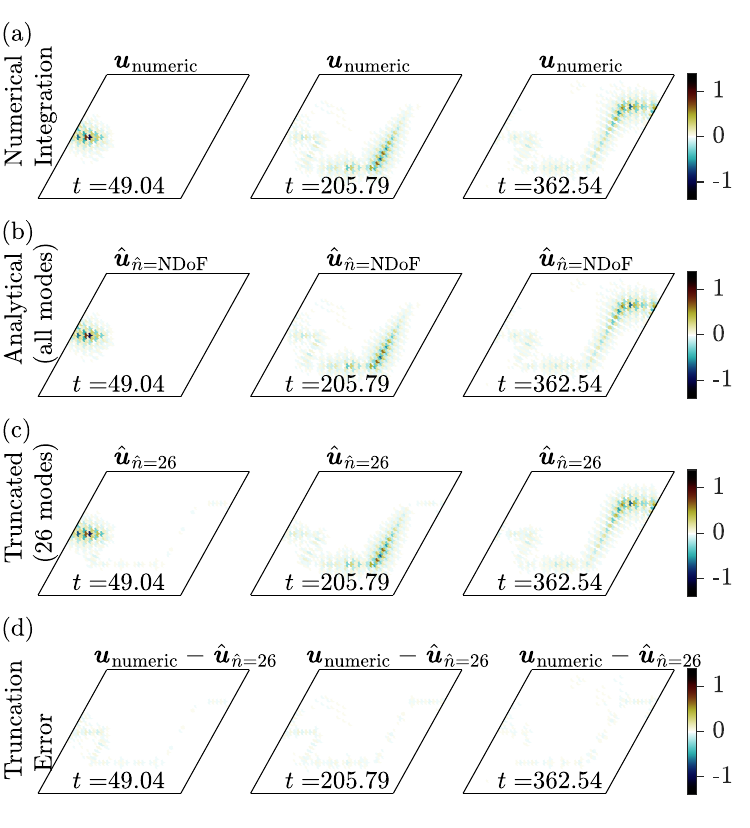}
		\end{subfigure} \hfill
		\begin{subfigure}{.475\linewidth}
			\includegraphics[width=\textwidth]{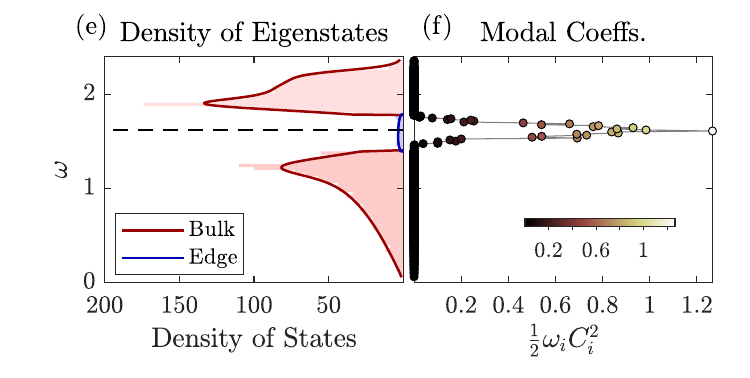}
			\includegraphics[width=\textwidth]{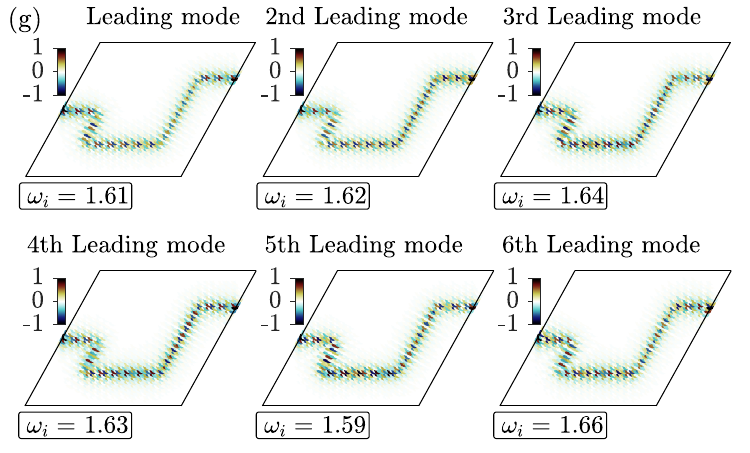}
		\end{subfigure}
		\caption{\color{rev1}{The modal decomposition of a topological wave in the double zigzagged lattice of dimension $36\times36$. (a) The numerical solution for the finite system, as well as (b) the analytical solution using all modes, and (c) a truncated analytical solution using only 26 modes; (d) depicts the difference between the numerical and truncated solutions.
				(e) The histogram of eigenfrequencies denoting the ``density'' of the eigenstates across the spectrum  with a black dashed line denoting excitation frequency, and (f) the energy associated with each eigenmode computed with solution~\eqref{EQ:expansion} with respect to their placement in the spectrum.
				%(e) The modal velocities of the numerical solution with envelop color depicting contribution to the dynamics. Also shown is (f) the modal frequency versus mode number of the analytical solution~\eqref{EQ:expansion} with color depicting contribution to the dynamics. 
				(g) The leading 6 eigenmodes used to construct the propagating topological wave as determined by their relative contributions in modal energy.}}
		\label{FIG:modal_decomp_sys3}
	\end{figure}

	% ---------------------------------------------------
	\textcolor{rev1}{The main text offered a detailed study of the decomposition of a zigzagged interface (termed configuration 2). To further support our claim that the topological wave propagation can be accurately described by a very narrow set of interface eigenmodes irrespective of the domain boundary path, we include here the same analysis applied to a double zigzagged system. 
		Fig~\ref{FIG:modal_decomp_sys3} depicts the same decomposition and truncated solution for the double zigzagged system as was described for Figs~\ref{FIG:modal_decomp_trivial} and~\ref{FIG:modal_decomp} of the main text. 
		As was the case for the single zigzag model, the topological wave propagation across the double zigzagged boundary is described by a limited subset of eigenmodes that are spatially localized along the domain boundary and spectrally localized in the band gap. Namely, the topological band gap of Fig~\ref{FIG:modal_decomp_sys3} exhibits the same sparsity, and the majority of the energy is possessed by eigenmodes localized at the domain boundary. Moreover, the truncation error using only 1\% of the total modes of the lattice is seemingly negligible, as was the case for the single zigzag model depicted in Fig~\ref{FIG:modal_decomp}.
	}
	%% ---------------------------------------------------

\begin{figure}[h!]
		\includegraphics[width=\linewidth]{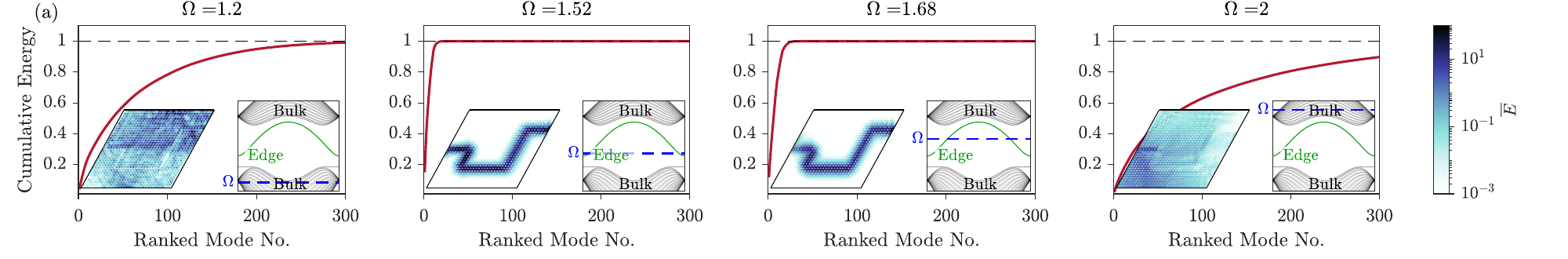}
		\includegraphics[width=\linewidth]{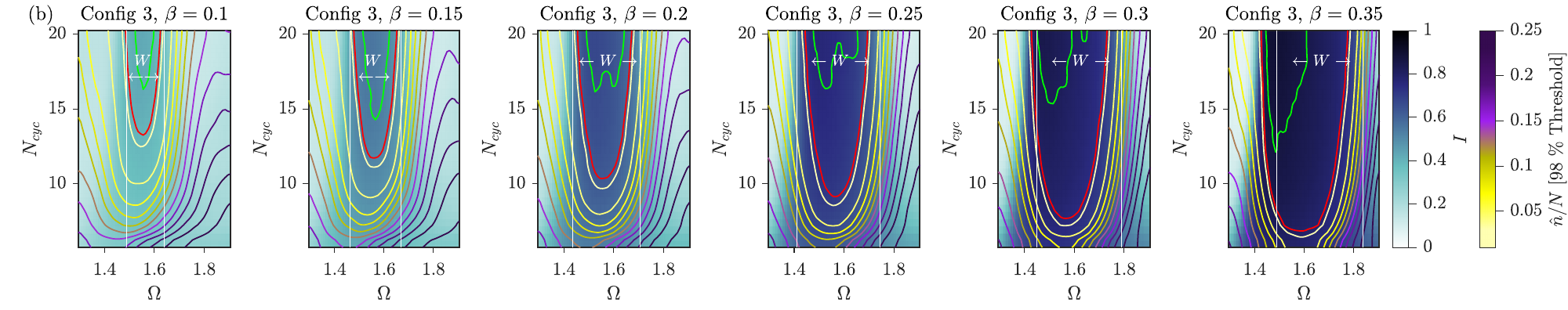}
		\caption{\textcolor{rev1}{(a) The cumulative variance of the ordered modal solution with respect to total modal energy accounted for by the leading 500 eigenmodes of the double zigzag model computed for excitation frequencies below ($\Omega<1.45$), inside ($1.45<\Omega<1.79$), and above ($\Omega>1.79$) the effective insulator bandwidth for a lattice with $\beta = 0.35$ with results depicted for finite constructions.
				The right insets depict a zoomed in view of the supercell band structure with a blue green line denoting the topological band, gray lines denoting bulk bands, and a dashed blue line denoting the location of the excitation frequency with respect to the band structure.
				(b) The energy intensities at the interface for various values of $\beta$ between 0.1 and 0.35 as a function of input signal bandwidth and frequency. The domains for which only 1\% and 2\% of the modes are required for the desired reconstructed energy are depicted by green and red lines, respectively.}}
		\label{FIG:EvBW_Config3}
	\end{figure}

	\textcolor{rev1}{The analysis described by Fig~\ref{FIG:EvBW_Config3} was conducted over a range of excitation frequencies and bandwidths, as well as a range of contrast parameters in order to directly confirm that the result of Fig~\ref{FIG:E_vs_BW} of the main text hold true for the double zigzag model as well. Fig~\ref{FIG:EvBW_Config3}(a) depicts examples of the cumulative modal energy of the rank-ordered eigenmodes as the excitation frequency passes from the lower bulk bands, thorough the topological band gap, and into the upper bulk bands, with insets depicting the time-averaged energy across the domain. Clearly a sharp corner is formed for the cumulative reconstruction once the topological wave is excited, which corresponds directly to the concentrated energy along the domain boundary. Fig~\ref{FIG:EvBW_Config3}(b) depicts the energy intensity at the interface $I$ as well as the number of modes required for a 98\% energy reconstruction. Very similar findings are reported here as were already stated for Fig~\ref{FIG:E_vs_BW} in section~\ref{Subsec: Modal Decomp} of the main text. Namely, that very few modes comprise the wave propagation if the excitation sufficiently excites the topological state and the contrast is high enough for sufficient spectral isolation of the interface states.
		%%%
		This directly supports the proposition that so long as the topological wave packet propagates along the domain boundary without significant backscatter, the wave can be accurately described by the spatially and spectrally  localized subset of interface modes possessing the appropriate phase differences to produce a spatially localized propagating wave packet along the domain boundary.
		%%%
		Hence, Fig~\ref{FIG:EvBW_Config3} further supports the finding that the modal description of topological wave propagation is seeming independent of the complexity of the domain boundary.
	}
	% ---------------------------------------------------
	% ---------------------------------------------------

	%% ---------------------------------------------------
	%\newpage
	%\section{Decomposition of Bulk Waves}
	%\label{APX: Bulk}
	%% ---------------------------------------------------

	% ---------------------------------------------------
	\clearpage
	\newpage
	\section{Trivial Waveguide}
	\label{APX: Trivial}
	% ---------------------------------------------------
	
	% \subsection{Trivial Waveguide}
	% ---------------------------------------------------
	\begin{figure}[b!]
		\begin{subfigure}{.475\linewidth}
			\includegraphics[width=\textwidth]{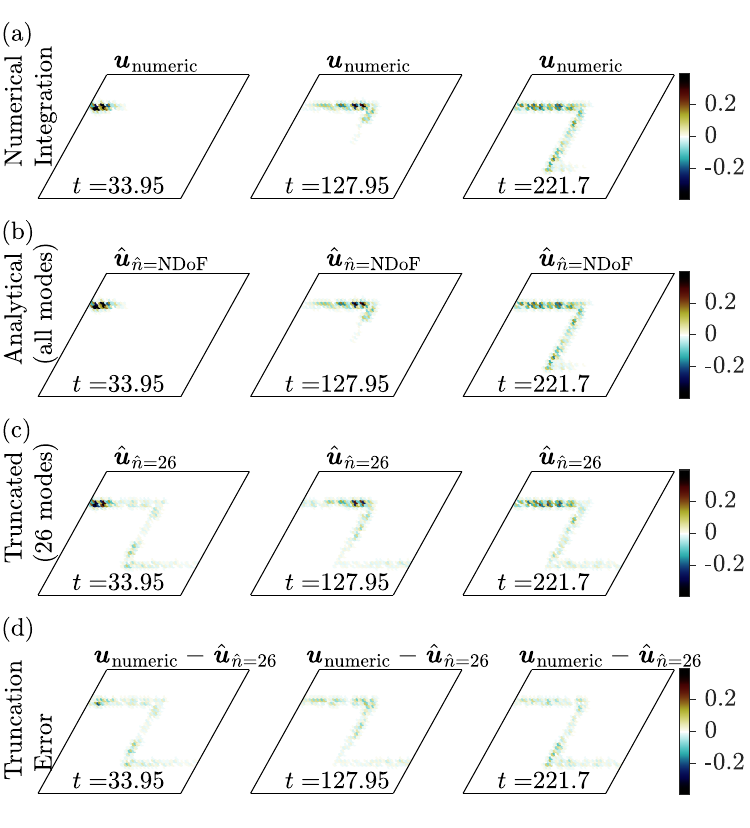}
		\end{subfigure} \hfill
		\begin{subfigure}{.475\linewidth}
			\includegraphics[width=\textwidth]{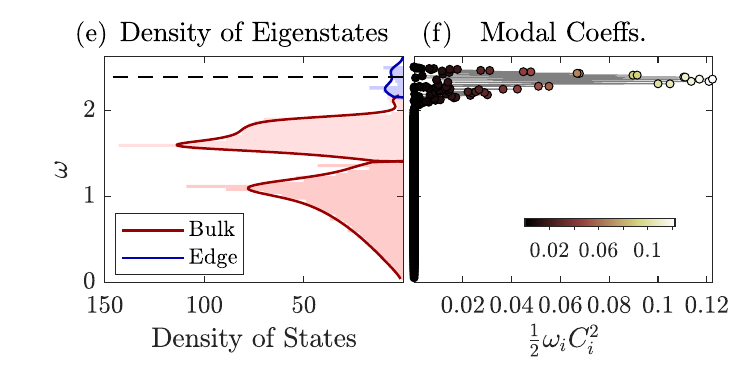}
			\includegraphics[width=\textwidth]{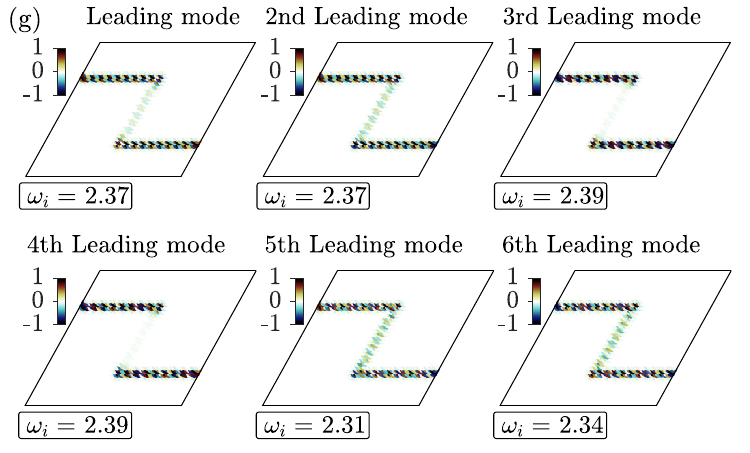}
		\end{subfigure}
		\caption{\textcolor{rev1}{The modal decomposition of the propagating trivial zigzagged wave guide. (a) The numerical solution for the finite system as well as (b) the analytical solution using all modes and (c) the truncated analytical solution using only 10 modes. (d) Depicts the difference between the numerical and truncated solution. 
				(e) The histogram of eigenfrequencies denoting the ``density'' of the eigenstates across the spectrum  with a black dashed line denoting excitation frequency, and (f) the energy associated with each eigenmode computed with solution~\eqref{EQ:expansion} with respect to their placement in the spectrum.
				%(e) The modal velocities of the numerical solution with envelop color depicting contribution to the dynamics. Also shown is (f) the modal frequency versus mode number of the analytical solution~\eqref{EQ:expansion} with color depicting contribution to the dynamics. 
				(g) The leading 6 eigenmodes used to construct the propagating topological wave as determined by the relative contribution of modal energy.}}
		\label{FIG:modal_decomp_waveguide}
	\end{figure}
	% ---------------------------------------------------
	\textcolor{rev1}{
		The main text focused primarily on the modal decomposition of topological waveguides within the context of reconstructing the propagating topological wave using a limited subspace of the modal spectrum. However, it should be stated that the acoustics of any linear waveguide can be reconstructed using the modal spectrum, as every linear system obeys modal superposition. Therefore, we compare here the same analysis applied to a trivial (non-topological) wave that propagates along the same paths as configurations 1 and 2 (cf.~Fig~\ref{FIG:SysDescription}(e,f))  discussed in the main text. The goal is to shed light on to how the decomposition (and truncated modal solutions) of localized topological states differ from the decomposition of trivial localized waveguides.
		To do so, we construct the trivial waveguide by considering the masses of domains 1 and 2 ($\Omega_1$ and $\Omega_2$ of Fig~\ref{FIG:SysDescription}(e,f)) to all be \textit{heavy} masses $(m_a=m_b=m+\beta)$, and the masses in the interface domain $\mathcal{I}$ to be light masses $(m_a=m_b=m-\beta)$, with the same nominal mass $m=1.25$ and spring stiffness $k=1$. This results in localized modes above the bulk-bands, which can be excited at the lattice boundary. 
	}
	
	\textcolor{rev1}{
		We begin with an example of the decomposition of the trivial waveguide applied to the zigzagged domain (configuration 2).
		Fig~\ref{FIG:modal_decomp_waveguide} depicts the same analysis described for Figs~\ref{FIG:modal_decomp_trivial} and~\ref{FIG:modal_decomp} of the main text applied to the trivial waveguide. Localized wave motion can clearly be observed at the onset of the wave propagation for both the numerical (Fig~\ref{FIG:modal_decomp_waveguide}(a)) and full analytical (Fig~\ref{FIG:modal_decomp_waveguide}(b)) solutions. However, when using the reduced basis comprised of the leading 1\% of modes (Fig~\ref{FIG:modal_decomp_waveguide}(c)), the localized wave packet is accompanied by energy dispersed throughout the spatial extent of the waveguide, leading to appreciable 50\% reconstruction error (Fig~\ref{FIG:modal_decomp_waveguide}(d)). For comparison, the same error quantity is only 10\% for the topological counterpart of the same lattice configuration.
	}
	
	\textcolor{rev1}{	
		Considering Fig~\ref{FIG:modal_decomp_waveguide}(e), it is apparent that the trivial edge states which form above the bulk bands are spectrally localized and sparse as compared to the bulk spectrum, although not nearly as spectrally sparse as what was seen for the equivalent topological lattice (cf.~Fig~\ref{FIG:modal_decomp}).  Fig~\ref{FIG:modal_decomp_waveguide}(f) shows that the distribution of excited eigenmodes is not nearly as narrow band. Moreover, there is no clear distribution of excited modes as is the case for topological wave guides. 
		Despite the delocalization of the excited spectrum, the leading eigenmodes themselves are spatially localized around the domain boundary (Fig~\ref{FIG:modal_decomp_waveguide}(g)).
		However, since backscatter is realized when the wave reaches the boundary (Fig~\ref{FIG:modal_decomp_waveguide}(a)), these modes alone will not reconstruct the complicated acoustics, as the modes would have to undergo a phase shift for the superimposed motion to transition from localized wave packet at the interface (with one dominant direction) to a scattered burst of energy that propagates both directions across the lattice. %Such a phase shift is not possible in the LTI system. 
		%%
		%Furthermore, while it is possible for few modes to reconstruct a traveling wave packet, the scattering behavior is far more complex.
		Therefore it is of little surprise that the truncated modal reconstruction of a trivial waveguide with a complex domain boundary cannot capture the wave propagation to nearly the same effectiveness as what was demonstrated for a topological wave guide with the same wave path.
	}
	
	\begin{figure}[t!]
		\includegraphics[width=.5\linewidth]{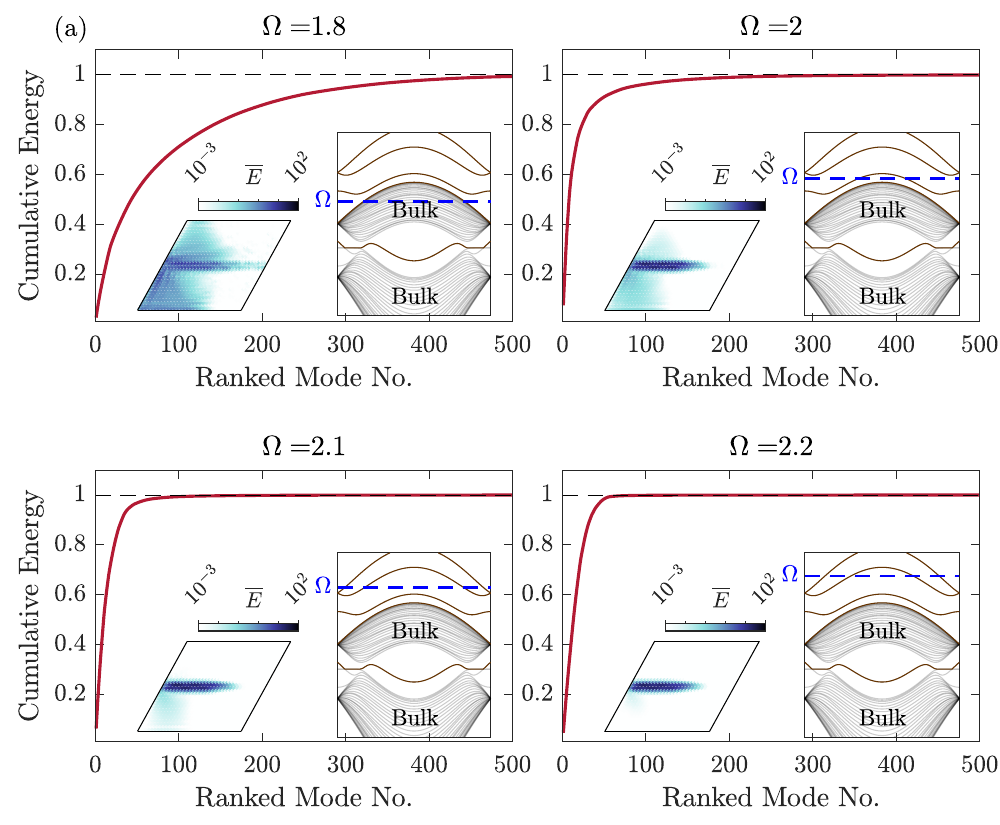}%
		\includegraphics[width=.5\linewidth]{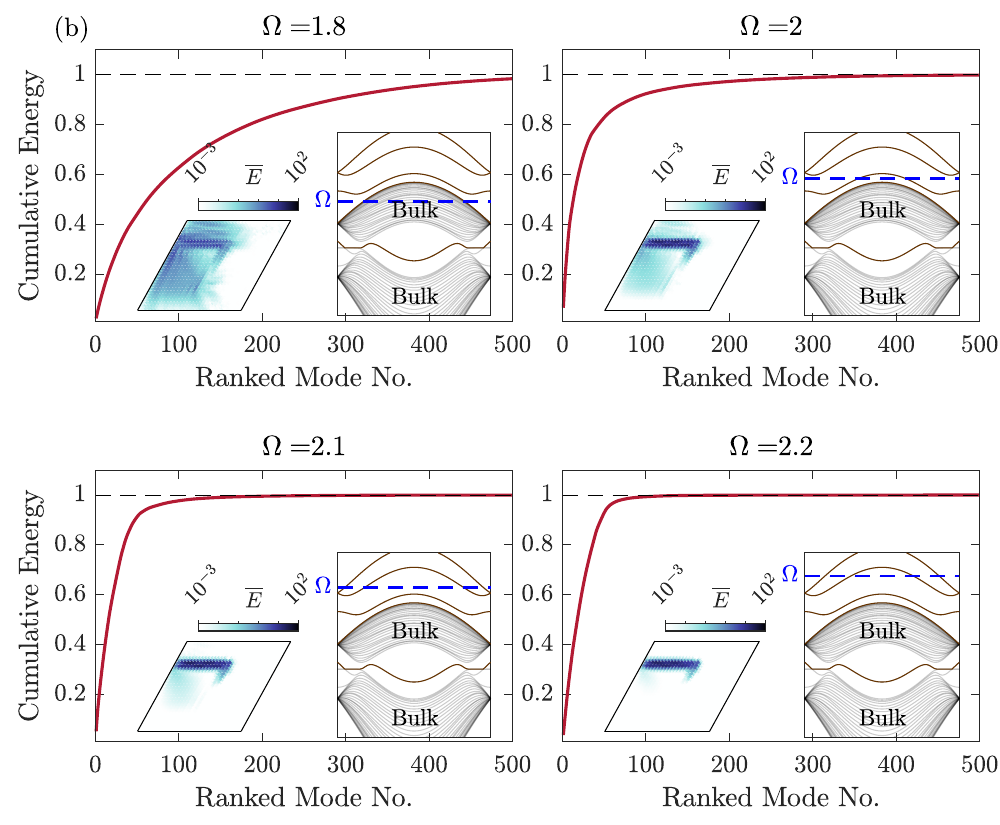}
		\includegraphics[width=.5\linewidth]{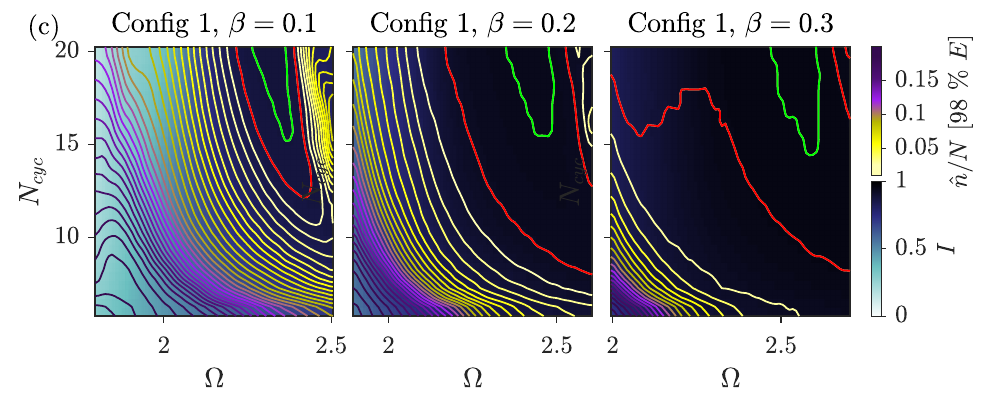}%
		\includegraphics[width=.5\linewidth]{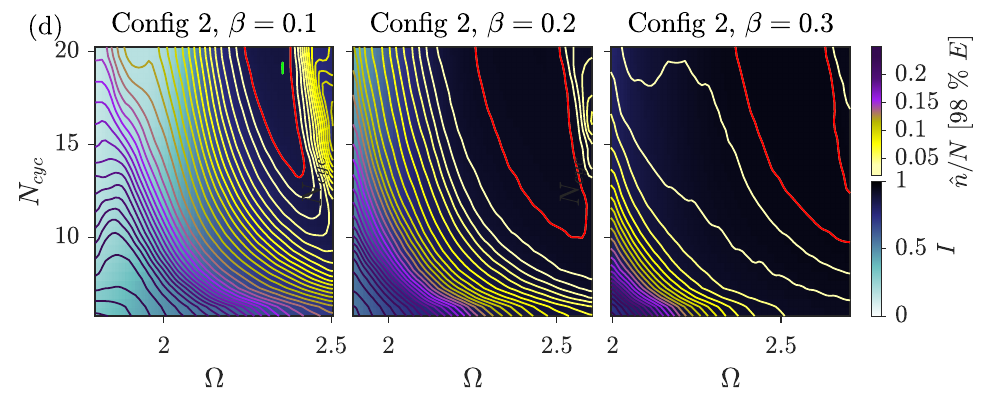}
		\caption{\textcolor{rev1}{Examples of the cumulative modal energy reconstruction for (a) configuration 1 and (b) configuration 2 of a trivial interlace model. 
				%Insets depict the time-averaged energy across the lattice as well as the supercell dispersion of the trivial models with brown lines denoting the localized edge bands and a blue dashed line denoting excitation frequency.
				The energy quantities (a) and (b) are normalized with respect to the total energy of the system.  The left insets depict the time-averaged energy for each oscillator. 
				The right insets depict a zoomed in view of the supercell band structure with brown lines denoting the trivial localized bads, gray lines denoting bulk bands, and a dashed blue line denoting the location of the excitation frequency with respect to the band structure.
				(c) and (d): The energy intensities at the interface for various $\beta$ between 0.1 and 0.3 as a function of input signal bandwidth and frequency for the trivial construction of configuration 1 and configuration 2, respectively.
				The superimposed contours depict the number of bulk modes of the modal decomposition  required  to reconstruct 98\% of the wave propagation. Green lines of (c,d) depict the parameter domain for which 1\% of modes account for the reconstruction, while red lines corresponding to  2\% of modes accounting for the reconstruction.}}
		\label{FIG:EvBW_triv}
	\end{figure}

	\textcolor{rev1}{
		The modal reconstruction properties of the trivial waveguide were further explored by considering a parametric study of excitation frequency and bandwidth, as well as the contrast parameter $\beta$, to provide a direct comparison to the results of section~\ref{Subsec: Modal Decomp}. Fig~\ref{FIG:EvBW_triv} depicts the cumulative energy of the rank ordered eigenmodes for trivial renditions of both configuration 1 and configuration 2 over a range  of frequencies. 
		The insets depict the trivial supercell, with the effective bandwidth of the trivial interface states appearing above the bulk bands. While the cumulative variance reaches unity much faster in the interface bandwidth, it is apparent that the energy accumulates less rapidly compared to the topological systems (cf.~Fig~\ref{FIG:E_vs_BW}). Moreover, the domain boundary seemingly now plays a role in the reconstruction as configuration 2 requires noticeably more modes than configuration 1 to converge.
	}

	\textcolor{rev1}{
		Figs~\ref{FIG:EvBW_triv}(c) and (d) depict the interface energy concentration along with the number of modes required to reconstruct the acoustics. 
		Contours depict the number of modes required to reconstruct 98\% of the modal energy for each solution, with green and red lines indicating the excitation domains for which only 1\% and 2\% of modes are required for the reconstruction, respectively.
		The interface energy concentration is clearly much higher for this trivial system than what was seen for a topological system, but this comes at the cost of backscattering around the corners of the domain boundary. 
		Moreover, the number of modes required to construct the dynamics of the trivial waveguide is generally higher than what was found for the topological systems (cf.~Fig~\ref{FIG:E_vs_BW}(a,b)). 
		Nevertheless, the straight-edge wave guide does in fact reach excitation domains whereby a 1\% modal reconstitution is sufficient. However, we note that this region is substantially smaller for the trivial waveguide. 
		More important is the fact that this same level of reconstruction is unachievable by the trivial configuration 2 waveguide, which is not fully describable by so few modes, irrespective of the excitation parameters.
	}

	\textcolor{rev1}{Figs~\ref{FIG:modal_decomp_waveguide} and \ref{FIG:EvBW_triv} allow us to draw a very important comparison to the results in section~\ref{Subsec: Modal Decomp}. 
		While the simplest of wave paths (configuration 1) yield comparable results between the trivial and topological systems, this is a special case for the waveguide with the simplest conceivable path. 
		Moreover, the accurate reconstruction of the trivial waveguides acoustics across \textit{arbitrarily complex} wave paths typically requires many more modes to sufficiently reconstruct the propagation, even when the wave motion along the domain boundary is highly confined as is the case for Figs~\ref{FIG:modal_decomp_waveguide} and \ref{FIG:EvBW_triv}.
		This is due in part to the effects of backscattering and delocalization of the wave around corners (delocalization in the sense that the confined wave packet breaks down, even if energy remains fixed at the domain boundaries), an effect which is not easily reproduced unless many modes are considered.
		In contrast, backscatter is of little concern for topological systems, and the wave retains its compactness and directionality around corners; therefore allowing the propagating wave packets in \textit{complex domains} to be described by a subset of closely spaced modes very accurately.
		\textit{Hence, while the findings reported in section~\ref{Subsec: Modal Decomp} may hold for specific cases of trivial wave guides, they are more generally applicable to systems with a topological band structure which gives rise to highly predictable localized waves that are immune to backscatter and delocalization. }
	}

	\clearpage
	\newpage
	\section{Density of Finite Modal Spectrum: Topological versus Trival Waveguide}
	\label{APX: DOS}
	% ---------------------------------------------------
	%\textcolor{rev1}{To confirm the spectral localization of the topological edge states, the distribution of modal frequencies was computed for each configuration. This allows us to confirm the already well-known result that the density of eigenfreqeuncies is far higher in the bulk spectrum than is for the topological edge states located in the band gap.%, as depicted in Fig~\ref{Fig:DOS} for the trivial configuration 0 as well as the nontrivial configurations 1 and 2. 
	%Fig~\ref{Fig:DOS} depicts the density of eigenfrequencies for the trivial bulk lattice ($\beta =0$) as well as for the topological configurations of configurations 1 and 2 with bulk modes indicated by red shading and interface modes with blue. The topological interface states are clearly localized in the band gap and with a relatively low spectral density for both configurations. This is key to reconstructing the interface states with so few eigenmodes, as the topological propagating wave requires effectively only the modes shaded in blue to accurately reconstruct the traveling wave packet. Notably, this is even for the system with zigzagged boundaries (configuration 2), so long as the wave propagates along the interface domain with relatively little backscatter.}

	\begin{figure}[b!]\centering
		\includegraphics[width=.48\linewidth]{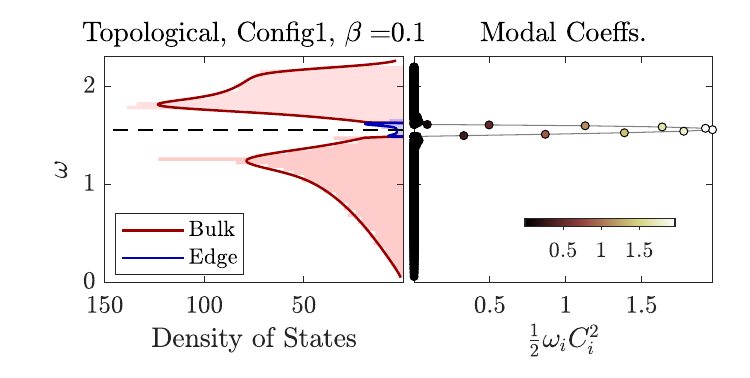}
		\includegraphics[width=.485\linewidth]{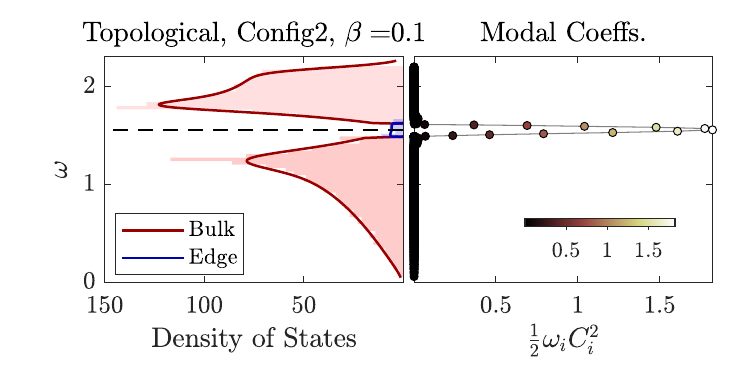}
		\includegraphics[width=.485\linewidth]{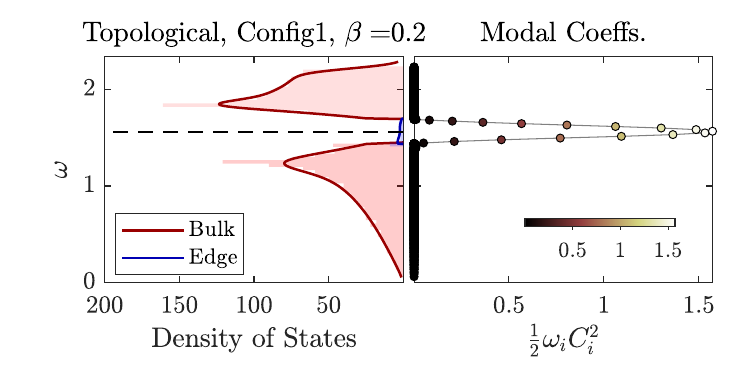}
		\includegraphics[width=.485\linewidth]{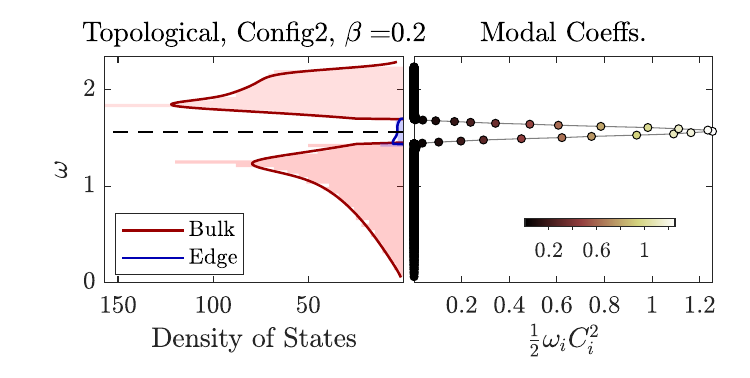}
		\includegraphics[width=.485\linewidth]{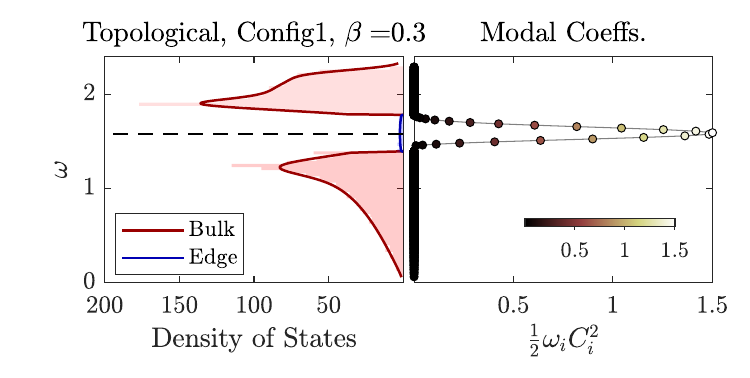}
		\includegraphics[width=.485\linewidth]{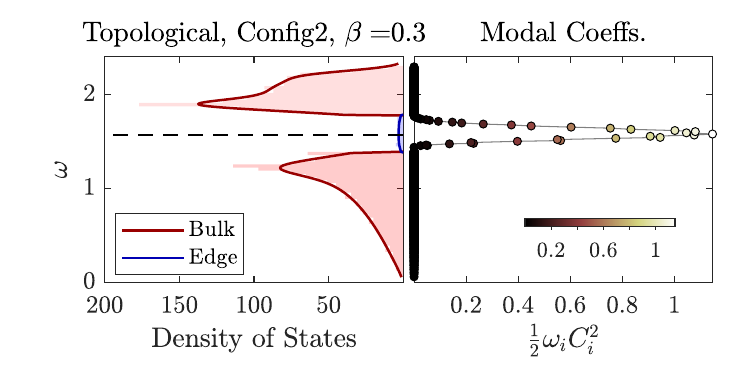}
		\caption{The density of eigenstates and resulting modal coefficients computed for topological wave propagation for $\beta=0.1$ and $\beta = 0.3$ subject to an excitation centered in the topological band gap (denoted with a black dashed lined) and $N_{cyc} = 12$.}
		\label{DOS:Top}
	\end{figure}
	
	\textcolor{rev1}{
		Fig~\ref{FIG:modal_decomp}(e,f) of the main text demonstrated the sparse density of eigenfrequencies within the topological band gap as well as the narrow band of excited modal coefficients for the topological lattice of configuration 2 (zigzagged domain). 
		For completeness, we have performed the same analysis for the configuration 1 (straight-edge) domain for contrast parameters of $\beta = 0.1$, 0.2, and 0.3. Moreover, the same analysis was also performed for the trivial waveguides discussed in~\ref{APX: Trivial} for a direct comparison.}
	%The modal decomposition (and moreover truncated modal reconstructions) of the topological wave travel are unique due to the spatial and spectral localization of the topological states can be further explored by comparing directly the spectrum of excited modes (similar to Figs~\ref{FIG:modal_decomp_trivial}(e,f) and \ref{FIG:modal_decomp}(e,f)}) over a range of contrast $\beta$ and for both straight edge and zigzagged domains.
	% This is done for the trivial waveguide discussed in~\ref{APX: Trivial} as well in order to juxtapose the 
	
	\textcolor{rev1}{
		Fig~\ref{DOS:Top} depicts the results for the topological system. For $\beta = 0.1$, the topological band gaps are obviously much smaller for both systems, however the density of interface states within this band gap is still substantially lower when compared to the bulk states, indicating that far fewer modes will be required to account for the total energy (if the excitation is sufficiently localized in this band gap).
		As $\beta$ grows, the localization becomes more pronounced, and the density of modes stays exceptionally low. Moreover, this is consistent between both configurations. 
		Looking at the modal coefficients that are recovered for an input frequency centered in the topological band gap to excite the topological wave, a clear narrow band distribution is noted with maximum modal energy centered at the excitation frequency, together with a sharp decline in energy for states not in the topological band gap.
		This drop off of energy upon entering the bulk bands is more pronounced for higher $\beta$ values, corresponding directly to an increase in spectral localization of the edge states. 
		This is again seen for both configurations, further confirming that the narrow-band modal reconstruction of the topological wave propagation is indifferent to the complexity of the domain boundary. 
	}

\begin{figure}[h!]\centering
		\includegraphics[width=.485\linewidth]{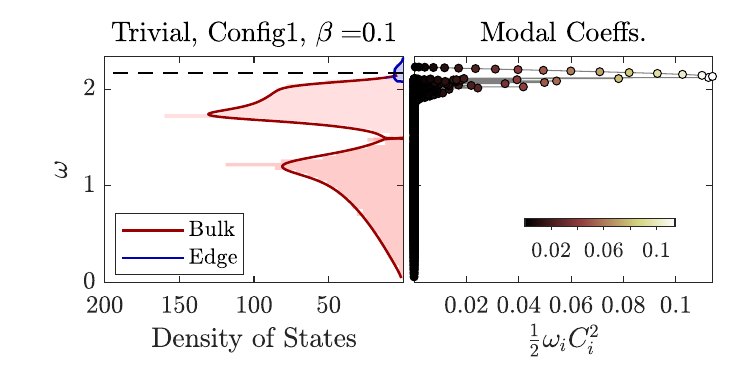}
		\includegraphics[width=.485\linewidth]{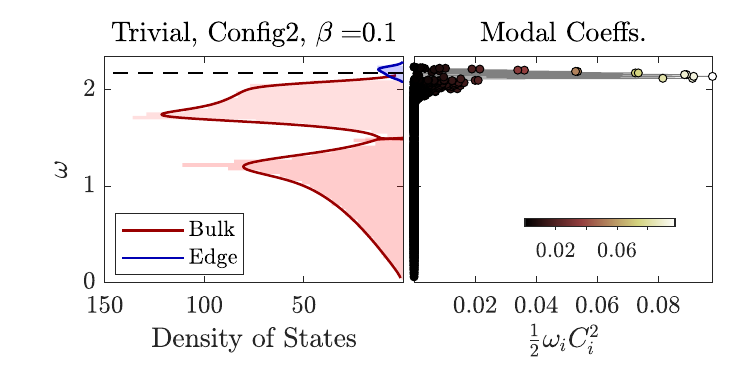}
		\includegraphics[width=.485\linewidth]{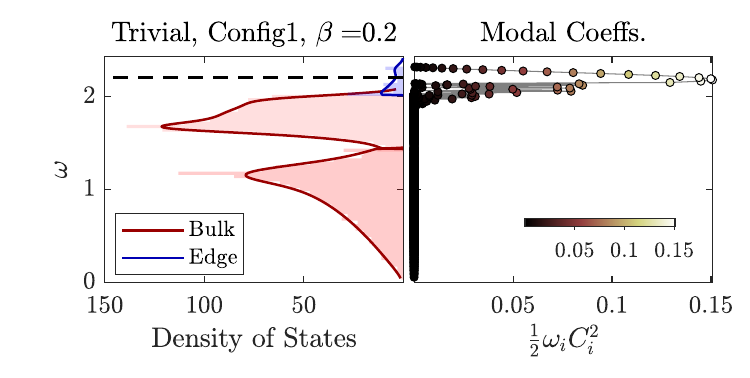}
		\includegraphics[width=.485\linewidth]{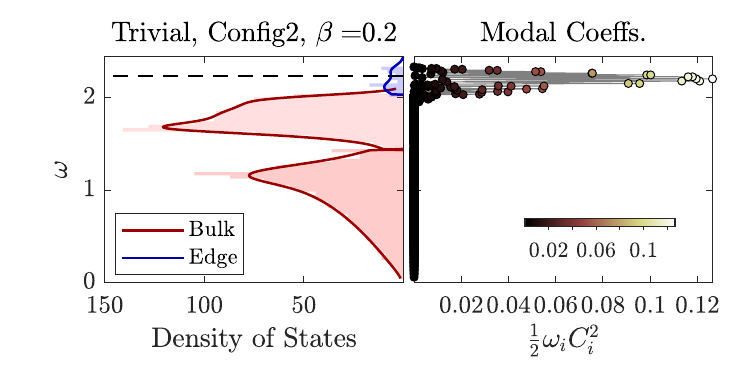}
		\includegraphics[width=.485\linewidth]{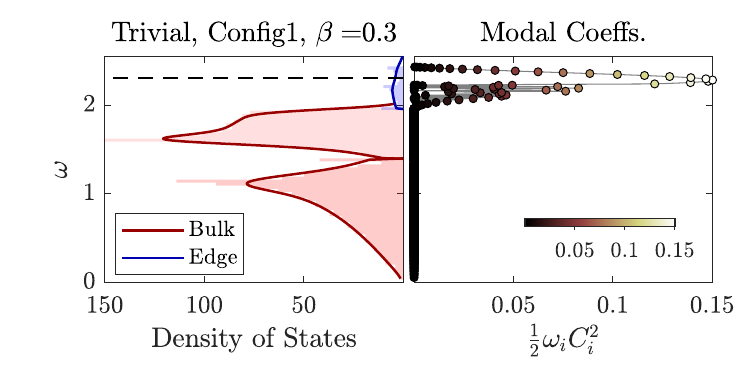}
		\includegraphics[width=.485\linewidth]{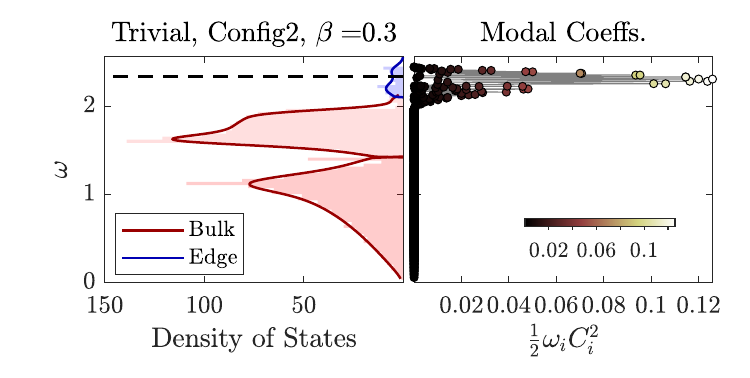}
		\caption{The density of eigenstates and resulting modal coefficients computed for trivial wave propagation for $\beta=0.1$ and $\beta = 0.3$ subject to an excitation frequency  centered on the subset of edge states existing above the bulk bands (denoted by a black dashed line) and $N_{cyc}=12$.}
		\label{DOS:Triv}
	\end{figure}

	\textcolor{rev1}{
		%These findings may be directly compared to 
		Fig~\ref{DOS:Triv} depicts the same analysis applied to the trivial waveguide configurations discussed in~\ref{APX: Trivial}.
		A main difference is that now the edge states form above the bulk bands which result from the geometric configuration of the lattice, and not from the topology of the band structure.  
		While the density of the eigenstates above the bulk bands is far lower than the density of bulk states, there is still an appreciably higher density than what was found for topological states within the band gap. 
		The modal coefficients recovered for Fig~\ref{DOS:Triv} were computed for inputs designed to excite the trivial localized wave by centering the input frequency about the mid point of the eigenfrequencies corresponding to trivial topological states. 
		It is clear that the  bandwidth of modal energy recovered for the trivial systems is larger as there is clearly more energy leaking into the bulk. Furthermore, no clearly recognizable energy distribution is recovered for the trivial modal spectrum.
		Moreover, this is exacerbated substantially for the zigzagged waveugide (configuration 2), indicating that the properties of the modal reconstruction are \textit{not} indifferent to the wave path for a trivial waveguide.
	}
	
	\textcolor{rev1}{
		The findings of Figs~\ref{DOS:Top} and~\ref{DOS:Triv}  are in agreement with what is reported in the main text and~\ref{APX: Trivial} where it has been shown that topological wave propagation can be accurately reconstructed with a very narrow subset of modes irrespective of the domain boundary (see section~\ref{Subsec: Modal Decomp}), and that trivial waveguides do not possess the same properties. 
		%%%
		More modes are required to reconstruct traveling waves in the trivial system, and  the path of the trivial waveguide plays a key role in the the  spectral energy distribution as backscattering introduces greater complexity (see~\ref{APX: Trivial}).
		Hence, Figs~\ref{DOS:Top} and~\ref{DOS:Triv} serve to provide key supporting evidence that the methods discussed in the main text are only \textit{generally applicable} to localized waves resulting from topological band properties.}
	
	% ---------------------------------------------------
	\newpage
	\section{Computational Advantages of the Truncated Modal Solution}
	\label{APX: CompTime}
	% ---------------------------------------------------

	\begin{figure}[b!]
		\includegraphics[width=\linewidth]{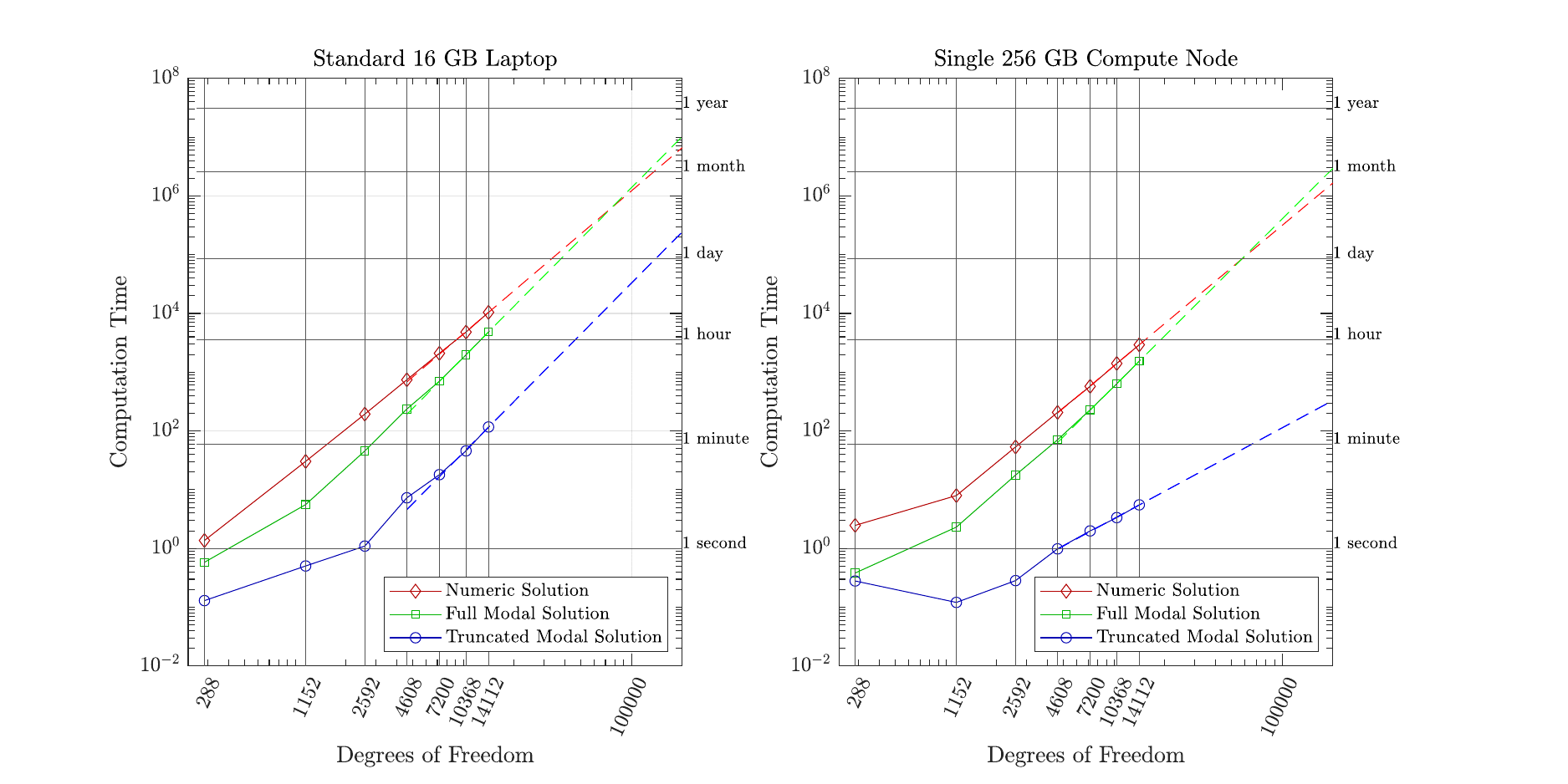}%
		\caption{\textcolor{rev1}{A comparison of time taken to solve the topological wave propagating through a discrete domain versus the lattice dimension using numerical integration, the full modal spectrum, and the truncated modal spectrum for both a standard laptop, and a single 256 GB compute node.}}
		\label{Fig:CompTime}
	\end{figure}

	\textcolor{rev1}{A unique advantage to describing the propagating topological wave by a small subset of eigenmodes comes in the form of greatly increased computational efficiency. Surely, for generic simulations of propagating waves, nearly all commercial software and research codes employ numerical integration schemes to resolve the  propagating wave packet with respect to time. While time integration may seem costly in comparison to solving the modal problem, it is in fact faster for realistically large systems ($O(10^5)$ to $O(10^6)$ DoF) than solving the entire eigenproblem and computing each coefficient for the modal solution. However, if it is known beforehand that only a \textit{narrow band} of the spectrum will be required to resolve the system, then it is in fact far more efficient to use the modal solution; it is for this reason that many steady-state excitation codes will in fact use the modal solution to resolve standing wave simulations.  This is because solving for only a portion of the modal spectrum (using numerical linear algebra tools) centered around the frequency of interest, and then summing only over the modes of the specified band, is far less costly for large systems than it is to solve the total modal solution or to perform direct numerical integration of the equations of motion. Typically, this is not possible for propagating wave solutions, as too much of the spectrum is required for this to be feasibly faster than direct integration.
	}

	\textcolor{rev1}{We provide here a brief computational study to emphasize this point in the context solving propagating topological wave packets in the discrete graphene model described in the main text. The finite lattice described my configuration 2 was considered for various dimensions between $12\times12$ to $84\times 84$ (288 DoF to 14112 DoF). Numerical solutions were recovered with the ODE45 routine in \texttt{\textsc{matlab}}. Full modal solutions were computed by directly solving the eigenvalue problem, and then solving for each coefficient, whereas the truncated modal solutions were solved by computing 5\% of the modes centered about the excitation frequency using the built-in \texttt{\textsc{matlab}} Krylov solver available with the \texttt{eigs} command. The required computation time to solve was then recorded for each solution using a standard Dell XPS laptop with 16 GB of RAM and an Intel(R) Core(TM) i7-9750H CPU @ 2.60GHz 2.59 GHz processor, as well as a standard compute node of the Illinois Campus Cluster with 256 GB of RAM. The computational times were extrapolated to larger systems ($O(10^5)$) using standard regression. Fig~\ref{Fig:CompTime} depicts the results, where it is confirmed that the modal solution indeed becomes more costly at higher DoFs, but that the truncated modal solution is always \textit{significantly} faster than either alternative. Using trivial (non-optimized) codes, we find that solutions of $O(10^5)$ can be resolved on the order of minutes with the truncated modal solution, whereas the full modal solution and numerical integration could take on the order of months for the same task. Hence, we demonstrate here the clear computational advantage of the truncated modal solutions presented in the main text in the context of rapidly solving large topological systems for propagating wave travel.
	}

	%\textcolor{rev1}{The same measure are computed for a trivial wave guide to demonstrate that the spectral localization is not the only factor at.}
	
	%%%%%%%%%%%%%%%%%%%%%%%%%%%%%%%%%%%%%%%%%%%%%%%%%%%%%%%%%%%%%%%%
	\bibliographystyle{elsarticle-num} 
	\bibliography{QVH_VG_R1}

\begin{thebibliography}{10}
\expandafter\ifx\csname url\endcsname\relax
  \def\url#1{\texttt{#1}}\fi
\expandafter\ifx\csname urlprefix\endcsname\relax\def\urlprefix{URL }\fi
\expandafter\ifx\csname href\endcsname\relax
  \def\href#1#2{#2} \def\path#1{#1}\fi

\bibitem{Ma2019}
G.~Ma, M.~Xiao, C.~T. Chan, Topological phases in acoustic and mechanical
  systems, Nature Reviews Physics 1~(4) (2019) 281--294.
\newblock \href {https://doi.org/10.1038/s42254-019-0030-x}
  {\path{doi:10.1038/s42254-019-0030-x}}.

\bibitem{Hasan2010}
M.~Z. Hasan, C.~L. Kane, \textit{Colloquium}: Topological insulators, Reviews
  of Modern Physics 82~(4) (2010) 3045--3067.
\newblock \href {https://doi.org/10.1103/revmodphys.82.3045}
  {\path{doi:10.1103/revmodphys.82.3045}}.

\bibitem{Suesstrunk2016}
R.~Süsstrunk, S.~D. Huber, Classification of topological phonons in linear
  mechanical metamaterials, Proceedings of the National Academy of Sciences
  113~(33) (aug 2016).
\newblock \href {https://doi.org/10.1073/pnas.1605462113}
  {\path{doi:10.1073/pnas.1605462113}}.

\bibitem{Vila2019}
J.~Vila, G.~H. Paulino, M.~Ruzzene, Role of nonlinearities in topological
  protection: Testing magnetically coupled fidget spinners, Physical Review B
  99~(12) (2019) 125116.
\newblock \href {https://doi.org/10.1103/physrevb.99.125116}
  {\path{doi:10.1103/physrevb.99.125116}}.

\bibitem{Pal2017}
R.~K. Pal, M.~Ruzzene, Edge waves in plates with resonators: an elastic
  analogue of the quantum valley hall effect, New Journal of Physics 19~(2)
  (2017) 025001.
\newblock \href {https://doi.org/10.1088/1367-2630/aa56a2}
  {\path{doi:10.1088/1367-2630/aa56a2}}.

\bibitem{Chen2018}
H.~Chen, H.~Nassar, G.~Huang, A study of topological effects in 1d and 2d
  mechanical lattices, Journal of the Mechanics and Physics of Solids 117
  (2018) 22--36.
\newblock \href {https://doi.org/10.1016/j.jmps.2018.04.013}
  {\path{doi:10.1016/j.jmps.2018.04.013}}.

\bibitem{Vila2017}
J.~Vila, R.~K. Pal, M.~Ruzzene, Observation of topological valley modes in an
  elastic hexagonal lattice, Physical Review B 96~(13) (2017) 134307.
\newblock \href {https://doi.org/10.1103/physrevb.96.134307}
  {\path{doi:10.1103/physrevb.96.134307}}.

\bibitem{Huo2021}
S.-Y. Huo, J.-J. Chen, H.-B. Huang, Y.-J. Wei, Z.-H. Tan, L.-Y. Feng, X.-P.
  Xie, Experimental demonstration of valley-protected backscattering
  suppression and interlayer topological transport for elastic wave in
  three-dimensional phononic crystals, Mechanical Systems and Signal Processing
  154 (2021) 107543.
\newblock \href {https://doi.org/10.1016/j.ymssp.2020.107543}
  {\path{doi:10.1016/j.ymssp.2020.107543}}.

\bibitem{Yang2015}
Z.~Yang, F.~Gao, X.~Shi, X.~Lin, Z.~Gao, Y.~Chong, B.~Zhang, Topological
  acoustics, Physical Review Letters 114~(11) (2015) 114301.
\newblock \href {https://doi.org/10.1103/physrevlett.114.114301}
  {\path{doi:10.1103/physrevlett.114.114301}}.

\bibitem{Kliewer2021}
E.~Kliewer, A.~Darabi, M.~J. Leamy, Additive manufacturing of channeled
  acoustic topological insulators, The Journal of the Acoustical Society of
  America 150~(4) (2021) 2461--2468.
\newblock \href {https://doi.org/10.1121/10.0006452}
  {\path{doi:10.1121/10.0006452}}.

\bibitem{Fleury2016}
R.~Fleury, A.~B. Khanikaev, A.~Al{\`{u}}, Floquet topological insulators for
  sound, Nature Communications 7~(1) (jun 2016).
\newblock \href {https://doi.org/10.1038/ncomms11744}
  {\path{doi:10.1038/ncomms11744}}.

\bibitem{Suesstrunk2015}
R.~Süsstrunk, S.~D. Huber, Observation of phononic helical edge states in a
  mechanical topological insulator, Science 349~(6243) (2015) 47--50.
\newblock \href {https://doi.org/10.1126/science.aab0239}
  {\path{doi:10.1126/science.aab0239}}.

\bibitem{Wang2015}
P.~Wang, L.~Lu, K.~Bertoldi, Topological phononic crystals with one-way elastic
  edge waves, Physical Review Letters 115~(10) (2015) 104302.
\newblock \href {https://doi.org/10.1103/physrevlett.115.104302}
  {\path{doi:10.1103/physrevlett.115.104302}}.

\bibitem{Nash2015}
L.~M. Nash, D.~Kleckner, A.~Read, V.~Vitelli, A.~M. Turner, W.~T.~M. Irvine,
  Topological mechanics of gyroscopic metamaterials, Proceedings of the
  National Academy of Sciences 112~(47) (2015) 14495--14500.
\newblock \href {https://doi.org/10.1073/pnas.1507413112}
  {\path{doi:10.1073/pnas.1507413112}}.

\bibitem{Miniaci2018}
M.~Miniaci, R.~Pal, B.~Morvan, M.~Ruzzene, Experimental observation of
  topologically protected helical edge modes in patterned elastic plates,
  Physical Review X 8~(3) (2018) 031074.
\newblock \href {https://doi.org/10.1103/physrevx.8.031074}
  {\path{doi:10.1103/physrevx.8.031074}}.

\bibitem{Babaa2020}
H.~A. Ba'ba'a, K.~Yu, Q.~Wang, Elastically-supported lattices for tunable
  mechanical topological insulators, Extreme Mechanics Letters 38 (2020)
  100758.
\newblock \href {https://doi.org/10.1016/j.eml.2020.100758}
  {\path{doi:10.1016/j.eml.2020.100758}}.

\bibitem{Xia2018}
J.-P. Xia, D.~Jia, H.-X. Sun, S.-Q. Yuan, Y.~Ge, Q.-R. Si, X.-J. Liu,
  Programmable coding acoustic topological insulator, Advanced Materials
  30~(46) (2018) 1805002.
\newblock \href {https://doi.org/10.1002/adma.201805002}
  {\path{doi:10.1002/adma.201805002}}.

\bibitem{Sirota2021}
L.~Sirota, D.~Sabsovich, Y.~Lahini, R.~Ilan, Y.~Shokef, Real-time steering of
  curved sound beams in a feedback-based topological acoustic metamaterial,
  Mechanical Systems and Signal Processing 153 (2021) 107479.
\newblock \href {https://doi.org/10.1016/j.ymssp.2020.107479}
  {\path{doi:10.1016/j.ymssp.2020.107479}}.

\bibitem{Tian2020}
Z.~Tian, C.~Shen, J.~Li, E.~Reit, H.~Bachman, J.~E.~S. Socolar, S.~A. Cummer,
  T.~J. Huang, Dispersion tuning and route reconfiguration of acoustic waves in
  valley topological phononic crystals, Nature Communications 11~(1) (feb
  2020).
\newblock \href {https://doi.org/10.1038/s41467-020-14553-0}
  {\path{doi:10.1038/s41467-020-14553-0}}.

\bibitem{Zhuo2021}
L.~Zhuo, H.~He, R.~Huang, S.~Su, Z.~Lin, W.~Qiu, B.~Huang, Q.~Kan, Group
  velocity modulation and light field focusing of the edge states in chirped
  valley graphene plasmonic metamaterials, Nanomaterials 11~(7) (2021) 1808.
\newblock \href {https://doi.org/10.3390/nano11071808}
  {\path{doi:10.3390/nano11071808}}.

\bibitem{ZangenehNejad2019}
F.~Zangeneh-Nejad, R.~Fleury, Topological analog signal processing, Nature
  Communications 10~(1) (may 2019).
\newblock \href {https://doi.org/10.1038/s41467-019-10086-3}
  {\path{doi:10.1038/s41467-019-10086-3}}.

\bibitem{Su1979}
W.~P. Su, J.~R. Schrieffer, A.~J. Heeger, Solitons in polyacetylene, Physical
  Review Letters 42~(25) (1979) 1698--1701.
\newblock \href {https://doi.org/10.1103/physrevlett.42.1698}
  {\path{doi:10.1103/physrevlett.42.1698}}.

\bibitem{Yin2018}
J.~Yin, M.~Ruzzene, J.~Wen, D.~Yu, L.~Cai, L.~Yue, Band transition and
  topological interface modes in 1d elastic phononic crystals, Scientific
  Reports 8~(1) (may 2018).
\newblock \href {https://doi.org/10.1038/s41598-018-24952-5}
  {\path{doi:10.1038/s41598-018-24952-5}}.

\bibitem{Tempelman2021}
J.~R. Tempelman, K.~H. Matlack, A.~F. Vakakis, Topological protection in a
  strongly nonlinear interface lattice, Physical Review B 104~(17) (2021)
  174306.
\newblock \href {https://doi.org/10.1103/physrevb.104.174306}
  {\path{doi:10.1103/physrevb.104.174306}}.

\bibitem{Hu2022}
G.~Hu, C.~Lan, L.~Tang, Y.~Yang, Deep-subwavelength interface states in
  mechanical systems, Mechanical Systems and Signal Processing 169 (2022)
  108598.
\newblock \href {https://doi.org/10.1016/j.ymssp.2021.108598}
  {\path{doi:10.1016/j.ymssp.2021.108598}}.

\bibitem{Chen2021a}
C.-W. Chen, R.~Chaunsali, J.~Christensen, G.~Theocharis, J.~Yang, Corner states
  in a second-order mechanical topological insulator, Communications Materials
  2~(1) (jun 2021).
\newblock \href {https://doi.org/10.1038/s43246-021-00170-x}
  {\path{doi:10.1038/s43246-021-00170-x}}.

\bibitem{SerraGarcia2018}
M.~Serra-Garcia, V.~Peri, R.~Süsstrunk, O.~R. Bilal, T.~Larsen, L.~G.
  Villanueva, S.~D. Huber, Observation of a phononic quadrupole topological
  insulator, Nature 555~(7696) (2018) 342--345.
\newblock \href {https://doi.org/10.1038/nature25156}
  {\path{doi:10.1038/nature25156}}.

\bibitem{Chen2021}
Y.~Chen, F.~Meng, X.~Huang, Creating acoustic topological insulators through
  topology optimization, Mechanical Systems and Signal Processing 146 (2021)
  107054.
\newblock \href {https://doi.org/10.1016/j.ymssp.2020.107054}
  {\path{doi:10.1016/j.ymssp.2020.107054}}.

\bibitem{Rosa2022}
M.~I.~N. Rosa, M.~J. Leamy, M.~Ruzzene, Amplitude-dependent edge states and
  discrete breathers in nonlinear modulated phononic lattices (Jan. 2022).
\newblock \href {http://arxiv.org/abs/2201.05526} {\path{arXiv:2201.05526}}.

\bibitem{Chaunsali2021}
R.~Chaunsali, H.~Xu, J.~Yang, P.~G. Kevrekidis, G.~Theocharis, Stability of
  topological edge states under strong nonlinear effects, Physical Review B
  103~(2) (2021) 024106.
\newblock \href {https://doi.org/10.1103/physrevb.103.024106}
  {\path{doi:10.1103/physrevb.103.024106}}.

\bibitem{Pal2018}
R.~K. Pal, J.~Vila, M.~Leamy, M.~Ruzzene, Amplitude-dependent topological edge
  states in nonlinear phononic lattices, Physical Review E 97~(3) (2018)
  032209.
\newblock \href {https://doi.org/10.1103/physreve.97.032209}
  {\path{doi:10.1103/physreve.97.032209}}.

\bibitem{Miniaci2019}
M.~Miniaci, R.~K. Pal, R.~Manna, M.~Ruzzene, Valley-based splitting of
  topologically protected helical waves in elastic plates, Physical Review B
  100~(2) (2019) 024304.
\newblock \href {https://doi.org/10.1103/physrevb.100.024304}
  {\path{doi:10.1103/physrevb.100.024304}}.

\bibitem{Wang2019}
W.~Wang, B.~Bonello, B.~Djafari-Rouhani, Y.~Pennec, Topological valley,
  pseudospin, and pseudospin-valley protected edge states in symmetric pillared
  phononic crystals, Physical Review B 100~(14) (2019) 140101.
\newblock \href {https://doi.org/10.1103/physrevb.100.140101}
  {\path{doi:10.1103/physrevb.100.140101}}.

\bibitem{ZangenehNejad2020}
F.~Zangeneh-Nejad, A.~Al{\`{u}}, R.~Fleury, Topological wave insulators: a
  review, Comptes Rendus. Physique (2020) 1--33\href
  {https://doi.org/10.5802/crphys.3} {\path{doi:10.5802/crphys.3}}.

\bibitem{Miniaci2021}
M.~Miniaci, R.~K. Pal, Design of topological elastic waveguides, Journal of
  Applied Physics 130~(14) (2021) 141101.
\newblock \href {https://doi.org/10.1063/5.0057288}
  {\path{doi:10.1063/5.0057288}}.

\bibitem{Neto2009}
A.~H.~C. Neto, F.~Guinea, N.~M.~R. Peres, K.~S. Novoselov, A.~K. Geim, The
  electronic properties of graphene, Reviews of Modern Physics 81~(1) (2009)
  109--162.
\newblock \href {https://doi.org/10.1103/revmodphys.81.109}
  {\path{doi:10.1103/revmodphys.81.109}}.

\bibitem{Berry1984}
M.~Berry, Quantal phase factors accompanying adiabatic changes, Proceedings of
  the Royal Society of London. A. Mathematical and Physical Sciences 392~(1802)
  (1984) 45--57.
\newblock \href {https://doi.org/10.1098/rspa.1984.0023}
  {\path{doi:10.1098/rspa.1984.0023}}.

\bibitem{Thouless1982}
D.~J. Thouless, M.~Kohmoto, M.~P. Nightingale, M.~den Nijs, Quantized hall
  conductance in a two-dimensional periodic potential, Physical Review Letters
  49~(6) (1982) 405--408.
\newblock \href {https://doi.org/10.1103/physrevlett.49.405}
  {\path{doi:10.1103/physrevlett.49.405}}.

\bibitem{Rudner2013}
M.~S. Rudner, N.~H. Lindner, E.~Berg, M.~Levin, Anomalous edge states and the
  bulk-edge correspondence for periodically driven two-dimensional systems,
  Physical Review X 3~(3) (2013) 031005.
\newblock \href {https://doi.org/10.1103/physrevx.3.031005}
  {\path{doi:10.1103/physrevx.3.031005}}.

\bibitem{Du2020}
Z.~Du, H.~Chen, G.~Huang, Optimal quantum valley hall insulators by rationally
  engineering berry curvature and band structure, Journal of the Mechanics and
  Physics of Solids 135 (2020) 103784.
\newblock \href {https://doi.org/10.1016/j.jmps.2019.103784}
  {\path{doi:10.1016/j.jmps.2019.103784}}.

\bibitem{Christiansen2019}
R.~E. Christiansen, F.~Wang, O.~Sigmund, Topological insulators by topology
  optimization, Physical Review Letters 122~(23) (2019) 234502.
\newblock \href {https://doi.org/10.1103/physrevlett.122.234502}
  {\path{doi:10.1103/physrevlett.122.234502}}.

\bibitem{Zhang2020}
Z.~Zhang, F.~Li, J.~Lu, T.~Liu, X.~Heng, Y.~He, H.~Liang, J.~Gan, Z.~Yang,
  Broadband photonic topological insulator based on triangular-holes array with
  higher energy filling efficiency, Nanophotonics 9~(9) (2020) 2839--2846.
\newblock \href {https://doi.org/10.1515/nanoph-2020-0086}
  {\path{doi:10.1515/nanoph-2020-0086}}.

\bibitem{Ma2021}
J.~Ma, C.~Ouyang, L.~Niu, Q.~Wang, J.~Zhao, Y.~Liu, L.~Liu, Q.~Xu, Y.~Li,
  J.~Gu, Z.~Tian, J.~Han, W.~Zhang, Topological edge state bandwidth tuned by
  multiple parameters in two-dimensional terahertz photonic crystals with
  metallic cross structures, Optics Express 29~(20) (2021) 32105.
\newblock \href {https://doi.org/10.1364/oe.440121}
  {\path{doi:10.1364/oe.440121}}.

\bibitem{Zeuner2015}
J.~M. Zeuner, M.~C. Rechtsman, Y.~Plotnik, Y.~Lumer, S.~Nolte, M.~S. Rudner,
  M.~Segev, A.~Szameit, Observation of a topological transition in the bulk of
  a non-hermitian system, Physical Review Letters 115~(4) (2015) 040402.
\newblock \href {https://doi.org/10.1103/physrevlett.115.040402}
  {\path{doi:10.1103/physrevlett.115.040402}}.

\bibitem{Zhu2018}
H.~Zhu, T.-W. Liu, F.~Semperlotti, Design and experimental observation of
  valley-hall edge states in diatomic-graphene-like elastic waveguides,
  Physical Review B 97~(17) (2018) 174301.
\newblock \href {https://doi.org/10.1103/physrevb.97.174301}
  {\path{doi:10.1103/physrevb.97.174301}}.

\bibitem{Xiong2016}
Y.~Xiong, T.~Wang, P.~Tong, The effects of dissipation on topological
  mechanical systems, Scientific Reports 6~(1) (sep 2016).
\newblock \href {https://doi.org/10.1038/srep32572}
  {\path{doi:10.1038/srep32572}}.

\bibitem{Auld1971}
B.~Auld, G.~Kin, Normal mode theory for acoustic waves and its application to
  the interdigital transducer, {IEEE} Transactions on Electron Devices 18~(10)
  (1971) 898--908.
\newblock \href {https://doi.org/10.1109/t-ed.1971.17303}
  {\path{doi:10.1109/t-ed.1971.17303}}.

\bibitem{Ditri1994}
J.~J. Ditri, J.~L. Rose, Excitation of guided waves in generally anisotropic
  layers using finite sources, Journal of Applied Mechanics 61~(2) (1994)
  330--338.
\newblock \href {https://doi.org/10.1115/1.2901449}
  {\path{doi:10.1115/1.2901449}}.

\bibitem{Auld}
B.~A. Auld, Acoustic Fields and Waves in Solids, 2 Vol. Set, Krieger Publishing
  Company.

\bibitem{Xin2020}
L.~Xin, Y.~Siyuan, L.~Harry, L.~Minghui, C.~Yanfeng, Topological mechanical
  metamaterials: A brief review, Current Opinion in Solid State and Materials
  Science 24~(5) (2020) 100853.
\newblock \href {https://doi.org/10.1016/j.cossms.2020.100853}
  {\path{doi:10.1016/j.cossms.2020.100853}}.

\bibitem{Qian2018}
K.~Qian, D.~J. Apigo, C.~Prodan, Y.~Barlas, E.~Prodan, Topology of the
  valley-chern effect, Physical Review B 98~(15) (2018) 155138.
\newblock \href {https://doi.org/10.1103/physrevb.98.155138}
  {\path{doi:10.1103/physrevb.98.155138}}.

\bibitem{Eisenberg2019}
Y.~Eisenberg, Y.~Barlas, E.~Prodan, Valley chern effect with $lc$ resonators: A
  modular platform, Physical Review Applied 11~(4) (2019) 044077.
\newblock \href {https://doi.org/10.1103/physrevapplied.11.044077}
  {\path{doi:10.1103/physrevapplied.11.044077}}.

\bibitem{Rose2000}
J.~Rose, Guided wave nuances for ultrasonic nondestructive evaluation, {IEEE}
  Transactions on Ultrasonics, Ferroelectrics and Frequency Control 47~(3)
  (2000) 575--583.
\newblock \href {https://doi.org/10.1109/58.842044}
  {\path{doi:10.1109/58.842044}}.

\bibitem{Hussein2010}
M.~I. Hussein, M.~J. Frazier, Band structure of phononic crystals with general
  damping, Journal of Applied Physics 108~(9) (2010) 093506.
\newblock \href {https://doi.org/10.1063/1.3498806}
  {\path{doi:10.1063/1.3498806}}.

\end{thebibliography}
	%%%%%%%%%%%%%%%%%%%%%%%%%%%%%%%%%%%%%%%%%%%%%%%%%%%%%%%%%%%%%%%%

\end{document}